%% file: main_article.tex
\documentclass[3p,onecolumn, final, times]{elsarticle}

\usepackage{amssymb}
\usepackage{amsmath}
\usepackage{color}
\usepackage{lineno}
\usepackage{physics}
\usepackage{hyperref}
\usepackage{multicol}
\usepackage{framed}
\usepackage{nomencl}
\usepackage{tikz}
\usepackage{mathdots}
\usepackage{yhmath}
\usepackage{cancel}
\usepackage{float}
\usepackage{color}
\usepackage{siunitx}
\usepackage{array}
\usepackage{multirow}
\usepackage{amssymb}
\usepackage{amsfonts}
\usepackage{tabularx} 
\usepackage{extarrows}
\usepackage{booktabs} 
\usepackage{comment}
\usepackage{cleveref}
\usepackage{float} 
\usepackage{subcaption}
\usepackage{tablefootnote}
\usepackage{tikzit}
\input{figures/st.tikzstyles}
\usepackage{etoolbox}
\renewcommand\nomgroup[1]{%
  \item[\bfseries
  \ifstrequal{#1}{P}{Subscripts and superscripts}{%
  {%
  }}%
]}
\makenomenclature
\renewcommand*\nompreamble{\begin{multicols}{1}}
\renewcommand*\nompostamble{\end{multicols}}

\usetikzlibrary{fadings}
\usetikzlibrary{patterns}
\usetikzlibrary{shadows.blur} 
\usetikzlibrary{shapes}
\journal{ Journal of Computational Physics}
\begin{document}

\begin{frontmatter}






\title{Multiscale Unsteady Conjugate Transfer via Modal Projection} 

\author[label1]{Yann Dreze\corref{cor1}} 
\cortext[cor1]{yann.dreze@engs.ox.ac.uk} 
\author[label1]{Muting Hao}
\author[label1]{Luca di Mare}

\affiliation[label1]{organization={Department of Engineering Science, Oxford University}, 
            addressline={ Oxford Thermofluids Institute, Southwell Building },  
            city={Oxford},
            postcode={OX20ES}, 
            country={United Kingdom}}


\begin{abstract} 
  This paper presents a multiscale methodology for efficient unsteady conjugate heat transfer simulations. 
  The solid domain is modelled by coupling a global representation of the  temperature field, based on the eigenfunctions of the unsteady heat conduction equation, with a local, fine-scale-resolving solution of the heat conduction equation at the conjugate interface. 
  To address the disparate time scales and enhance convergence, the decoupled modal equations are leveraged to enable targeted acceleration of the longest thermal time scales. 
  One-dimensional analyses validate the  properties of the scheme, while scale-resolving simulations demonstrate its practical application for steady and unsteady problems. Notably, the method achieves up to a fourfold reduction in computational time to reach steady thermal conditions compared to conventional conjugate simulations, without introducing significant computational overhead or error, offering an accurate and accelerated framework for unsteady thermal analysis.

\end{abstract}

\begin{highlights}
  \item  A modal decomposition of the solid temperature field is proposed, which allows for an efficient representation of the unsteady heat conduction problem.
  \item The modal representation reduces the solid domain to a set of uncoupled, single-degree-of-freedom ordinary differential equations, which require only surface integrals to be computed
  \item To address the timescale challenge, the thermal transients of the 
  modal equations are accelerated individually based on the respective time constants of the modes. 
  \item The method shows a fourfold reduction in time to steady state for the slowest modes, with an associated error of 6\%.
\end{highlights}
  
 
\end{frontmatter}
 
 
\nomenclature{$\rho$}{Density}
\nomenclature{$c_p$}{Specific heat capacity}
\nomenclature{$\mu$}{Dynamic viscosity}
\nomenclature{$\kappa$}{Thermal conductivity}
\nomenclature{$T$}{Temperature}
\nomenclature{$\sigma$}{Acceleration factor}
\nomenclature{$\delta$}{Channel half-width}
\nomenclature{$\lambda$}{Eigenvalue}
\nomenclature{$L_x$}{Streamwise length}
\nomenclature{$L_z$}{Spanwise length}
\nomenclature{$s$}{Laplace variable}
\nomenclature{$H$}{Laplace transfer function}
\nomenclature{$L_y$}{Transversal length}
\nomenclature{$\tau$}{Time scale}
\nomenclature{$\delta_b$}{Boundary layer thickness}
\nomenclature{$\omega$}{Angular frequency}
\nomenclature{$\text{Re}_\tau$}{Reynolds number based on friction velocity}
\nomenclature{$\text{M}$}{Mach number}
\nomenclature{$\text{Pr}$}{Prandtl number} 
\nomenclature{$\Delta y^+$}{Wall-normal grid spacing in viscous sub-layer}
\nomenclature{St}{Stanton number}
\nomenclature{$\text{Re}_D$}{Reynolds number based on pipe diameter}
\nomenclature{$T_{\infty}$}{Outer flow temperature}
\nomenclature{$\chi, \eta, \zeta$}{Local geodesic coordinates}
\nomenclature{$x, y, z$}{Global Cartesian coordinates}
\nomenclature{$R_c$}{Radius of curvature}
\nomenclature{$\phi$}{Angle}
\nomenclature{$\alpha$}{Thermal diffusivity}
\nomenclature{$F$}{Shape function}
\nomenclature{$\Omega$}{Domain}
\nomenclature{$\xi$}{Dimensionless coordinate}
\nomenclature{$\Gamma$}{Boundary}
\nomenclature{$Q_s$}{Heat source}
\nomenclature{$\chi_a$}{scaling factor for SFD}
\nomenclature{$\vec{n}$}{Normal vector}
\nomenclature{CFD}{Computational Fluid Dynamics}
\nomenclature{CHT}{Conjugate Heat Transfer}
\nomenclature{DNS}{Direct Numerical Simulation}
\nomenclature{LES}{Large Eddy Simulation}
\nomenclature{SFD}{Selective Frequency Damping}
\nomenclature{$U$}{Modal coefficients}
\nomenclature{$\alpha$}{Thermal diffusivity}
\nomenclature{$\Theta$}{Transformed temperature}
\nomenclature{$F_{\text{str}}$}{Strouhal Frequency}
\nomenclature{Str}{Strouhal number}
\nomenclature{N-S}{Navier-Stokes equations}
\nomenclature{$\mathcal{L}$}{Laplace operator}
\nomenclature{$\textit{FE}$}{Finite Element Solution}
\nomenclature{$\textit{MFE}$}{Modal Finite Element Solution}
\nomenclature{$\textit{CFE}$}{Combined Finite Element Solution}
\nomenclature{$\textit{CFEA}$}{Combined Finite Element Accelerated}
\nomenclature{URANS}{Unsteady Reynolds-averaged N-S}
\nomenclature{$\Delta r^+$}{Wall-normal grid spacing in radial direction}
\nomenclature{$C_{f,\theta}$}{Local skin friction coefficient}
\nomenclature{Nu$_{\theta}$}{Local Nusselt number}
\nomenclature{R$_{\text{in}}$,R$_{\text{out}}$}{Inner and outer radius}

\begin{table*}[!t]
  \begin{framed}
    \printnomenclature
  \end{framed}
\end{table*}

\section{Introduction}
\input{intro.tex}

\section{Method}
\input{method.tex}

\input{timeintegration.tex}

\section{Results}
\input{results.tex}

\section{Conclusion}
In this work, we have presented the theoretical foundation of a  multiscale framework for efficient unsteady high-fidelity conjugate transfer simulations.  
The proposed method addresses the common challenge of mismatched time and length scales in CHT problems by combining a modal decomposition with a scale-resolving mesh at the fluid-solid interface. 

This hybrid approach enables the separation of long and short scale thermal fluctuations within the solid domain. Large-scale thermal behavior is captured through the modal projection, while finer, near-interface features are resolved using a scale-resolving  mesh.
Continuity of both temperature and heat flux is maintained across the fluid-solid interface, as well as across the region where the scale-resolving mesh overlaps with the modal projection, ensuring the  accuracy of the method.

The timescale mismatch is addressed by leveraging the uncoupled modal equations, the approach to steady state of the overall solution can be accelerated without significant error by only altering the mode whose approach to steady-state is longer than the allowable simulation time.   
In this study, the acceleration strategy is implemented by adjusting the coefficients of the modal equations; however, it is worth noting that other acceleration techniques can also be integrated effectively within the modal decomposition framework.
The method has been rigorously validated against DNS data for a turbulent channel flow and a pipe in cross flow.
Results from the pipe case demonstrate that the acceleration technique can reduce the time to steady state for the slowest modes by up to a factor of 8, with the associated error kept within 16\%.
\section{Acknowledgements}
The authors are grateful for the support from Rolls-Royce 
PLC and the EPSRC Center for Doctoral Training in Future 
Propulsion and Power. The authors would also like to acknowledge the use of the Cirrus UK National Tier-2 HPC Service at
EPCC (\url{http://www.cirrus.ac.uk}) funded by the University of Edinburgh and EPSRC (EP/P020267/1).
\appendix

\section{ Heat conduction in thin layers}\label{sec:thinlayer}
\input{thinlayer.tex}
\bibliographystyle{elsarticle-num-names} 
\bibliography{main}
%


%
%
%
%
\end{document}

%% file: figures/st.tikzstyles
\tikzstyle{rotated}=[fill=none, draw=none, rotate=-6.3]
\tikzstyle{Blue_point}=[fill={rgb,255: red,20; green,8; blue,148}, draw=none, shape=circle, inner sep=0pt, minimum size=3pt]
\tikzstyle{blue text}=[fill=none, draw=none, shape=circle]
\tikzstyle{Black node}=[fill={rgb,255: red,7; green,7; blue,7}, draw=none, shape=circle, inner sep=0pt, minimum size=4pt]

\tikzstyle{new edge style 0}=[-, fill={rgb,255: red,64; green,47; blue,255}, line width=1pt, draw opacity=0.3, nearly transparent]
\tikzstyle{new edge style 1}=[draw=black, fill=none, ->, line width=1pt]
\tikzstyle{Black edge}=[-, line width=1pt]
\tikzstyle{blue edge}=[-, draw={rgb,255: red,8; green,27; blue,149}]
\tikzstyle{gray edge}=[-, draw={rgb,255: red,131; green,131; blue,131}]
\tikzstyle{new edge style 2}=[-, line width=1 pt]

%% file: intro.tex
Over the years, the development of computer capabilities has increased the popularity of large-eddy simulations (LES) and direct numerical simulations (DNS). For such scale-resolving techniques, while the precision is dictated by the computational resources available, the physical accuracy of the result is strongly influenced by the boundary conditions imposed.
Simple boundary conditions, such as constant Dirichlet or Neumann may not be sufficient to accurately model real-world problems which often involve interactions across multiple physical domains.
Indeed, in recent years, the scope of CFD has evolved recognising the importance of fluid-structure interactions \cite{ hou2024unsteady,FAN2024112584,doi:10.2514/6.2025-1236,mourato2022computational}.

In the context of the thermal boundary condition, most of the fluid prediction methods frequently used in research and in industry do not take into account conjugate heat transfer (CHT) — the thermal interaction between fluid and ajacent solid.
The boundary conditions between solid walls and the fluid domain are usually specified as a fixed heat flux or a fixed temperature. However, there is a direct interest in metal temperature distribution in presence of fluid flow with large temperature variations. This is because the heat transfer  and the temperature gradients  between the fluid and solid  can significantly influence the flow and the simple specifications commonly used are not accurate. A classic example is that of high pressure turbines rows, where accurate modelling of  the thermal operating conditions requires a conjugate heat transfer analysis as shown by  \cite{10.1115/98-GT-088,jj,Dunni}.
 
Since the pionerring works of \citet{PERELMAN1961293}, CHT has now become a critical aspect of many engineering applications. 
Its significance spans from microscopic levels, such as near-wall turbulence \cite{SHARMA2019108430} or the use of nanofluids for improved heat transfer \cite{osti_196525}, to macroscale systems such as thermal management in spacecraft, insulation in nuclear reactors, cooling of turbine blades, and thermal regulation in battery technology, \cite{REvhs, john2018applied}.



\label{sec:challenges}
Coupling domains with distinct governing equations presents significant modeling challenges. These challenges stem from the different physical processes, conduction dominated in solids and convection dominated in most fluids, which operate on different spatial and temporal scales. This disparity complicates the achievement of accurate and efficient simulations.

Firstly, in transient CHT problems, approach to a steady-state takes place at signficiantly different rates in the solid and fluid domain, the mismatch in time scales being potentially very large.   
\cite{YANN} presented a dimensional analysis of the time scales in CHT problems. 
The analysis demonstrates that the timescale ratio $\tau_s/\tau_f$ between convection and conduction is given by the following \autoref{eq:time_scale_ratio}:
\begin{equation}
    \label{eq:time_scale_ratio}
     \frac{\tau_s}{\tau_f}= \frac{\kappa_f\rho_f c_{p,f}}{\kappa_s\rho_s c_{p,s}}  \text{St} \,\text{Re}^a \Bigg/ \left({\frac{\text{Re}^{-1+2a}}{\text{Pr}}}\right) 
 \end{equation}
The timescale ratio $\tau_s/\tau_f$ is a function of the thermal effusivity $\kappa \rho c_p$, which characteristices the rate at which heat is absorbed, stored, and conducted away from an interface, \cite{schultz1973heat, tiselj2012dns}. The timescale ratio also depends  on the Stanton number (St), the Prandtl number (Pr). The properties of the boundary layer near the solid/fluid interface also affect the timescale ratio through the scaling law of the thickness of the thermal boundary layer (Re$^a$). For a typical air-steel system, $\tau_s/\tau_f \approx 10^4$. Similarly, large values are found in most gas/metal interfaces of practical interest. 
Large values of the ratio $\tau_s/\tau_f$ indicate that progress towards the attainement of a steady temperature distribution in the conjugate system is dominated by the thermal transient in the metal. In conjugate heat transfer simulations with large solid domains, this may require very long simulation times for a true steady state to be reached \cite{Hickling,KOREN2017340}. Reconciling these timescales to ensure a statistical steady state remains problematic. 
  
Secondly, the temperature distribution in the solid and fluid domains may exhibit drastically different length scales. The thermal boundary layers thicknesses in the solid and fluid domains stand in a ratio dictated by the ratio of the thermal conductivities \cite{schultz1973heat,YANN}, as shown in \autoref{eq:interfacefinal}:
\begin{equation}
    \label{eq:interfacefinal}
  \frac{\delta_s}{\delta_f} \simeq  \mathcal{O}\left( \frac{\kappa_s}{\kappa_f}\right)
\end{equation}
As an example, for the same air-steel system the ratio is about 100, higlighting the differences in thermal boundary layer heights ($  \frac{\delta_s}{\delta_f}$).
The concept of thermal penetration depth is also widely used in the literature to describe the spatial scales of temperature variations in the solid  domain for unsteady CHT problems. There are various definitions of the thermal penetration depth, but for periodic cases the most common is the distance over which the temperature fluctuations from a harmonic forcing decrease by a factor of $1/e$. The thermal penetration depth is given by \autoref{eq:thermal_penetration_depth}, where $\omega$ is the angular frequency of the temperature oscillations.
\begin{equation}
    \label{eq:thermal_penetration_depth}
    \delta_P = \sqrt{2 \alpha_s / \omega}   
\end{equation}
Because most current simulation methods rely on a time-marching approach, such as URANS or LES, the small timesteps required, combined with the strict grid requirements for the solid domain, make these simulations resource-intensive when conducted over timescales relevant to the solid. 

From a computational implementation perspective, coupling strategies are used to manage how the heat transfer information is exchanged at the interface between the fluid and solid regions during a simulation. There are two primary categories of coupling strategies: weakly coupled and strongly coupled (or fully coupled). The coupling strategy can lead to additional modelling error, such as interpolation errors if the grids for both methods are different and interpolation is needed. 

\subsection{Strategies to efficiently solve CHT problems}
\label{sec:strategies}
The fundamental challenge of the disparity in timescales between the fluid and solid domains and its impact on the complexity of initialising and conducting unsteady conjugate heat transfer analyses is widely recognised \cite{YANN,Hickling,he2011unsteady}. Various methods have been proposed to accelerate the initial transient towards the statistical steady-state, while still enabling accurate time-dependent solutions for temperature fluctuations in the solid domain.


One of the most straightforward approaches is to alter the solid properties to try to realign the fluid and solid time scales. 
\citet{10.1115/1.4050111} focused on a ribbed cooling passage using LES with the immersed boundary method. They showed that the timescale disparity can be overcome by using an artificially high solid thermal diffusivity while maintaining a constant Biot number. The higher Fourier number allowed for a faster approach to statistical steady-state. Once statistical steady-state is reached, the solid thermal properties were changed back to their original values and the simulation ran until stationary steady-state was achieved again with the original values.  Their predictions are compared with experiments and other LES studies, however, the simulation time remains high due to the convergence of two successive steady-states separated by a discontinuous change in solid properties.
Similarly, \citet{SHI2021209}  modified the solid thermal properties and calculation time based on a Biot and Fourier number scaling of the equations. While the scaling is correct for the standalone unsteady heat equations, the scaling breaks  the similarity principle for fluid convection. It allowed for a reduction of  the simulation time by an arbitrary factor, but lead to errors in the predictions of both the mean and fluctuating temperature fields.

Another type of approach relies on frequency-based decomposition  to address the timescale mismatch between the fluid and solid domains.
\citet{he2011unsteady} implemented a hybrid coupling approach based on a time-marching technique for the fluid  domain and a frequency-based for the solid domain, with a continuously updating Fourier transform implemented at the interface. 
This method also has the advantage of directly answering the timescale mismatch by solving the solid region in the frequency domain. 
Since this work,  frequency based approaches have been used frequently for unsteady CHT simulations. \citet{Knapke2015} used a harmonic balance approach to with a quasi-Newton solver for CHT simulations. They showed that harmonic balance is an effective technique for performing accurate conjugate heat transfer problems with periodic unsteady simulations.  
This was confirmed by \citet{Hodges}, who presented a similar method and validated it for an internally cooled turbine blade. Another frequency-based decomposition, the non-linear harmonic method has also been used  for CHT simulations.  
\citet{ids} extended an existing commercial harmonic code to conjugate heat transfer. They used an updated harmonic equation with the addition of the harmonic source term to update the wall temperature at each timestep and on each side of the thermal interface. Further studies including \cite{He2019, Hickling} refined the method and adapted it for multiscale thermal systems. 

Aside from frequency-based methods, other decomposition approaches have been used. \citet{PODHeat} proposed a POD decomposition to tackle more efficiently transient heat transfer problems with fixed thermal boundary conditions. 
They used a combination of a time marching technique for the initial transient and then POD decomposition is used to reduce the dimensionality of the problem.  
\citet{10.1115/1.4032453} applied a similar reduced order model to the complete conjugate heat transfer problem. In addition, discrete Green function approaches have been applied to decouple conjugate heat transfer problems with any temperature variations as done in \cite{HACKER1997131, 10.1115/1.4048992}. Discrete Green functions decompose the temperature field into a set of functions independent of the thermal boundary condition.   
The functions describe the relationship between surface temperature and convective heat transfer, relating heat transfer from each element of the source surface to temperature rise on all other elements of the target surface. This approach allows direct, non-iterative calculation of heat transfer for any temperature distribution, irrespective of thermal boundary conditions.

Eigenanalysis has also been applied for to tackle heat transfer problems. \citet{doi:10.1080/01495728308963097} applied it  to transient heat conduction, demonstrating its accuracy and computational efficiency compared to classical implicit and explicit time-marching numerical schemes, particularly for long-duration and large-domain transient problems.
\citet{QUEMENER20121197} used modal analysis on advection diffusion problems with time-dependent parameters, achieving a significant computational time reduction compared to the finite elements model, by efficiently selecting influential modes and minimising the error between the reduced and physical models. Other applications can be found in \cite{GERSTENMAIER2001801,zhong1992finite,palmieri1978cave3}. Eigenanalysis has been also applied to CHT problems by \citet{knupp2020conjugate}. They used an integral transform approach to the solution of the problem on conjugate heat transfer. They achieved a significant improvement in convergence rate for a transient two-dimensional incompressible channel flow case.

Finally, when the behavior of the long thermal transient is of interest, the coupling conditions can be loosened  to obtain efficient results. \citet{10.1115/1.3147105} proposed a method based on the consideration  that for these transients the fluid flow time scales are much shorter than those for the solid heat conduction and therefore the influence of unsteadiness in fluid regions is negligible on the longer thermal transients. Their technique employs iterative procedures and steady CFD calculations to ensure continuity of temperature and heat flux. The procedure allows for defining CFD models at key time points and offers a "frozen flow" option for improved computational efficiency. 
\citet{10.1115/1.4040997} developed a loosely coupled CHT methodology using a source-term based modelling approach and adaptive time stepping. The technique demonstrated comparable accuracy to fully coupled unsteady simulations, but with significantly reduced computational costs. The technique was tested on predictions of turbine thermal loads during fast startup/shutdown cycles. 


As demonstrated in the literature overview, achieving efficient CHT simulations is a complex task. The key challenge lies in addressing the interplay between time and length scales in both the metal and the fluid. Simple dimensional arguments indicate that these scales are inherently linked, making it impossible to treat spatial and temporal scale separation independently. In most practical cases, what is desired is an acceleration of the large-scale, slow varying transients of the solid temperature field. Simple techniques such as alterations of the solid properties inevitably affect all length scales simultaneously and ultimately compromise either computational efficiency or physical accuracy. Methods based on orthogonal decompositions of the temperature field seeem better placed to achieve the twin goals of preserving accuracy and improving computational performance of unsteady CHT simulations.



In this paper, we build on the work of \cite{GERSTENMAIER2001801,QUEMENER20121197,knupp2020conjugate} by employing a modal decomposition approach for the solid temperature field. A modal basis can 
represent a given temperature field within a prescribed error with the smallest number of degrees of freedom. A modal basis also inherently preserves the natural relationships between large-scale, slow-evolving features and small-scale, fast-evolving features of the temperature distribution \cite{vlase2019eigenvalue}. Since the modes are formally mutually uncoupled, a modal representation reduces the solid domain to a set of uncoupled, single-degree-of-freedom ordinary differential equations, which require only surface integrals to be computed. 
Additionally, by recognising the linear nature of heat conduction,  we show that eigendecomposition provides an effective alternative for handling different scales individually—both in the numerical scheme and in the deliberate manipulation of time constants.  We demonstrate a method to  selectively modify the behavior of the modes responsible for the time/length scale reconciliation problem. We show that only a small subset of modes should be accelerated and establish an appropriate modal truncation criterion. Finally, we introduce a turbulent resolving grid at the interface to capture the remaining fluctuations



%% file: method.tex
This section develops the theoretical framework for the proposed strategy to efficiently solve an unsteady conjugate heat transfer problem.
\subsection{Problem specification and governing equations}
Consider a conjugate heat transfer problem over a domain $\Omega$ composed of the union of subdomains $\Omega_s $  
and $\Omega_f$. 
 The  interface between the subdomains $\Omega_s$ and $\Omega_f$ is denoted as $\Gamma$ and the exterior boundaries are $\Gamma_{\text{ext}}$.  
The governing equations are the compressible Navier-Stokes (N-S) equations on $\Omega_f$ and the unsteady heat conduction equation for $\Omega_s$.
This leads to the formal formulation of the CHT problem:
\begin{equation}
    \left\{
            \begin{aligned}
                &\text{N-S equations}  &\quad \text{on } \Omega_f \\
                \rho_s c_{p,s} \frac{\partial T}{\partial t} &= \nabla \cdot (\kappa_s  \nabla T) + Q_s\quad &\text{on } \Omega_s\\
                T_f &= T_s \quad &\text{on }\Gamma \\
                \kappa_f \nabla T_f \cdot \vec{n} &= \kappa_s \nabla T_s \cdot \vec{n} \quad &\text{on }\Gamma\\
                g(\vec{x},T, \nabla T) &= 0 \quad &\text{on }\Gamma_{\text{ext}}\\
                T(\vec{x},0) &= T_0(\vec{x}) \quad &\text{on }\Omega
            \end{aligned}
    \right. 
    \label{eq:system_coupled}
\end{equation}
Where $Q_s$ is a heat source, $g$ the boundary condition on the exterior boundary and the subscripts $f$ and $s$ refer to   subdomains $\Omega_f$ and $\Omega_s$, respectively. The initial condition is given by $T_0$.

\subsection{Modal representation of the heat conduction problem}
 \label{sec:modal_theory}
The unsteady heat conduction problem in \autoref{eq:system_coupled}  can be written in a finite element formulation:
\begin{equation} 
    {\bf M}\dfrac{d{\bf T}}{dt}=-{\bf K}{\bf T}+{\bf G}(t)
\label{eq:heat_conduction}
\end{equation}
Considering only 
natural boundary conditions 
gives the following expression for the matrices :
\begin{equation}
    { M}_{ij}=\int_{\Omega_s} \rho c_{p} F_{i} F_{j} \,d S 
\end{equation}
\begin{equation}
    { K}^h_{ij}=\int_{\Omega_s}\kappa\dfrac{\partial F_i}{\partial x_k}\dfrac{\partial F_j}{\partial x_k}dS    
\end{equation}
\begin{equation}
    { G}^{ij}=\int_{\Gamma,\Gamma_{ext}}F_i \kappa \dfrac{\partial F_j}{\partial n}dS+\int_{\Omega_s} F_i Q_s \, dS
\end{equation}
Where $F_i$ are the shape functions.\\ 
For the purpose of the following developments, it is useful to state certain general assumptions made about the structure of the fluid solution.
It is assumed that the fluid solution is obtained through a finite volume solver with cell centred variables, so that a straightforward algebraic relation exists
between the numerical surface heat flux and linear combinations of a small number of solid surface temperatures with a small number of gas near-wall temperatures.
It should be stressed that whereas this assumption appears in the diagrams and in some of the derivations, it is by no means essential. A different structure of the fluid solution,
e.g. a discontinuous Galerkin solver with non-comformal discretisation at the element boundary would still express the surface flux as a linear combination of surface solid and fluid temperature, but the expression would involve a larger number of degrees of freedom.\\
Invoking a  general mixed type boundary condition in \autoref{eq:bc}. With $h(x)$ being the thermal law of the wall coefficient and $q_r(x,t)$ is a heat flux that is not influenced by the presence of the gas (e.g. radiation)  and $T_g$ being the boundary temperature from subdomain $\Omega_f$, yields \autoref{eq:heat_conduction2}.
\begin{equation}
    -\kappa \dfrac{\partial T}{\partial n}=h(x) \left(T-T_g(x,t)\right)+ \beta(x,t)\qquad \text{on }\Gamma_{\text{ext}}  \text{ and on }  \Gamma 
    \label{eq:bc}
 \end{equation}
 \begin{equation} 
    {\bf M}\dfrac{d{\bf T}}{dt}    =-\left( \bf{K}^h + \bf{K}^b \right) \bf{T}+\bf{G}(t)
\label{eq:heat_conduction2}
\end{equation}
\begin{equation}
    {K}^b_{ij}=\int_{\Gamma,\Gamma_{ext}} h(x) F_i F_j dS  
\end{equation}
\begin{equation}
    {G}_{i}=\int_{\Omega_s} F_i Q_s \, dS+ \int_{\Gamma, \Gamma_{ext}} \left(h(x)  T_g(x,t)-q_r(x,t)\right)F_i \, dS 
\end{equation}
This shows that the matrix ${\bf K}$ is composed of symmetric positive blocks. \\
Considering the homogeneous part of the system \autoref{eq:heat_conduction2}, solutions of the following type are sought:
\begin{equation}
    {\bf T}={\bf Z}e^{-\lambda t}
\end{equation}
where ${\bf Z}$ is time independent.
This yields to the modified eigenvalue problem for the homogeneous part of the system. 
\begin{equation}
    \lambda {\bf M}{\bf Z}= {\bf K}{\bf Z}
    \label{eq:eigen}
\end{equation}
The eigenvectors ${\bf Z}$ can be normalised in such a way that
\begin{align}
    {\bf Z}^\top {\bf M} {\bf Z}&={\bf I}\\
    {\bf Z}^\top {\bf K} {\bf Z}&={\bf \Lambda}\\
    {\bf Z}{\bf U}&= {\bf T}
\end{align}
With ${\bf U}$ being  the modal amplitude.\\
Now returning to the inhomogeneous problem of \autoref{eq:heat_conduction}
\begin{eqnarray}
   {\bf M}\dfrac{d{\bf T}}{dt}&=&-{\bf K}{\bf T}+{\bf G}(t)\\
   {\bf Z}^\top{\bf M}{\bf Z}\dfrac{d{\bf U}}{dt}&=&-{\bf Z}^\top{\bf K}{\bf Z}{\bf U}+{\bf Z}^\top{\bf G}(t)\\
   \dfrac{d{\bf U}}{dt}&=&-{\bf \Lambda}{\bf U}+{\bf Z}^\top{\bf G}(t)
   \label{eq:modal_1}
\end{eqnarray}
Since the matrix $\bf{\Lambda}$ is diagonal, the equations describing the evolution of the modal amplitudes are decoupled. 
For each mode, a scalar equation needs to be solved for the modal amplitude, with ${\bf \Lambda}$ and  ${\bf Z}^\top$ as inputs. 

In practical implementation, the modes are only needed at the boundaries where the scalar product ${\bf Z}^\top{\bf G}(t)$ will not vanish. 
 This greatly reduces the memory requirements for a simulation as only the boundary values have to be stored  instead of the full eigenvector matrix.
 Additionally, depending on the coupling methodology used, the solution to the eigenproblem of \autoref{eq:eigen} is usually only needed once, for instance if the coupling is done through an inhomogeneous Neumann boundary condition. However, in some cases the influence of the coupling on the mass and conductance matrix is expected to vary greatly during the simulation span, for instance through a varying heat transfer coefficient. 
For these cases, if the variation of properties with temperature is mild and can be represented with a linearized relation, then the structure of the eigenvalue problem is unchanged, which limits the need to recompute the eigendecomposition again. 
\subsection{Acceleration techniques}
\label{sec:acctech}

Modal decomposition (\autoref{sec:modal_theory}) decouples the heat conduction equation, inherently enabling targeted acceleration strategies. 
Previous methods (discussed in \autoref{sec:strategies}) often introduced errors by altering solid characteristics or accelerating time integration across all scales. 
In contrast, the present approach leverages the modal basis to selectively accelerate only the slowest thermal modes within the solid domain—those corresponding to long low-energy time scales. 
This targeted modification minimises the impact on solution accuracy compared to prior techniques while still providing significant computational acceleration.

Concerning the practical ways \autoref{eq:modal_1} can be accelerated to statistical steady-state, most of the  techniques mentioned in \autoref{sec:strategies} can be applied to the modal equations. For example, a harmonic or Fourier formulation  can be done on the individual modes. If nonlinear interactions between the modes are neglected it would make the assumptions 
of the Fourier transform stronger when applied on the decoupled modal equations. 
While it would be interesting to test the performance of these various methods in the modal context, the current paper will focus mostly on one strategy that arise naturally from the decomposition. 

The following subections introduce the proposed acceleration technique and assesses its performance on simple test cases. This is followed by a discussion on the properties of the temperature eigenmodes and the possible truncation of the modal basis.



\subsubsection{One dimensional - Single mode analysis}
\label{sec:1D}
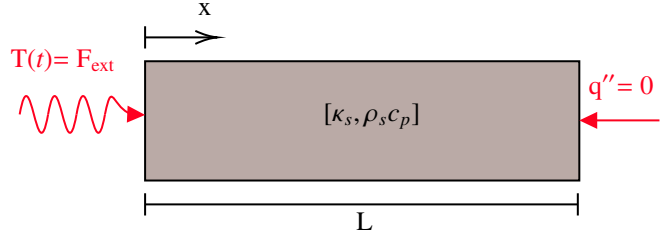
\begin{figure}
    \centering
    \input{figures/1Dsolid.tex} 
    \caption{Sketch of the 1D domain with boundary conditions and coordinate system}
    \label{fig:1Dsolid}
\end{figure}

To illustrate how the proposed method tackles the problem of the disparity in timescales between the fluid and solid domains, we consider a simple 1D solid domain of length $L$ with  constant thermal diffusivity $\alpha_s$. 
On one end the solid is submitted to external thermal excitation, similar to the thermal Stokes problem and adiabatic conditions are set on the other end. A sketch of the domain can be found in \autoref{fig:1Dsolid}. The  domain is discretised into $N$ elements using a finite element approach and temporal advancement is performed using an implicit Euler scheme. 

\begin{figure*}[hbt!]  
    \centering
    \begin{subfigure}[b]{0.45\textwidth}
        \centering
        \scalebox{.9}{\input{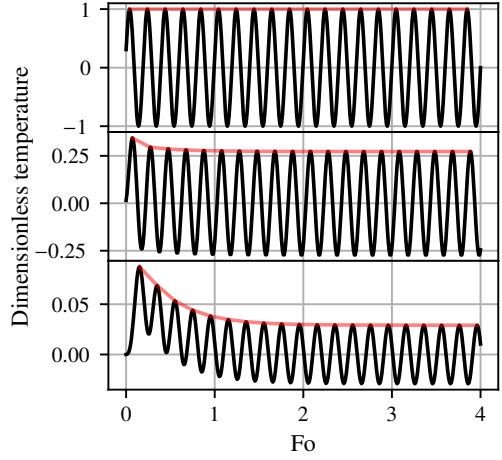}}
        \caption{Time trace of the temperature (black line) and upper envelope (red line) at selected locations. From top to bottom: $\xi =[ 0, 0.3,0.9]$}
        \label{fig:1DTem}
    \end{subfigure}
    \hfill
\begin{subfigure}[b]{0.45\textwidth}
    \centering
    \scalebox{.9}{\input{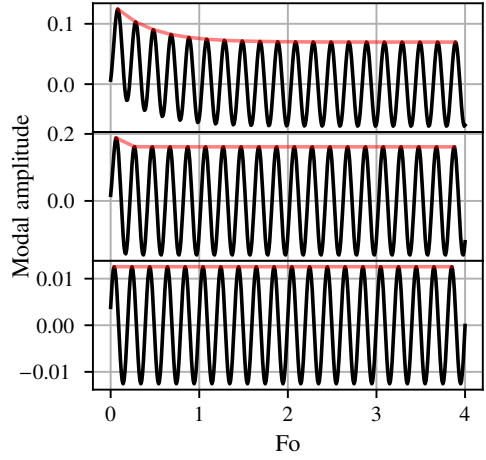} }
    \caption{Time trace of the modal amplitude (black line) and upper envelope (red line) for selected modes. From top to bottom: $U_1, U_2, U_{30}$.}
    \label{fig:1Dmodes}
\end{subfigure}
\caption{Time evolution of the temperature and modal amplitudes for the 1D isolated domain}
\label{fig:1Dtime}
\end{figure*}

The unsteady heat conduction equation (\autoref{eq:system_coupled}) can be written in  non-dimensional form by introdusing the Fourier number Fo, the normalised temperature $\Theta = (T - T_m)/ (T_{max}-T_m)$ and the dimensionless coordinate $\xi = x/L$ where $T_m$ is the mean temperature and $T_{\text{max}}$ the maximum temperature.
\begin{equation}
    \dfrac{\partial\Theta}{\partial\text{Fo}} =  \dfrac{\partial^2 \Theta}{\partial \xi^2}
\end{equation}
With the prescribed boundary conditions, the finite element system can be written 
as:
\begin{equation}
   {\bf M} \dfrac{d{\bf \Theta}}{d \text{Fo}}=-{\bf K^h}{\bf \Theta}+{\bf G}(t)
\end{equation}
And the modal equation is: 
\begin{equation}
    \frac{d{\bf U}}{d\text{Fo}}={\bf \Lambda }{\bf U}+{\bf Z}^{T}{\bf G}(t) \qquad \text{with : } {\bf \Theta}=\sum_i {\bf z}_i U_i
    \label{eq:1Dsolid_modal}
\end{equation}
In the first test case, the prescribed temperature is represented by a sine wave with zero mean, an amplitude of $T_{\text{max}}$ and a dimensionless angular frequency $\tilde{\omega} =10\pi$. 
The  temperature response of the system at different depths is plotted  in  \autoref{fig:1DTem}, with the initial condition set to the time-averaged value. The curves exhibit the classical exponential decay found in transient heat conduction in a slab, \cite{lienhard2005heat}. As explained by \citet{Hickling}, due to the nature of the external excitation (a step function modulated by a harmonic function), the initial response of the solid exhibits an overshoot before gradually approaching a steady state. 
The closer to the interface, the faster the convergence until $\xi = 1 $ that takes approximately 1 Fo to reach the steady-state. This behavior is seen also in  \autoref{fig:1Dmodes}, which shows the time evolution of the modal amplitudes, 
$U_i$. Notably, only the first few modes, those associated with the longest time constants, converge to a steady state over an extended period. It can be seen that the first mode shape needs approximately 1 Fo to reach steady-state, similarly to the direct simulation, in \autoref{fig:1DTem}. Thus, in practice, only a limited subset of modes, those with the longest timescales, require acceleration.
 
\begin{figure*}
    \centering
      \begin{subfigure}[b]{0.39\textwidth}
        \centering
        \scalebox{.9}{\input{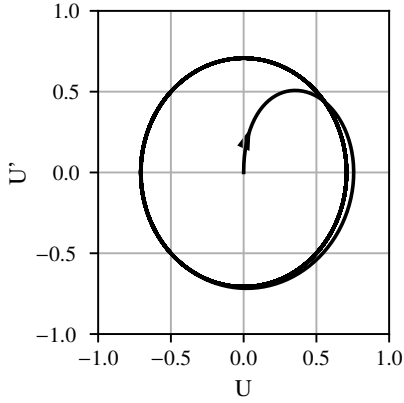}}
        \caption{Phase portrait of \autoref{eq:model_eq_1d} with initial condition $U(0) = 0$\\\quad } 
        \label{fig:1DCircle}
    \end{subfigure} 
    \centering
      \begin{subfigure}[b]{0.3\textwidth}
        \centering
        \scalebox{.9}{\input{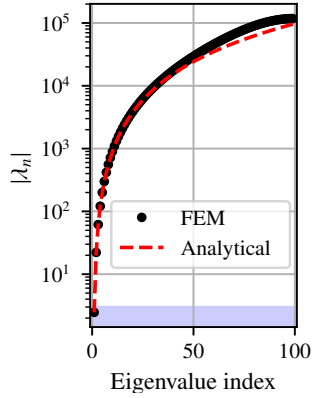}
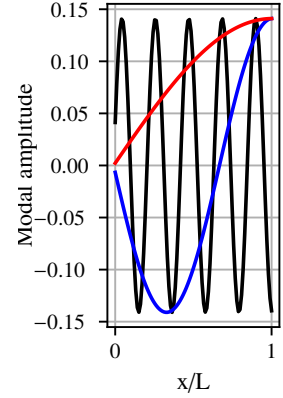}
        \caption{Eigenvalues and the light blue zone corresponds to $|\lambda_n| < 3$}
        \label{fig:sub_eigvals}
    \end{subfigure} 
    \hfill
    \begin{subfigure}[b]{0.3\textwidth}
        \centering
        \scalebox{.9}{\input{figures/mode.pgf}}
        \caption{Selected mode shapes. Red: first mode, blue: second mode, black: $20^{\text{th}}$ mode}
        \label{fig:firstmode}
    \end{subfigure}
    \caption{Analysis for the 1D domain test case}
    \label{fig:wholefigure}
\end{figure*}

To gain a deeper understanding of how does the unsteady heat conduction equation behave, we seek an analytical solution to \autoref{eq:1Dsolid_modal} considering a single mode.
As the external heat flux ${ G}(t)$ is a sine wave of angular frequency $\tilde{\omega}$ with zero mean, \autoref{eq:1Dsolid_modal} becomes \autoref{eq:model_eq_1d} where $\beta$, $\sigma$ are scaling factors and $\phi$ an arbitrary phase. The general solution of the ODE in \autoref{eq:model_eq_1d}   is given by \autoref{eq:modal_solution}. The subscript has been omitted for clarity. 
\begin{equation}
    \frac{dU}{d\text{Fo}} = -\lambda U + G(\text{Fo}) = -\lambda U + \sin(\tilde{\omega} \text{Fo}+\phi)
    \label{eq:model_eq_1d} 
\end{equation}  
The transfer function $H(s)$ of the system is given by \autoref{eq:transfer_function}. For sinusoidal inputs the system acts as a type of low-pass filter with a cut-off frequency of $\lambda$.
\begin{equation}
    H(s) = \frac{U(s)}{G(s)} = \frac{1}{\lambda + s}
    \label{eq:transfer_function}
\end{equation}
The solution of \autoref{eq:model_eq_1d} to a harmonic forcing of angular frequency $\tilde{\omega}$ with zero mean is given by \autoref{eq:modal_solution}.
\begin{align}
    U(\text{Fo}) =&  \frac{\tilde{\omega} \cos(\phi)- \lambda \sin ( \phi)
     }{ \omega ^2+ \lambda^2}e^{ -\lambda \text{Fo}} + \frac{\lambda \sin(\tilde{\omega} \text{Fo}+\phi) - \omega\cos (\tilde{\omega} \text{Fo}+\phi) }{ \omega ^2+ \lambda^2} 
    \label{eq:modal_solution}
\end{align}
Inspecting the terms, the exponential term  is the initial transient response similar to the classical transient heat equation problem. 
The decaying behavior arises from the initial condition because to have steady state a balance in the function's value and its first derivative is necessary. A phase portrait is shown in \autoref{fig:1DCircle}, it has been initialised with $U(\text{Fo}=0) = 0 $ and $  \lambda = 1$. It shows the solution path slowly converging  $\{U(0),U'(0)\} =\{0,0\}$ to its steady-state solution on the circle.  \autoref{eq:modal_solution} explains the slow steady-state convergence of modes with the lowest eigenvalues (longest time constants), as illustrated in \autoref{fig:1Dmodes}, and notably reveals that their decay constants are independent of excitation frequency.

\Cref{fig:sub_eigvals,fig:firstmode} 
show the eigenvalues and selected eigenvectors of the problem. For a slab problem formulated in dimensionless variables with the prescribed boundary conditions, the analytical eigenvalues of the Laplace operator  are given by \cite{gottlieb1985eigenvalues}:
\begin{equation} 
    \lambda_n = \left(\frac{(2n-1)\pi}{2}\right)^2  \quad n = 1,2,3,...
\label{lambdas}
\end{equation}
\autoref{fig:sub_eigvals} shows that the finite element space accurately captures the eigenmodes that are well resolved by the mesh, while the higher-frequency modes are less well resolved, resulting in inaccuracies in the corresponding eigenvalues.
\autoref{eq:modal_solution} showed that the decay of the transient behavior for each mode depends solely on the product of its eigenvalue and the Fourier number. Therefore, the eigenvalues provide insight into which modes require acceleration based on the Fourier number of the simulation, which characterises the allowed simulation time.  For example, with an allowable Fourier number of 1 and a target threshold for the decay is 95 \% (approximately exp$(-3)$), any mode with an eigenvalue less than 3 Fo will not have decayed sufficiently, this is illustrated by the light blue region in \autoref{fig:sub_eigvals}. 
Consequently, in the present case, only the first eigenmode has not decayed to the threshold level by Fo = 1, as also visible by  \autoref{eq:modal_solution}.
This type of analysis allows one to determine a priori which modes need to be accelerated, depending on the total simulation time and a threshold number for the transient to decay to acceptable levels. This is the key advantage of the current targeted acceleration approach; typically, only a few modes require acceleration because the eigenvalues increase rapidly with mode number $\lambda_n \propto n^2$.
Furthermore, as shown in \autoref{fig:firstmode}, lower eigenvalues correspond to large-scale spatial patterns, confirming that larger temperature fluctuations take longer to approach steady state.

\begin{figure*}[h!]
    \centering
    \begin{subfigure}[b]{0.49\textwidth}
        \centering
        \input{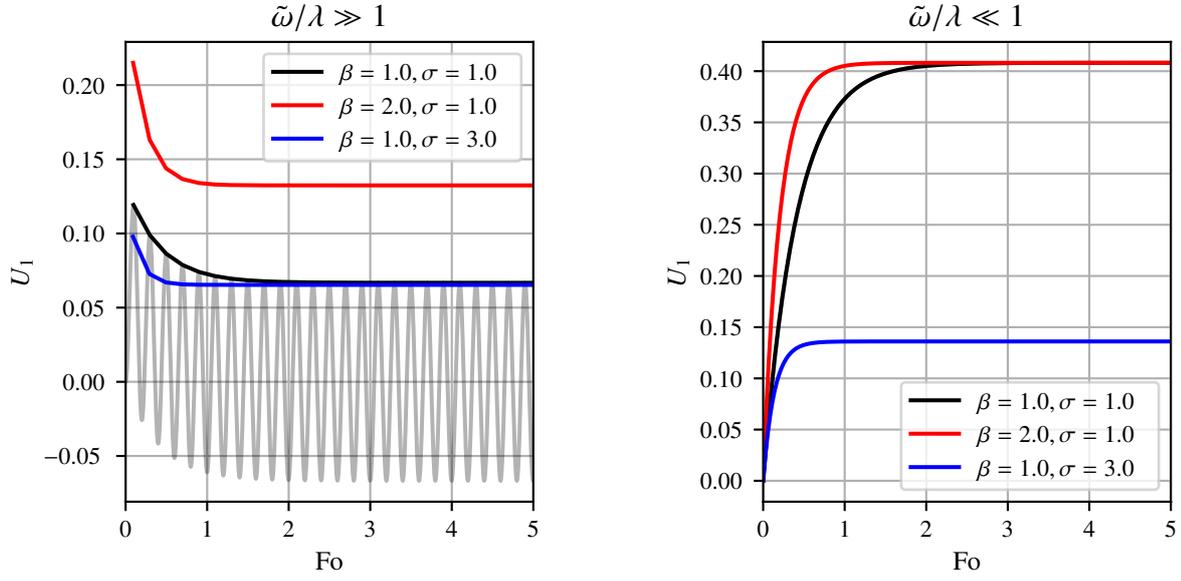}
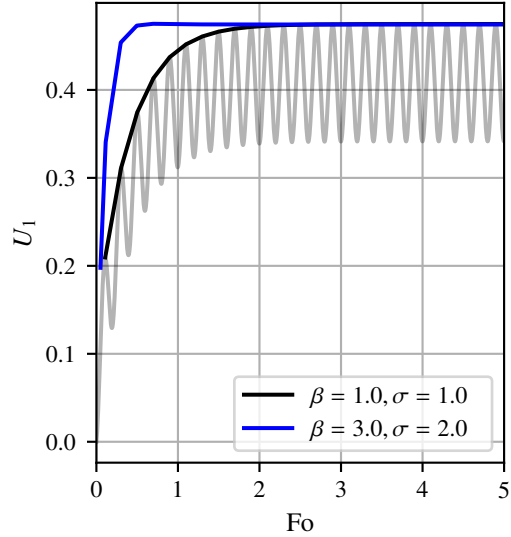
        \caption{Illustration of different acceleration techniques for $\tilde{\omega}/ \lambda \gg 1$ via upper envelope, with signal time trace for reference case.}
        \label{fig:acc_type}
    \end{subfigure}
    \hfill
    \begin{subfigure}[b]{0.49\textwidth}
        \centering
        \input{figures/acc_type_mean_only.pgf}
        \caption{Illustration of different acceleration techniques for $\tilde{\omega}/ \lambda \ll 1$.\\ \quad}
        \label{fig:acc_mean}
    \end{subfigure}
    \begin{subfigure}[b]{0.49\textwidth}
        \centering
        \input{figures/acc_type_with_mean.pgf}
        \caption{Illustration of  the SFD technique via upper envelope, with signal time trace for reference case. The forcing signal has a nonzero mean and oscillating component at $\tilde{\omega}/\lambda \gg1$.}
        \label{fig:acc_type_with_mean}
    \end{subfigure}
    \caption{Illustrations of acceleration techniques for the 1D domain test case}
\end{figure*}

To accelerate the approach to steady state, two direct techniques  can be inferred from \autoref{eq:modal_solution}. The first involves scaling the time update  by an arbitrary factor $\beta$, as done  \citet{KOREN2017340}. 
The second option is to artificially increase the eigenvalue by a factor $\sigma$. 
\autoref{eq:modal_eq_mean} shows the modified equation for a periodic excitation which has now a non-zero mean, $\Psi$, where $\beta$ and $\sigma$ are scaling factors. The solution of \autoref{eq:modal_eq_mean} is given in \autoref{eq:modal_solution_mean}, where $C$ is a constant.
From \autoref{eq:modal_solution_mean}, it can be concluded that both techniques amplify heat flux fluctuations received by the solid interface, as both frequency and response amplitude are modified proportionally to  $\beta$ or $\sigma$. 
However, there are regimes where one technique is more advantageous than the other.
\begin{equation}
    \frac 1\beta\frac{dU}{d\text{Fo}} = -\sigma\lambda U + \left(\Psi +\sin(\tilde{\omega} +\phi)\right)
    \label{eq:modal_eq_mean}
\end{equation}
\begin{align}  
    U(\text{Fo}) =& \frac{\Psi}{\sigma \lambda}+ C e^{ -\sigma\lambda\beta\text{Fo}}+
    \frac{ \sigma\lambda \sin (\tilde{\omega} \text{Fo}+\phi)-\tilde{\omega} /\beta \cos ( \tilde{\omega} \text{Fo}+\phi) 
    }{ \tilde{\omega} ^2/\beta^2+ \sigma^2\lambda^2} 
    \label{eq:modal_solution_mean}
\end{align}
Typically,  the eigenvalues of the  modes that require acceleration are smaller than $\tilde{\omega}$.
The lowest dimensionless frequency resolved by the simulation is $\tilde{\omega}= 1/$Fo and the modes that require acceleration have $\lambda$ Fo $\ll1$, which leads to  $\lambda\ll \tilde{\omega}$. 
Therefore, altering the eigenvalue through the $\sigma$ factor has lower influence on the temporal solution once statistical steady-state has been reached compared to modifying $\beta$, according to \autoref{eq:modal_solution_mean}. 
\autoref{fig:acc_type} illustrates the comparative performance of a $\beta$ and a $\sigma$ scaling using $\lambda_1$ from \autoref{lambdas} and $\tilde{\omega}= 10 \pi$, leading to $\tilde{\omega}/\lambda_1 \gg1$. 
Analysis of the upper envelope of the signal demonstrates that while both $\beta$ and $\sigma$  modifications improve convergence rates, $\beta-$alterations proportionally modifies the amplitude of the response's leading term, for the chosen ratio $\tilde{\omega}/\lambda_1$.  In contrast, $\sigma$-modifications primarily influence the solution's minor term as long as $\tilde{\omega}$ is greater than $\lambda$ which is precisely the modal regime targeted for acceleration, as depicted \autoref{fig:sub_eigvals}.

On the other hand, when the forcing signal has some content at lower frequency than the eigenvalue ($\lambda> \tilde{\omega}$) or if the forcing has a nonzero mean, 
the steady state time-averaged solution $\bar{U}$ is given in \autoref{eq:modal_solution_mean_avg}. 
\begin{equation}
    \bar{U} = 
    \frac{\Psi}{\sigma\lambda}
    \label{eq:modal_solution_mean_avg}
\end{equation}
\autoref{fig:acc_mean} compares $\beta$ and $\sigma$ alterations for a case when the forcing signal is constant in time, $\tilde{\omega}/\lambda \ll1$. 
As expected, altering the stiffness properties of the system through the eigenvalue will impact the steady-state time-averaged solution. 

To avoid this issue, a Selective Frequency Damping  (SFD) approach is used. SFD is a technique coming from system control theory that aims to accelerate the transient decay of specific frequency components in a system while preserving the dynamics of others. SFD has notably been used in the context of CFD simulations to accelerate the convergence of the flow equations, \cite{cunha2015optimization,10.1063/1.2211705,CASACUBERTA2018481}. The approach consists of  adding a linear forcing term to \autoref{eq:model_eq_1d} in order to achieve frequency-selective damping via coupled low-pass filtering and high-frequency feedback. The modified equation is given in \autoref{eq:modal_eq_mean_feedback} where $U_{LP}$ is a low-pass filtered solution.  
The modified differential equation written for an arbitrary forcing $G(\text{Fo}) $ with mean $\Psi$ is given in \autoref{eq:modal_eq_mean_feedback}.
\begin{equation}
\frac{dU}{d\text{Fo}} = -\lambda U + G(\text{Fo}) - \chi_a\left(U -U_{LP}\right) 
\label{eq:modal_eq_mean_feedback}
\end{equation}
Here, \(\chi_a > 0\) is a feedback gain parameter, effectively increasing the effective decay  rate of the higher frequencies to \(\lambda + \chi_a\), thereby speeding up the approach to the steady state. The steady state remains unchanged because the additional term vanishes when considering the time-averaged system. To keep a consistent notation we  will write the feedback parameter as  $\sigma  =1-\chi_a/\lambda$. The low-pass filtered solution $U_{LP}$ can be accelerated through the $\beta$ parameter as done in \autoref{eq:modal_eq_mean}.
\autoref{fig:acc_type_with_mean} plots the different acceleration techniques for a forcing with a  finite mean amplitude
superimposed with a higher frequency signal and the initial condition is kept at $U(0) = 0$.
From this point onward, the $\beta$ modifications are applied exclusively to the low-pass filtered solution $U_{LP}$, whereas the $\sigma$ modifications are applied to the high-pass component, defined as $U - U_{LP}$.
It can be seen that the unmodified system needs similar time to reach the steady state compared to \autoref{fig:acc_type} as expected because the decay constant is not a function of frequency. The modified equations converge significantly faster than the unmodified system without loss in accuracy in both the mean and fluctuating components. 

\subsubsection{Coupled domains - Complete thermal field analysis}
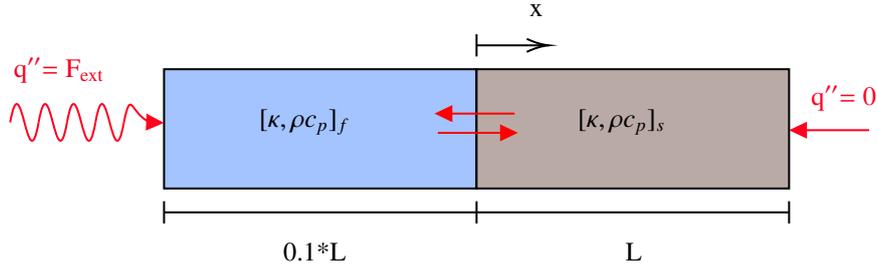
\begin{figure*}[h!] 
    \centering
    \input{figures/1D2solids.tex} 
    \caption{Sketch of the system of two solids with boundary conditions and coordinate system} 
    \label{fig:sketch_system}
\end{figure*}

The next test case involves two coupled one-dimensional domains, each governed by the unsteady heat equation, as illustrated in \autoref{fig:sketch_system}. This configuration has been previously analyzed by \citet{KOREN2017340}, and the thermal properties of both domains, shown in \autoref{tab:properties2}, match those presented in their study. The domains are discretised using finite elements and explicit time  integration is used. 
This test case evaluates whether the proposed acceleration technique influences a coupled system with significantly lower thermal capacity. The outer boundaries are subject to a Neumann boundary conditions. 
Here, the coupling is modeled as convective, with the heat flux at the interface computed using Newton's law of cooling, assuming a heat transfer coefficient of 
$ h = 10$ W/(m$^2$K). This condition makes the system similar to real life scenarios, such as a turbine with a metal casing surrounded by insulating material. 
Focusing on the solid $s$ in \autoref{fig:sketch_system}, the combination of Robin and Neumann boundary conditions makes the   domain less stiff, leading to lower eigenvalues  compared to the case in \autoref{sec:1D}, which was subjected to a Dirichlet boundary condition. These lower eigenvalues are expected to require a longer time to reach steady-state than those in \autoref{sec:1D}, making the need for acceleration more pronounced in this case.
\begin{table}[h!]
    \centering
    \caption{Thermal properties of the coupled domains}
    \begin{tabular}{ccccc}
        &$\kappa$ [W/(mK)]& $\rho c_p $[J/(m$^3$K)] & L [m] & N$[-]$\\
        \hline
        Domain $s$ &7.3 & 4500*570 & 0.005 & 100\\
        \hline
        Domain $f$ &0.158 & 1738*3.65 & 0.0005 & 100\\
        \hline
    \end{tabular}
    \label{tab:properties2} 
\end{table}

The signal imposed at the boundary of domain $f$ is more realistic than the one used in \autoref{sec:1D}. It consists of a series of harmonics with  random phase. The spectrum is made of a plateau until 1 Hz after that a $-5/3$ power law is applied to the higher frequencies. As shown in \autoref{tab:properties2}, the length of the domain where the excitation force is applied has also been divided by 10, minimising the damping of high-frequency content reaching the interface. The system has one eigenvalue lower than the highest dominant forcing frequency that is 1 Hz. The corresponding eigenvector is the constant mode, which will therefore require acceleration. The constant solid temperature mode has a non vanishing eigenvalue because of the Robin boundary condition a the interface with the fluid domain.
Two acceleration levels have been tested using the SFD technique, one with the acceleration factors $\sigma$ and $\beta$ set to $1/\lambda_1$ and the other with an intermediate factor set to $0.1/\lambda_1$, where $\lambda_1$ is the eigenvalue of the constant mode.  

\begin{figure*}[!htbp]    
    \centering
    \scalebox{.91}{\input{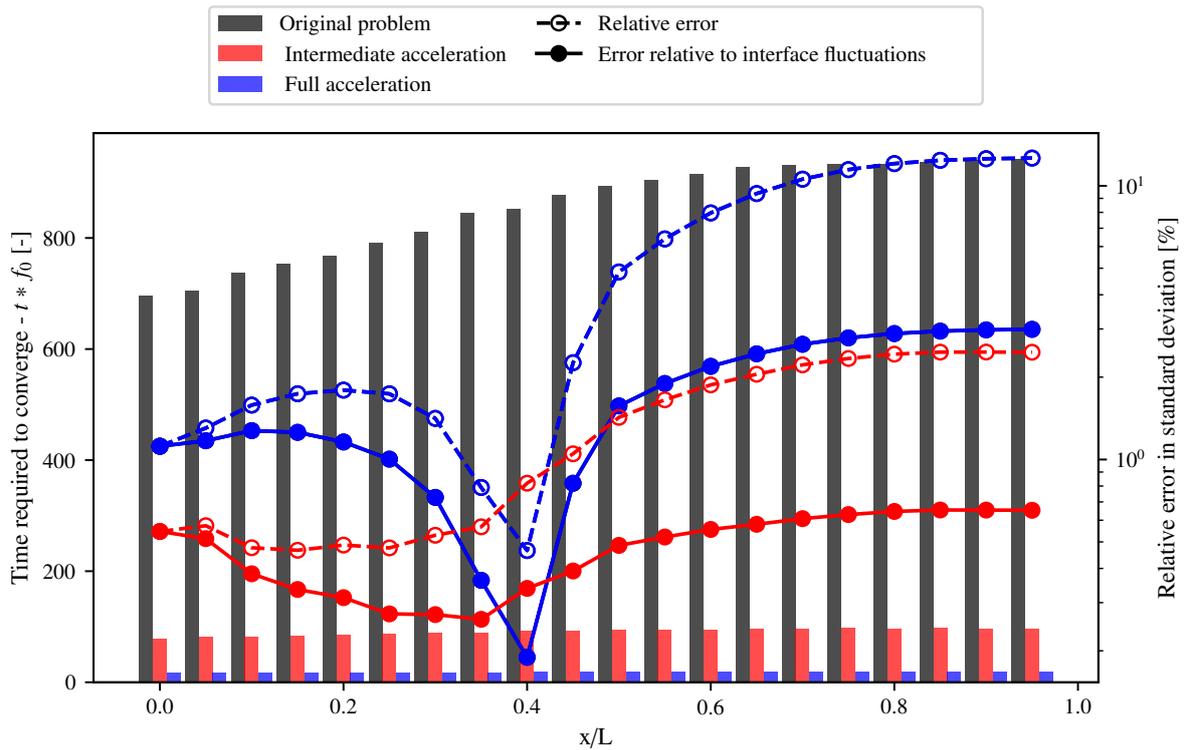} }
    \caption{Convergence time and relative error in standard deviation for the solid $s$ of the  system of two solids. The interface corresponds to $x/L =0$. The bar levels represent the convergence time for the thermal field at different depths within the solid $s$ (corresponding axis is on the left-hand side). The dashed lines correspond to the relative error while the solid line is the error in the standard deviation relative to the interface standard deviation (corresponding axis is on the right-hand side). 
    } 
    \label{fig:systemresponse}
\end{figure*}

The results are shown in \autoref{fig:systemresponse}. The bar levels represent the  time required to reach steady-state for the thermal field at different levels within the solid domain. The interface corresponds to $x/L =0$. Three simulations are compared, the first one is the original  non-accelerated system, in dark grey. In blue is the accelerated system, with acceleration factor set to $1/\lambda$. Finally, in red is the intermediate acceleration level. It can be seen that the acceleration factor has a significant impact on the approach to steady-state of the system. The  time required to reach steady-state is reduced by a factor of 10 and 70 for the intermediate and high acceleration levels, respectively. 

 The error made in the standard deviation through the acceleration factors is shown in the line plots in \autoref{fig:systemresponse}. The dashed lines correspond to the relative error in the standard deviation while the solid line is the error in the standard deviation relative to the standard deviation at the interface, i.e. the maximum amplitude. The relative error has a maximum of 2.5 \% and 13 \% for the intermediate and high acceleration levels, respectively. The error is more pronounced towards the far end of the solid domain, where the low-frequency modes are more dominant. The error relative to the interface fluctuations is less than 3 \% for the high acceleration level and less than 0.7 \% for the intermediate acceleration level. Showing that compared to the interface fluctuations, the error remains low.

In conclusion, this section showed briefly some capabilities of the proposed approach in terms of reducing the time required to reach steady-state. It proved to have a low impact on the steady-state mean and fluctuating components of the solution while improving the convergence time greatly, as long as most of the energy is not contained close to the pole of the transfer function. At the pole, the error made directly correlates with the acceleration factor. As stated at the beginning of this section, the modal approach is not limited to the acceleration technique presented here. It can be used with most of the acceleration techniques presented in \autoref{sec:strategies}. The main advantage of the modal decomposition is that it allows for a targeted acceleration. This is in contrast to the other acceleration techniques, which modify the entire system.

\subsection{Solving  complex multiscale heat conduction problems} 
Having demonstrated the advantages of the modal decomposition to speed up the approach to steady-state of unsteady CHT problems, it is important to understand the properties of the temperature eigenmodes. This section will delve into the spatial distribution of the eigenmodes and the practical implementation of the modal method for more complex scenarios beyond the simple 1D cases discussed in \autoref{sec:acctech}.

\begin{figure*}[h]
    \begin{subfigure}[b]{0.32\textwidth}
        \includegraphics[width=\textwidth]{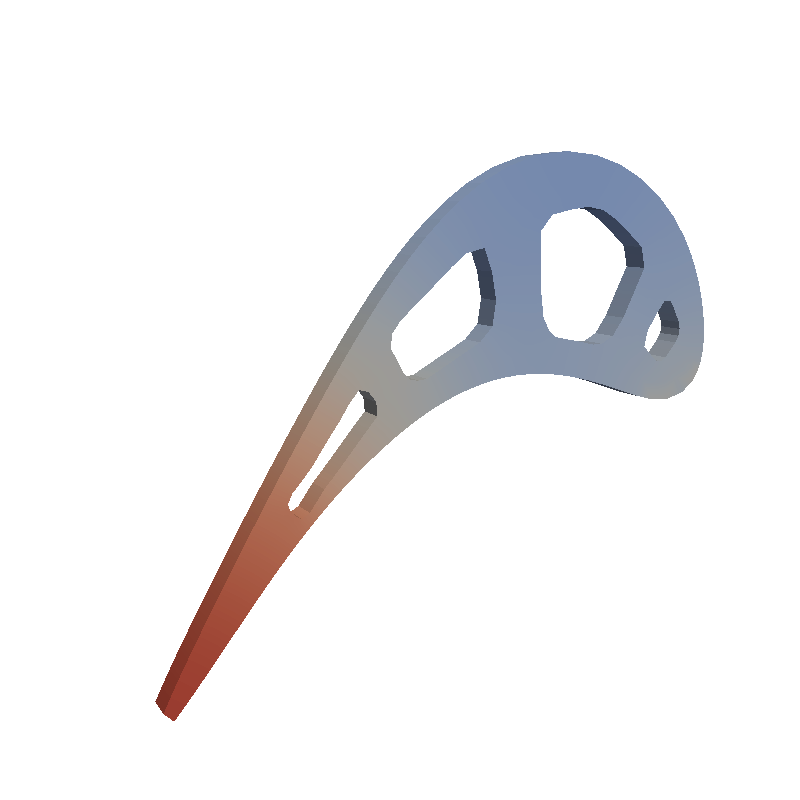}
        \caption{}
        \label{fig:mode1}
    \end{subfigure}
    \begin{subfigure}[b]{0.32\textwidth}
        \includegraphics[width=\textwidth]{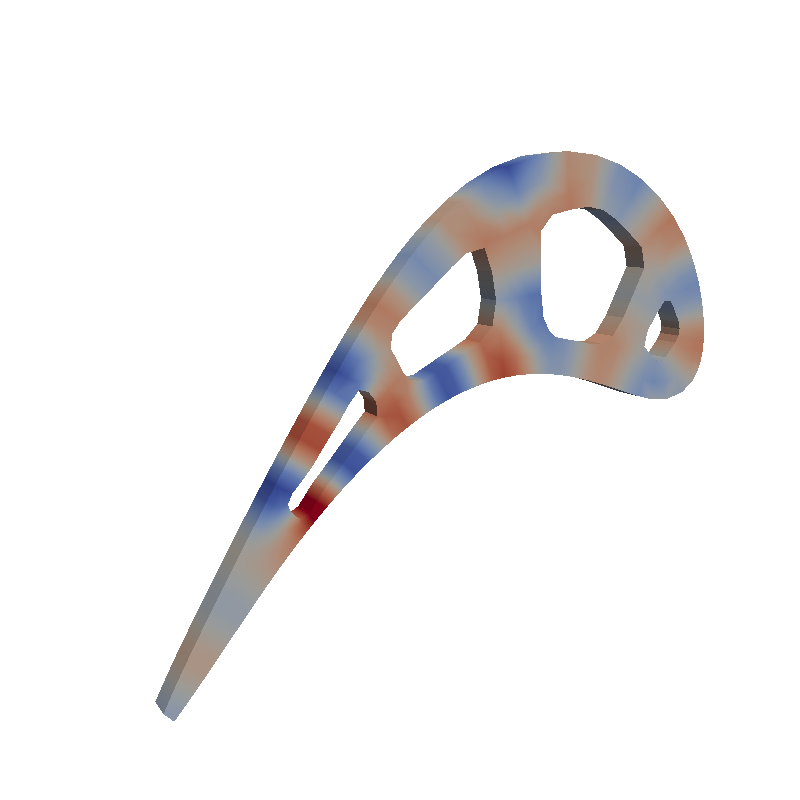}
        \caption{ }
    \end{subfigure}
    \begin{subfigure}[b]{0.32\textwidth}
        \includegraphics[width=\textwidth]{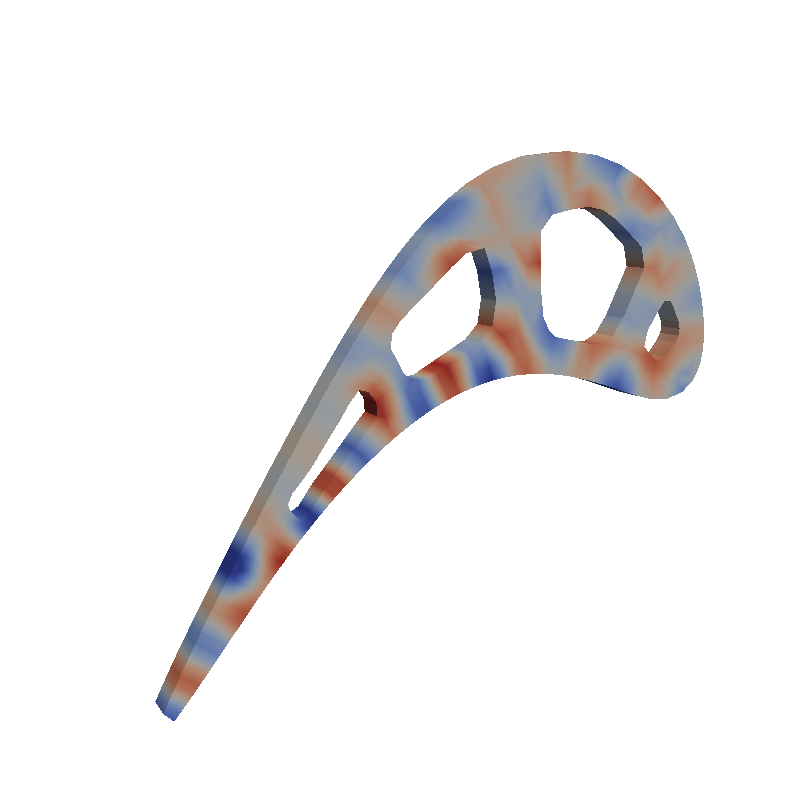}
        \caption{ }
        \label{fig:mode3}
    \end{subfigure}
    \caption{Temperature modes for a typical turbine section. The eigenvalue increases from left to right and the color scale is symmetrical around zero with red being positive and blue negative.} 
    \label{fig:turbinemodes}
\end{figure*}

\subsubsection{Timescale selection and truncation of the modal basis}
\autoref{fig:turbinemodes} displays three eigenmodes for a typical turbine blade section, with the eigenvalue magnitude increasing from left to right. Each eigenmode corresponds to a specific timescale, which is inversely proportional to its eigenvalue. The timescale determines the spatial extent of the mode, directly related to its penetration depth. The mode in \autoref{fig:mode3} exhibits more localized features, while the mode  in \autoref{fig:mode1} demonstrates a broader, global trend.
In practical scenarios involving complex three-dimensional geometries, it is neither feasible nor necessary to retain a large modal basis during a simulation. 
Additionally, high-frequency modes are often physically unrealisable because they are dominated by the discretisation error of the method that was used to build the mass and conductivity matrix.  
 Therefore, to achieve optimal efficiency it is beneficial to operate in practice with a truncated the modal basis. 
Truncating the modal basis comes with two main advantages. First, it reduces interpolation errors, particularly for high-frequency or purely numerical modes that are discarded. Second, when performing eigenanalysis on a large system, various algorithms can efficiently compute a subset of the eigenvalues and eigenvectors, such as those with the lowest magnitude.

The selection of time scales and truncation of the modal basis is a critical step in the analysis process. Only the scales that can be resolved by the simulation should be considered.
One important factor to consider is the time evolution of slow modes. These modes might exhibit  time scales longer that the simulation and may have minimal impact on the overall behaviour of the system within the given computation length. Consequently, including them in the analysis may not be necessary for obtaining accurate results. 
By carefully considering these factors, an appropriate selection and truncation of the modal basis can be achieved. This enables an efficient and accurate analysis of the system while balancing computational constraints and the need for reliable results.
    
\subsubsection{Treatment of the residual modal flux}
\begin{figure*}
    \centering
    \input{figures/new_thin_layer}  
    \caption{Partition of solid degrees of freedom near an interface boundary in a conjugate heat transfer problem. $\Gamma $ is the interface between $\Omega_s$ and $\Omega_f$. $\Xi $ is the interface between $\Omega_s$ and $\Omega_{s,\text{FE}}$. }
    \label{fig:conjugate}
\end{figure*}
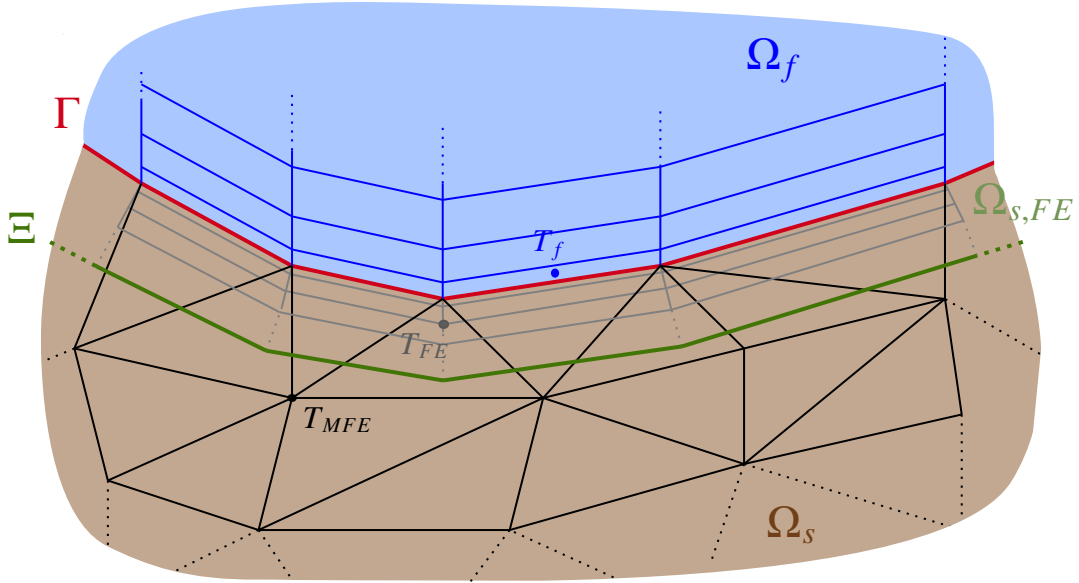
The truncation of the modal basis results in higher-order frequencies not being captured by the modal decomposition. This residual flux is expected to have a small penetration depth. Therefore, to capture it  we propose using a finite-element grid of relatively small thickness compared to the overall thickness of the solid domain.  This local grid is referred to as the \textit{FE} solution. \\
A sketch of the different grids used is available in \autoref{fig:conjugate}. The body-following grid (in blue) represents the classical cell-centered CFD mesh solving for  $T_f$, the coarser  grid  (black lines, $T_{MFE}$) corresponds to the node-based modal grid defined on $\Omega_s$, referred to as the \textit{MFE} solution, used to capture the large scale thermal field and drive it to statistical steady-state. The finer grid  (in light grey, $T_{FE}$) is the node-based \textit{FE} grid, defined on $\Omega_{s,\text{FE}}\in \Omega_s$. Finally, the  interface between $\Omega_f $ and $\Omega_s $ is noted $\Gamma$ while the remaining interface between $\Omega_s$ and $\Omega_{s,FE}$ is $\Xi$

 The \textit{FE} grid is directly extruded from the normal at the interface $\Gamma$ due to its conduction-dominated nature, see \cite{YANN}. This technique avoids the requirements of generating manually an additional mesh and as well as the overheads related to  separate data structures for the solid grid and provisions for interpolation of temperatures and fluxes between the two grids.   The \textit{FE} grid solves the unsteady heat equation using a finite element approach with local geodesic coordinates ($ \xi, \eta, \zeta$). The equivalent equations are available in  \autoref{thinlayereqn} (2D) and \autoref{thinlayereqn2} (3D), under the approximation of a small penetration depth and a smooth surface.  Further developments  and assumptions for the Laplace operator in geodesic coordinates are available in \ref{sec:thinlayer}.
\begin{align}
   \text{2D:} \qquad \dfrac{\partial T}{\partial t}&= \alpha\left(
    \dfrac{\partial^2 T}{\partial \chi^2}+
    \dfrac{\partial^2 T}{\partial \eta^2}+ \dfrac{1}{R_c}
    \dfrac{\partial T}{\partial \eta}\right)\label{thinlayereqn}\\
\label{thinlayereqn2}\text{3D:} \qquad    \dfrac{\partial T}{\partial t}&= \alpha\left(
    \dfrac{\partial^2 T}{\partial \chi^2}+
    \dfrac{\partial^2 T}{\partial \eta^2}+ \dfrac{\partial^2 T}{\partial \zeta^2}+\dfrac{1}{R_c}
    \dfrac{\partial T}{\partial \zeta}\right)
\end{align}
where $\frac 1R_c$ is the local harmonic mean curvature of the surface and $\alpha$ is the thermal diffusivity. 

For the domain $\Omega_{s,\text{FE}}$, the overlap between the \textit{MFE} and the \textit{FE} field is handled by solving the unsteady heat conduction equation for the combination $ \left(T_{FE}+T_{MFE}\right)$. Written in global coordinates for clarity, the compatibility equation is: 
\begin{equation}
    \rho c_p\frac{\partial \left(T_{FE}+T_{MFE}\right)}{\partial t}=
    \kappa\left( \frac{\partial^2}{\partial x^2}+\frac{\partial^2}{\partial y^2}+\frac{\partial^2}{\partial z^2}\right)\left(T_{FE}+T_{MFE}\right)
    \end{equation}
    The coupling strategy applies a Dirichlet boundary condition to the fluid side while a Robin boundary condition is applied on the solid solution. This approach was proven to be stable by \citet{Giles}.   This give the following boundary conditions at the interface $\Gamma$ between $\Omega_s$ and $\Omega_f$:
    \begin{equation}
        -\kappa \dfrac{\partial }{\partial n}\left(T_{FE}+T_{MFE}\right)=h(x) \left(\left(T_{FE}+T_{MFE}\right)-T_g(x,t)\right)+ \beta(x,t)\qquad \text{on } \Gamma 
     \end{equation}
    The weak form of the equation is obtained by multiplying by a test function $F_i$ and integrating over the \textit{FE}  domain $\Omega_{s,\text{FE}}$:
    \begin{equation}
    \rho c_p    \int_{\Omega_{s,\text{FE}}}{F_i \frac{\partial }{\partial t}\left(T_{FE}+T_{MFE}\right)d\Omega}= 
    \kappa \int_{\Omega_{s,\text{FE}}}{F_i \frac{\partial^2}{\partial x_k \partial x_k}\left(T_{FE}+T_{MFE}\right) d\Omega}
    \end{equation}
    Integrating by parts yields and describing the thermal field with the shape functions of the \textit{FE} grid:
    \begin{align}
    \rho c_p \left(   \int_{\Omega_{s,\text{FE}}}{F_i F_jd\Omega} \right) \frac{\partial T_{FE}^j}{\partial t} &= 
    -\kappa \left( \int_{\Omega_{s,\text{FE}}}{\frac{\partial F_i}{\partial x_k}\frac{\partial F_j}{\partial x_k} d\Omega}  \right) T_{FE}^j 
    +\kappa \int_{\Gamma} F_i \frac{\partial }{\partial n}\left( T_{FE}+T_{MFE}\right)dS  \, \nonumber\\
    & +\kappa \int_{\Xi} F_i \frac{\partial }{\partial n}\left( T_{FE}+T_{MFE}\right)dS -\rho c_p    \int_{\Omega_{s,\text{FE}}}{F_i \frac{\partial T_\text{MFE}}{\partial t}d\Omega}-
    \kappa \int_{\Omega_{s,\text{FE}}}{\frac{\partial F_i}{\partial x_k}\frac{\partial T_\text{MFE}}{\partial x_k} d\Omega} 
    \end{align}
    Substituting the boundary condition 
    \begin{align}
        \rho c_p \left(   \int_{\Omega_{s,\text{FE}}}{F_i F_jd\Omega} \right) \frac{\partial T_{FE}^j}{\partial t} &= 
        -\kappa \left( \int_{\Omega_{s,\text{FE}}}{\frac{\partial F_i}{\partial x_k}\frac{\partial F_j}{\partial x_k} d\Omega}  + \int_{\Gamma}h(x){F_i F_jdS}\right) T_{FE}^j 
        + \int_{\Gamma} F_i \left(h(x) \left(T_g-T_{MFE}-T_{FE}\right) +\beta \right) dS  \, \nonumber\\
        & +\kappa \int_{\Xi} F_i \frac{\partial }{\partial n}\left( T_{FE}+T_{MFE}\right)dS -\rho c_p    \int_{\Omega_{s,\text{FE}}}{F_i \frac{\partial T_\text{MFE}}{\partial t}d\Omega}-
        \kappa \int_{\Omega_{s,\text{FE}}}{\frac{\partial F_i}{\partial x_k}\frac{\partial T_\text{MFE}}{\partial x_k} d\Omega} 
        \label{eq:overlap}
    \end{align}

\autoref{eq:overlap} shows that the surface perturbation temperature field is driven by the gas temperature reduced by the modal contribution to the solid temperature. 
The left-hand side and the first two terms on the right-hand side of  \autoref{eq:overlap} represent the usual mass and conductivity matrix for the finite element problem in the surface layer $\Omega_{s,FE}$. The remaining terms contain the modal temperature field and its derivatives weighted by the shape functions of the surface finite element space and represent forcing terms. The functional form of these terms also indicates that the temperature mode shapes are only needed at the numerical integration points of the surface layer grid and its boundary.  
Alternatively, if the modal flux at the interface $\Xi$ is assumed to be exact, the correction terms could be neglected, allowing the thermal field within the thin layer domain to be entirely described by the \text{FE} solution, overriding the \textit{MFE} field. This assumption makes the $FE$ implementation more straightforward, as the modal field is only used to provide the boundary conditions for the \textit{FE} solution.

The extent of the \textit{FE} domain should be chosen according to the frequency range present at the interface as well as the frequency range that the remaining modal basis is able to capture. As an example, we can study the error made by a coarse truncated modal basis as a function of the surface layer extent $\delta_{\text{FE}}$. For that purpose, the problem of \autoref{sec:1D} is studied again with a harmonic external forcing. The reference solution is a finite element solution with a grid resolution able to capture well the forcing. On the other hand, the grid for the  modal basis is purposefully coarse, with five grid points within  $[0,\delta_P]$. Finally, the \textit{FE} grid has the same resolution as the reference solution.  
\autoref{fig:error_thin_layer} shows the relative energy error with the thin layer extend and the modal truncation level.
As expected, the error decreases with the thin layer extend. The slope of the error is greater closer to the interface where the higher density captures better the fluctuations until $\delta_{\text{FE}} /\delta_P \approx 1$ where the slope decreases. The error is also dependent on the truncation level of the modal basis. For this case, omitting half of the modes introduces a relative error less than 1\%. However, the error increases exponentially with further truncations. 
Finally, the case where the \textit{FE}  solution overrides the \textit{MFE} solution is plotted using the dotted line in \autoref{fig:error_thin_layer} with 70\% of the modes kept. When the \textit{FE} grid does not extend far enough within the solid, the error made by the modal solution invalidates the assumption and therefore the model performs worst. However, when $\delta_{FE}/\delta_P\gtrapprox3$, neglecting the additional terms offers similar level of performance.



%% file: figures/1Dsolid.tex
\tikzset{every picture/.style={line width=0.75pt}} 

\begin{tikzpicture}[x=0.75pt,y=0.75pt,yscale=-1,xscale=1]

\draw  [fill={rgb, 255:red, 186; green, 170; blue, 166 }  ,fill opacity=1 ] (204,96) -- (420.75,96) -- (420.75,156) -- (204,156) -- cycle ;
\draw [color={rgb, 255:red, 252; green, 2; blue, 31 }  ,draw opacity=1 ]   (460.75,125.75) -- (423.75,125.75) ;
\draw [shift={(420.75,125.75)}, rotate = 360] [fill={rgb, 255:red, 252; green, 2; blue, 31 }  ,fill opacity=1 ][line width=0.08]  [draw opacity=0] (8.93,-4.29) -- (0,0) -- (8.93,4.29) -- cycle    ;
\draw [color={rgb, 255:red, 255; green, 0; blue, 31 }  ,draw opacity=1 ]   (155.58,122.75) .. controls (158.38,106.92) and (160.49,114.54) .. (162.59,122.45) .. controls (164.75,130.56) and (166.89,138.97) .. (169.77,122.75) ;
\draw [color={rgb, 255:red, 255; green, 0; blue, 31 }  ,draw opacity=1 ]   (183.96,122.75) .. controls (188.16,102.3) and (187.33,123.35) .. (201.17,123.29) ;
\draw [shift={(204,123)}, rotate = 182.56] [fill={rgb, 255:red, 255; green, 0; blue, 31 }  ,fill opacity=1 ][line width=0.08]  [draw opacity=0] (8.93,-4.29) -- (0,0) -- (8.93,4.29) -- cycle    ;
\draw [color={rgb, 255:red, 255; green, 0; blue, 31 }  ,draw opacity=1 ]   (169.77,122.75) .. controls (175.44,90.69) and (178.28,154.8) .. (183.96,122.75) ;
\draw [color={rgb, 255:red, 255; green, 0; blue, 31 }  ,draw opacity=1 ]   (141.39,122.75) .. controls (144.19,106.92) and (146.3,114.54) .. (148.4,122.45) .. controls (150.56,130.56) and (152.7,138.97) .. (155.58,122.75) ;
\draw    (204,84) -- (240,84) ;
\draw [shift={(240,84)}, rotate = 180] [color={rgb, 255:red, 0; green, 0; blue, 0 }  ][line width=0.75]    (10.93,-3.29) .. controls (6.95,-1.4) and (3.31,-0.3) .. (0,0) .. controls (3.31,0.3) and (6.95,1.4) .. (10.93,3.29)   ;
\draw [shift={(204,84)}, rotate = 180] [color={rgb, 255:red, 0; green, 0; blue, 0 }  ][line width=0.75]    (0,5.59) -- (0,-5.59)   ;
\draw    (204,168) -- (240,168) -- (420,168) ;
\draw [shift={(420,168)}, rotate = 180] [color={rgb, 255:red, 0; green, 0; blue, 0 }  ][line width=0.75]    (0,5.59) -- (0,-5.59)   ;
\draw [shift={(204,168)}, rotate = 180] [color={rgb, 255:red, 0; green, 0; blue, 0 }  ][line width=0.75]    (0,5.59) -- (0,-5.59)   ;

\draw (135.5,88.4) node [anchor=north west][inner sep=0.75pt]   [align=left] {\textcolor[rgb]{1,0,0.12}{T}\textcolor[rgb]{1,0,0.12}{$\displaystyle ( t)$}\textcolor[rgb]{1,0,0.12}{= F}\textcolor[rgb]{1,0,0.12}{$\displaystyle _{\text{ext}}$}};
\draw (423.92,100.92) node [anchor=north west][inner sep=0.75pt]   [align=left] {\textcolor[rgb]{1,0,0.12}{q}\textcolor[rgb]{1,0,0.12}{$\displaystyle ''$}\textcolor[rgb]{1,0,0.12}{ = 0}};
\draw (291.5,115.75) node [anchor=north west][inner sep=0.75pt]   [align=left] {$\displaystyle [ \kappa _{s} ,\rho_sc_{p}]$};
\draw (229,65) node [anchor=north west][inner sep=0.75pt]   [align=left] {x};
\draw (308,170.33) node [anchor=north west][inner sep=0.75pt]   [align=left] {L};

\end{tikzpicture}

%% file: figures/mode.pgf
\begingroup%
\makeatletter%
\begin{pgfpicture}%
\pgfpathrectangle{\pgfpointorigin}{\pgfqpoint{2.000000in}{2.300000in}}%
\pgfusepath{use as bounding box, clip}%
\begin{pgfscope}%
\pgfsetbuttcap%
\pgfsetmiterjoin%
\definecolor{currentfill}{rgb}{1.000000,1.000000,1.000000}%
\pgfsetfillcolor{currentfill}%
\pgfsetlinewidth{0.000000pt}%
\definecolor{currentstroke}{rgb}{1.000000,1.000000,1.000000}%
\pgfsetstrokecolor{currentstroke}%
\pgfsetdash{}{0pt}%
\pgfpathmoveto{\pgfqpoint{0.000000in}{0.000000in}}%
\pgfpathlineto{\pgfqpoint{2.000000in}{0.000000in}}%
\pgfpathlineto{\pgfqpoint{2.000000in}{2.300000in}}%
\pgfpathlineto{\pgfqpoint{0.000000in}{2.300000in}}%
\pgfpathlineto{\pgfqpoint{0.000000in}{0.000000in}}%
\pgfpathclose%
\pgfusepath{fill}%
\end{pgfscope}%
\begin{pgfscope}%
\pgfsetbuttcap%
\pgfsetmiterjoin%
\definecolor{currentfill}{rgb}{1.000000,1.000000,1.000000}%
\pgfsetfillcolor{currentfill}%
\pgfsetlinewidth{0.000000pt}%
\definecolor{currentstroke}{rgb}{0.000000,0.000000,0.000000}%
\pgfsetstrokecolor{currentstroke}%
\pgfsetstrokeopacity{0.000000}%
\pgfsetdash{}{0pt}%
\pgfpathmoveto{\pgfqpoint{0.510000in}{0.379500in}}%
\pgfpathlineto{\pgfqpoint{1.500000in}{0.379500in}}%
\pgfpathlineto{\pgfqpoint{1.500000in}{2.254000in}}%
\pgfpathlineto{\pgfqpoint{0.510000in}{2.254000in}}%
\pgfpathlineto{\pgfqpoint{0.510000in}{0.379500in}}%
\pgfpathclose%
\pgfusepath{fill}%
\end{pgfscope}%
\begin{pgfscope}%
\pgfpathrectangle{\pgfqpoint{0.510000in}{0.379500in}}{\pgfqpoint{0.990000in}{1.874500in}}%
\pgfusepath{clip}%
\pgfsetrectcap%
\pgfsetroundjoin%
\pgfsetlinewidth{0.803000pt}%
\definecolor{currentstroke}{rgb}{0.690196,0.690196,0.690196}%
\pgfsetstrokecolor{currentstroke}%
\pgfsetdash{}{0pt}%
\pgfpathmoveto{\pgfqpoint{0.555000in}{0.379500in}}%
\pgfpathlineto{\pgfqpoint{0.555000in}{2.254000in}}%
\pgfusepath{stroke}%
\end{pgfscope}%
\begin{pgfscope}%
\pgfsetbuttcap%
\pgfsetroundjoin%
\definecolor{currentfill}{rgb}{0.000000,0.000000,0.000000}%
\pgfsetfillcolor{currentfill}%
\pgfsetlinewidth{0.803000pt}%
\definecolor{currentstroke}{rgb}{0.000000,0.000000,0.000000}%
\pgfsetstrokecolor{currentstroke}%
\pgfsetdash{}{0pt}%
\pgfsys@defobject{currentmarker}{\pgfqpoint{0.000000in}{-0.048611in}}{\pgfqpoint{0.000000in}{0.000000in}}{%
\pgfpathmoveto{\pgfqpoint{0.000000in}{0.000000in}}%
\pgfpathlineto{\pgfqpoint{0.000000in}{-0.048611in}}%
\pgfusepath{stroke,fill}%
}%
\begin{pgfscope}%
\pgfsys@transformshift{0.555000in}{0.379500in}%
\pgfsys@useobject{currentmarker}{}%
\end{pgfscope}%
\end{pgfscope}%
\begin{pgfscope}%
\definecolor{textcolor}{rgb}{0.000000,0.000000,0.000000}%
\pgfsetstrokecolor{textcolor}%
\pgfsetfillcolor{textcolor}%
\pgftext[x=0.555000in,y=0.282278in,,top]{\color{textcolor}{\rmfamily\fontsize{9.000000}{10.800000}\selectfont\catcode`\^=\active\def^{\ifmmode\sp\else\^{}\fi}\catcode`\%=\active\def
\end{pgfscope}%
\begin{pgfscope}%
\pgfpathrectangle{\pgfqpoint{0.510000in}{0.379500in}}{\pgfqpoint{0.990000in}{1.874500in}}%
\pgfusepath{clip}%
\pgfsetrectcap%
\pgfsetroundjoin%
\pgfsetlinewidth{0.803000pt}%
\definecolor{currentstroke}{rgb}{0.690196,0.690196,0.690196}%
\pgfsetstrokecolor{currentstroke}%
\pgfsetdash{}{0pt}%
\pgfpathmoveto{\pgfqpoint{1.455000in}{0.379500in}}%
\pgfpathlineto{\pgfqpoint{1.455000in}{2.254000in}}%
\pgfusepath{stroke}%
\end{pgfscope}%
\begin{pgfscope}%
\pgfsetbuttcap%
\pgfsetroundjoin%
\definecolor{currentfill}{rgb}{0.000000,0.000000,0.000000}%
\pgfsetfillcolor{currentfill}%
\pgfsetlinewidth{0.803000pt}%
\definecolor{currentstroke}{rgb}{0.000000,0.000000,0.000000}%
\pgfsetstrokecolor{currentstroke}%
\pgfsetdash{}{0pt}%
\pgfsys@defobject{currentmarker}{\pgfqpoint{0.000000in}{-0.048611in}}{\pgfqpoint{0.000000in}{0.000000in}}{%
\pgfpathmoveto{\pgfqpoint{0.000000in}{0.000000in}}%
\pgfpathlineto{\pgfqpoint{0.000000in}{-0.048611in}}%
\pgfusepath{stroke,fill}%
}%
\begin{pgfscope}%
\pgfsys@transformshift{1.455000in}{0.379500in}%
\pgfsys@useobject{currentmarker}{}%
\end{pgfscope}%
\end{pgfscope}%
\begin{pgfscope}%
\definecolor{textcolor}{rgb}{0.000000,0.000000,0.000000}%
\pgfsetstrokecolor{textcolor}%
\pgfsetfillcolor{textcolor}%
\pgftext[x=1.455000in,y=0.282278in,,top]{\color{textcolor}{\rmfamily\fontsize{9.000000}{10.800000}\selectfont\catcode`\^=\active\def^{\ifmmode\sp\else\^{}\fi}\catcode`\%=\active\def
\end{pgfscope}%
\begin{pgfscope}%
\definecolor{textcolor}{rgb}{0.000000,0.000000,0.000000}%
\pgfsetstrokecolor{textcolor}%
\pgfsetfillcolor{textcolor}%
\pgftext[x=1.005000in,y=0.115611in,,top]{\color{textcolor}{\rmfamily\fontsize{10.000000}{12.000000}\selectfont\catcode`\^=\active\def^{\ifmmode\sp\else\^{}\fi}\catcode`\%=\active\def
\end{pgfscope}%
\begin{pgfscope}%
\pgfpathrectangle{\pgfqpoint{0.510000in}{0.379500in}}{\pgfqpoint{0.990000in}{1.874500in}}%
\pgfusepath{clip}%
\pgfsetrectcap%
\pgfsetroundjoin%
\pgfsetlinewidth{0.803000pt}%
\definecolor{currentstroke}{rgb}{0.690196,0.690196,0.690196}%
\pgfsetstrokecolor{currentstroke}%
\pgfsetdash{}{0pt}%
\pgfpathmoveto{\pgfqpoint{0.510000in}{0.410735in}}%
\pgfpathlineto{\pgfqpoint{1.500000in}{0.410735in}}%
\pgfusepath{stroke}%
\end{pgfscope}%
\begin{pgfscope}%
\pgfsetbuttcap%
\pgfsetroundjoin%
\definecolor{currentfill}{rgb}{0.000000,0.000000,0.000000}%
\pgfsetfillcolor{currentfill}%
\pgfsetlinewidth{0.803000pt}%
\definecolor{currentstroke}{rgb}{0.000000,0.000000,0.000000}%
\pgfsetstrokecolor{currentstroke}%
\pgfsetdash{}{0pt}%
\pgfsys@defobject{currentmarker}{\pgfqpoint{-0.048611in}{0.000000in}}{\pgfqpoint{-0.000000in}{0.000000in}}{%
\pgfpathmoveto{\pgfqpoint{-0.000000in}{0.000000in}}%
\pgfpathlineto{\pgfqpoint{-0.048611in}{0.000000in}}%
\pgfusepath{stroke,fill}%
}%
\begin{pgfscope}%
\pgfsys@transformshift{0.510000in}{0.410735in}%
\pgfsys@useobject{currentmarker}{}%
\end{pgfscope}%
\end{pgfscope}%
\begin{pgfscope}%
\definecolor{textcolor}{rgb}{0.000000,0.000000,0.000000}%
\pgfsetstrokecolor{textcolor}%
\pgfsetfillcolor{textcolor}%
\pgftext[x=0.084462in, y=0.367332in, left, base]{\color{textcolor}{\rmfamily\fontsize{9.000000}{10.800000}\selectfont\catcode`\^=\active\def^{\ifmmode\sp\else\^{}\fi}\catcode`\%=\active\def
\end{pgfscope}%
\begin{pgfscope}%
\pgfpathrectangle{\pgfqpoint{0.510000in}{0.379500in}}{\pgfqpoint{0.990000in}{1.874500in}}%
\pgfusepath{clip}%
\pgfsetrectcap%
\pgfsetroundjoin%
\pgfsetlinewidth{0.803000pt}%
\definecolor{currentstroke}{rgb}{0.690196,0.690196,0.690196}%
\pgfsetstrokecolor{currentstroke}%
\pgfsetdash{}{0pt}%
\pgfpathmoveto{\pgfqpoint{0.510000in}{0.712740in}}%
\pgfpathlineto{\pgfqpoint{1.500000in}{0.712740in}}%
\pgfusepath{stroke}%
\end{pgfscope}%
\begin{pgfscope}%
\pgfsetbuttcap%
\pgfsetroundjoin%
\definecolor{currentfill}{rgb}{0.000000,0.000000,0.000000}%
\pgfsetfillcolor{currentfill}%
\pgfsetlinewidth{0.803000pt}%
\definecolor{currentstroke}{rgb}{0.000000,0.000000,0.000000}%
\pgfsetstrokecolor{currentstroke}%
\pgfsetdash{}{0pt}%
\pgfsys@defobject{currentmarker}{\pgfqpoint{-0.048611in}{0.000000in}}{\pgfqpoint{-0.000000in}{0.000000in}}{%
\pgfpathmoveto{\pgfqpoint{-0.000000in}{0.000000in}}%
\pgfpathlineto{\pgfqpoint{-0.048611in}{0.000000in}}%
\pgfusepath{stroke,fill}%
}%
\begin{pgfscope}%
\pgfsys@transformshift{0.510000in}{0.712740in}%
\pgfsys@useobject{currentmarker}{}%
\end{pgfscope}%
\end{pgfscope}%
\begin{pgfscope}%
\definecolor{textcolor}{rgb}{0.000000,0.000000,0.000000}%
\pgfsetstrokecolor{textcolor}%
\pgfsetfillcolor{textcolor}%
\pgftext[x=0.084462in, y=0.669337in, left, base]{\color{textcolor}{\rmfamily\fontsize{9.000000}{10.800000}\selectfont\catcode`\^=\active\def^{\ifmmode\sp\else\^{}\fi}\catcode`\%=\active\def
\end{pgfscope}%
\begin{pgfscope}%
\pgfpathrectangle{\pgfqpoint{0.510000in}{0.379500in}}{\pgfqpoint{0.990000in}{1.874500in}}%
\pgfusepath{clip}%
\pgfsetrectcap%
\pgfsetroundjoin%
\pgfsetlinewidth{0.803000pt}%
\definecolor{currentstroke}{rgb}{0.690196,0.690196,0.690196}%
\pgfsetstrokecolor{currentstroke}%
\pgfsetdash{}{0pt}%
\pgfpathmoveto{\pgfqpoint{0.510000in}{1.014745in}}%
\pgfpathlineto{\pgfqpoint{1.500000in}{1.014745in}}%
\pgfusepath{stroke}%
\end{pgfscope}%
\begin{pgfscope}%
\pgfsetbuttcap%
\pgfsetroundjoin%
\definecolor{currentfill}{rgb}{0.000000,0.000000,0.000000}%
\pgfsetfillcolor{currentfill}%
\pgfsetlinewidth{0.803000pt}%
\definecolor{currentstroke}{rgb}{0.000000,0.000000,0.000000}%
\pgfsetstrokecolor{currentstroke}%
\pgfsetdash{}{0pt}%
\pgfsys@defobject{currentmarker}{\pgfqpoint{-0.048611in}{0.000000in}}{\pgfqpoint{-0.000000in}{0.000000in}}{%
\pgfpathmoveto{\pgfqpoint{-0.000000in}{0.000000in}}%
\pgfpathlineto{\pgfqpoint{-0.048611in}{0.000000in}}%
\pgfusepath{stroke,fill}%
}%
\begin{pgfscope}%
\pgfsys@transformshift{0.510000in}{1.014745in}%
\pgfsys@useobject{currentmarker}{}%
\end{pgfscope}%
\end{pgfscope}%
\begin{pgfscope}%
\definecolor{textcolor}{rgb}{0.000000,0.000000,0.000000}%
\pgfsetstrokecolor{textcolor}%
\pgfsetfillcolor{textcolor}%
\pgftext[x=0.084462in, y=0.971342in, left, base]{\color{textcolor}{\rmfamily\fontsize{9.000000}{10.800000}\selectfont\catcode`\^=\active\def^{\ifmmode\sp\else\^{}\fi}\catcode`\%=\active\def
\end{pgfscope}%
\begin{pgfscope}%
\pgfpathrectangle{\pgfqpoint{0.510000in}{0.379500in}}{\pgfqpoint{0.990000in}{1.874500in}}%
\pgfusepath{clip}%
\pgfsetrectcap%
\pgfsetroundjoin%
\pgfsetlinewidth{0.803000pt}%
\definecolor{currentstroke}{rgb}{0.690196,0.690196,0.690196}%
\pgfsetstrokecolor{currentstroke}%
\pgfsetdash{}{0pt}%
\pgfpathmoveto{\pgfqpoint{0.510000in}{1.316750in}}%
\pgfpathlineto{\pgfqpoint{1.500000in}{1.316750in}}%
\pgfusepath{stroke}%
\end{pgfscope}%
\begin{pgfscope}%
\pgfsetbuttcap%
\pgfsetroundjoin%
\definecolor{currentfill}{rgb}{0.000000,0.000000,0.000000}%
\pgfsetfillcolor{currentfill}%
\pgfsetlinewidth{0.803000pt}%
\definecolor{currentstroke}{rgb}{0.000000,0.000000,0.000000}%
\pgfsetstrokecolor{currentstroke}%
\pgfsetdash{}{0pt}%
\pgfsys@defobject{currentmarker}{\pgfqpoint{-0.048611in}{0.000000in}}{\pgfqpoint{-0.000000in}{0.000000in}}{%
\pgfpathmoveto{\pgfqpoint{-0.000000in}{0.000000in}}%
\pgfpathlineto{\pgfqpoint{-0.048611in}{0.000000in}}%
\pgfusepath{stroke,fill}%
}%
\begin{pgfscope}%
\pgfsys@transformshift{0.510000in}{1.316750in}%
\pgfsys@useobject{currentmarker}{}%
\end{pgfscope}%
\end{pgfscope}%
\begin{pgfscope}%
\definecolor{textcolor}{rgb}{0.000000,0.000000,0.000000}%
\pgfsetstrokecolor{textcolor}%
\pgfsetfillcolor{textcolor}%
\pgftext[x=0.184384in, y=1.273347in, left, base]{\color{textcolor}{\rmfamily\fontsize{9.000000}{10.800000}\selectfont\catcode`\^=\active\def^{\ifmmode\sp\else\^{}\fi}\catcode`\%=\active\def
\end{pgfscope}%
\begin{pgfscope}%
\pgfpathrectangle{\pgfqpoint{0.510000in}{0.379500in}}{\pgfqpoint{0.990000in}{1.874500in}}%
\pgfusepath{clip}%
\pgfsetrectcap%
\pgfsetroundjoin%
\pgfsetlinewidth{0.803000pt}%
\definecolor{currentstroke}{rgb}{0.690196,0.690196,0.690196}%
\pgfsetstrokecolor{currentstroke}%
\pgfsetdash{}{0pt}%
\pgfpathmoveto{\pgfqpoint{0.510000in}{1.618755in}}%
\pgfpathlineto{\pgfqpoint{1.500000in}{1.618755in}}%
\pgfusepath{stroke}%
\end{pgfscope}%
\begin{pgfscope}%
\pgfsetbuttcap%
\pgfsetroundjoin%
\definecolor{currentfill}{rgb}{0.000000,0.000000,0.000000}%
\pgfsetfillcolor{currentfill}%
\pgfsetlinewidth{0.803000pt}%
\definecolor{currentstroke}{rgb}{0.000000,0.000000,0.000000}%
\pgfsetstrokecolor{currentstroke}%
\pgfsetdash{}{0pt}%
\pgfsys@defobject{currentmarker}{\pgfqpoint{-0.048611in}{0.000000in}}{\pgfqpoint{-0.000000in}{0.000000in}}{%
\pgfpathmoveto{\pgfqpoint{-0.000000in}{0.000000in}}%
\pgfpathlineto{\pgfqpoint{-0.048611in}{0.000000in}}%
\pgfusepath{stroke,fill}%
}%
\begin{pgfscope}%
\pgfsys@transformshift{0.510000in}{1.618755in}%
\pgfsys@useobject{currentmarker}{}%
\end{pgfscope}%
\end{pgfscope}%
\begin{pgfscope}%
\definecolor{textcolor}{rgb}{0.000000,0.000000,0.000000}%
\pgfsetstrokecolor{textcolor}%
\pgfsetfillcolor{textcolor}%
\pgftext[x=0.184384in, y=1.575352in, left, base]{\color{textcolor}{\rmfamily\fontsize{9.000000}{10.800000}\selectfont\catcode`\^=\active\def^{\ifmmode\sp\else\^{}\fi}\catcode`\%=\active\def
\end{pgfscope}%
\begin{pgfscope}%
\pgfpathrectangle{\pgfqpoint{0.510000in}{0.379500in}}{\pgfqpoint{0.990000in}{1.874500in}}%
\pgfusepath{clip}%
\pgfsetrectcap%
\pgfsetroundjoin%
\pgfsetlinewidth{0.803000pt}%
\definecolor{currentstroke}{rgb}{0.690196,0.690196,0.690196}%
\pgfsetstrokecolor{currentstroke}%
\pgfsetdash{}{0pt}%
\pgfpathmoveto{\pgfqpoint{0.510000in}{1.920760in}}%
\pgfpathlineto{\pgfqpoint{1.500000in}{1.920760in}}%
\pgfusepath{stroke}%
\end{pgfscope}%
\begin{pgfscope}%
\pgfsetbuttcap%
\pgfsetroundjoin%
\definecolor{currentfill}{rgb}{0.000000,0.000000,0.000000}%
\pgfsetfillcolor{currentfill}%
\pgfsetlinewidth{0.803000pt}%
\definecolor{currentstroke}{rgb}{0.000000,0.000000,0.000000}%
\pgfsetstrokecolor{currentstroke}%
\pgfsetdash{}{0pt}%
\pgfsys@defobject{currentmarker}{\pgfqpoint{-0.048611in}{0.000000in}}{\pgfqpoint{-0.000000in}{0.000000in}}{%
\pgfpathmoveto{\pgfqpoint{-0.000000in}{0.000000in}}%
\pgfpathlineto{\pgfqpoint{-0.048611in}{0.000000in}}%
\pgfusepath{stroke,fill}%
}%
\begin{pgfscope}%
\pgfsys@transformshift{0.510000in}{1.920760in}%
\pgfsys@useobject{currentmarker}{}%
\end{pgfscope}%
\end{pgfscope}%
\begin{pgfscope}%
\definecolor{textcolor}{rgb}{0.000000,0.000000,0.000000}%
\pgfsetstrokecolor{textcolor}%
\pgfsetfillcolor{textcolor}%
\pgftext[x=0.184384in, y=1.877357in, left, base]{\color{textcolor}{\rmfamily\fontsize{9.000000}{10.800000}\selectfont\catcode`\^=\active\def^{\ifmmode\sp\else\^{}\fi}\catcode`\%=\active\def
\end{pgfscope}%
\begin{pgfscope}%
\pgfpathrectangle{\pgfqpoint{0.510000in}{0.379500in}}{\pgfqpoint{0.990000in}{1.874500in}}%
\pgfusepath{clip}%
\pgfsetrectcap%
\pgfsetroundjoin%
\pgfsetlinewidth{0.803000pt}%
\definecolor{currentstroke}{rgb}{0.690196,0.690196,0.690196}%
\pgfsetstrokecolor{currentstroke}%
\pgfsetdash{}{0pt}%
\pgfpathmoveto{\pgfqpoint{0.510000in}{2.222765in}}%
\pgfpathlineto{\pgfqpoint{1.500000in}{2.222765in}}%
\pgfusepath{stroke}%
\end{pgfscope}%
\begin{pgfscope}%
\pgfsetbuttcap%
\pgfsetroundjoin%
\definecolor{currentfill}{rgb}{0.000000,0.000000,0.000000}%
\pgfsetfillcolor{currentfill}%
\pgfsetlinewidth{0.803000pt}%
\definecolor{currentstroke}{rgb}{0.000000,0.000000,0.000000}%
\pgfsetstrokecolor{currentstroke}%
\pgfsetdash{}{0pt}%
\pgfsys@defobject{currentmarker}{\pgfqpoint{-0.048611in}{0.000000in}}{\pgfqpoint{-0.000000in}{0.000000in}}{%
\pgfpathmoveto{\pgfqpoint{-0.000000in}{0.000000in}}%
\pgfpathlineto{\pgfqpoint{-0.048611in}{0.000000in}}%
\pgfusepath{stroke,fill}%
}%
\begin{pgfscope}%
\pgfsys@transformshift{0.510000in}{2.222765in}%
\pgfsys@useobject{currentmarker}{}%
\end{pgfscope}%
\end{pgfscope}%
\begin{pgfscope}%
\definecolor{textcolor}{rgb}{0.000000,0.000000,0.000000}%
\pgfsetstrokecolor{textcolor}%
\pgfsetfillcolor{textcolor}%
\pgftext[x=0.184384in, y=2.179362in, left, base]{\color{textcolor}{\rmfamily\fontsize{9.000000}{10.800000}\selectfont\catcode`\^=\active\def^{\ifmmode\sp\else\^{}\fi}\catcode`\%=\active\def
\end{pgfscope}%
\begin{pgfscope}%
\definecolor{textcolor}{rgb}{0.000000,0.000000,0.000000}%
\pgfsetstrokecolor{textcolor}%
\pgfsetfillcolor{textcolor}%
\pgftext[x=0.1048907in,y=1.316750in,,bottom,rotate=90.000000]{\color{textcolor}{\rmfamily\fontsize{10.000000}{12.000000}\selectfont\catcode`\^=\active\def^{\ifmmode\sp\else\^{}\fi}\catcode`\%=\active\def
\end{pgfscope}%
\begin{pgfscope}%
\pgfpathrectangle{\pgfqpoint{0.510000in}{0.379500in}}{\pgfqpoint{0.990000in}{1.874500in}}%
\pgfusepath{clip}%
\pgfsetrectcap%
\pgfsetroundjoin%
\pgfsetlinewidth{1.505625pt}%
\definecolor{currentstroke}{rgb}{0.000000,0.000000,0.000000}%
\pgfsetstrokecolor{currentstroke}%
\pgfsetdash{}{0pt}%
\pgfpathmoveto{\pgfqpoint{0.555000in}{1.566084in}}%
\pgfpathlineto{\pgfqpoint{0.564091in}{1.793590in}}%
\pgfpathlineto{\pgfqpoint{0.573182in}{1.979353in}}%
\pgfpathlineto{\pgfqpoint{0.582273in}{2.107109in}}%
\pgfpathlineto{\pgfqpoint{0.591364in}{2.165675in}}%
\pgfpathlineto{\pgfqpoint{0.600455in}{2.149924in}}%
\pgfpathlineto{\pgfqpoint{0.609545in}{2.061234in}}%
\pgfpathlineto{\pgfqpoint{0.618636in}{1.907370in}}%
\pgfpathlineto{\pgfqpoint{0.627727in}{1.701801in}}%
\pgfpathlineto{\pgfqpoint{0.636818in}{1.462524in}}%
\pgfpathlineto{\pgfqpoint{0.645909in}{1.210486in}}%
\pgfpathlineto{\pgfqpoint{0.655000in}{0.967750in}}%
\pgfpathlineto{\pgfqpoint{0.664091in}{0.755566in}}%
\pgfpathlineto{\pgfqpoint{0.673182in}{0.592511in}}%
\pgfpathlineto{\pgfqpoint{0.682273in}{0.492857in}}%
\pgfpathlineto{\pgfqpoint{0.691364in}{0.465329in}}%
\pgfpathlineto{\pgfqpoint{0.700455in}{0.512337in}}%
\pgfpathlineto{\pgfqpoint{0.709545in}{0.629766in}}%
\pgfpathlineto{\pgfqpoint{0.718636in}{0.807335in}}%
\pgfpathlineto{\pgfqpoint{0.727727in}{1.029500in}}%
\pgfpathlineto{\pgfqpoint{0.736818in}{1.276811in}}%
\pgfpathlineto{\pgfqpoint{0.745909in}{1.527619in}}%
\pgfpathlineto{\pgfqpoint{0.755000in}{1.759967in}}%
\pgfpathlineto{\pgfqpoint{0.764091in}{1.953515in}}%
\pgfpathlineto{\pgfqpoint{0.773182in}{2.091318in}}%
\pgfpathlineto{\pgfqpoint{0.782273in}{2.161313in}}%
\pgfpathlineto{\pgfqpoint{0.791364in}{2.157373in}}%
\pgfpathlineto{\pgfqpoint{0.800455in}{2.079842in}}%
\pgfpathlineto{\pgfqpoint{0.809545in}{1.935508in}}%
\pgfpathlineto{\pgfqpoint{0.818636in}{1.737006in}}%
\pgfpathlineto{\pgfqpoint{0.827727in}{1.501714in}}%
\pgfpathlineto{\pgfqpoint{0.836818in}{1.250229in}}%
\pgfpathlineto{\pgfqpoint{0.845909in}{1.004568in}}%
\pgfpathlineto{\pgfqpoint{0.855000in}{0.786236in}}%
\pgfpathlineto{\pgfqpoint{0.864091in}{0.614347in}}%
\pgfpathlineto{\pgfqpoint{0.873182in}{0.503948in}}%
\pgfpathlineto{\pgfqpoint{0.882273in}{0.464705in}}%
\pgfpathlineto{\pgfqpoint{0.891364in}{0.500051in}}%
\pgfpathlineto{\pgfqpoint{0.900455in}{0.606894in}}%
\pgfpathlineto{\pgfqpoint{0.909545in}{0.775880in}}%
\pgfpathlineto{\pgfqpoint{0.918636in}{0.992215in}}%
\pgfpathlineto{\pgfqpoint{0.927727in}{1.236961in}}%
\pgfpathlineto{\pgfqpoint{0.936818in}{1.488692in}}%
\pgfpathlineto{\pgfqpoint{0.945909in}{1.725370in}}%
\pgfpathlineto{\pgfqpoint{0.955000in}{1.926277in}}%
\pgfpathlineto{\pgfqpoint{0.964091in}{2.073824in}}%
\pgfpathlineto{\pgfqpoint{0.973182in}{2.155094in}}%
\pgfpathlineto{\pgfqpoint{0.982273in}{2.162974in}}%
\pgfpathlineto{\pgfqpoint{0.991364in}{2.096773in}}%
\pgfpathlineto{\pgfqpoint{1.000455in}{1.962286in}}%
\pgfpathlineto{\pgfqpoint{1.009545in}{1.771287in}}%
\pgfpathlineto{\pgfqpoint{1.018636in}{1.540497in}}%
\pgfpathlineto{\pgfqpoint{1.027727in}{1.290119in}}%
\pgfpathlineto{\pgfqpoint{1.036818in}{1.042073in}}%
\pgfpathlineto{\pgfqpoint{1.045909in}{0.818072in}}%
\pgfpathlineto{\pgfqpoint{1.055000in}{0.637728in}}%
\pgfpathlineto{\pgfqpoint{1.064091in}{0.516827in}}%
\pgfpathlineto{\pgfqpoint{1.073182in}{0.465953in}}%
\pgfpathlineto{\pgfqpoint{1.082273in}{0.489561in}}%
\pgfpathlineto{\pgfqpoint{1.091364in}{0.585583in}}%
\pgfpathlineto{\pgfqpoint{1.100455in}{0.745614in}}%
\pgfpathlineto{\pgfqpoint{1.109545in}{0.955644in}}%
\pgfpathlineto{\pgfqpoint{1.118636in}{1.197285in}}%
\pgfpathlineto{\pgfqpoint{1.127727in}{1.449386in}}%
\pgfpathlineto{\pgfqpoint{1.136818in}{1.689875in}}%
\pgfpathlineto{\pgfqpoint{1.145909in}{1.897699in}}%
\pgfpathlineto{\pgfqpoint{1.155000in}{2.054666in}}%
\pgfpathlineto{\pgfqpoint{1.164091in}{2.147033in}}%
\pgfpathlineto{\pgfqpoint{1.173182in}{2.166715in}}%
\pgfpathlineto{\pgfqpoint{1.182273in}{2.111988in}}%
\pgfpathlineto{\pgfqpoint{1.191364in}{1.987645in}}%
\pgfpathlineto{\pgfqpoint{1.200455in}{1.804569in}}%
\pgfpathlineto{\pgfqpoint{1.209545in}{1.578788in}}%
\pgfpathlineto{\pgfqpoint{1.218636in}{1.330067in}}%
\pgfpathlineto{\pgfqpoint{1.227727in}{1.080181in}}%
\pgfpathlineto{\pgfqpoint{1.236818in}{0.851005in}}%
\pgfpathlineto{\pgfqpoint{1.245909in}{0.662601in}}%
\pgfpathlineto{\pgfqpoint{1.255000in}{0.531463in}}%
\pgfpathlineto{\pgfqpoint{1.264091in}{0.469072in}}%
\pgfpathlineto{\pgfqpoint{1.273182in}{0.480889in}}%
\pgfpathlineto{\pgfqpoint{1.282273in}{0.565879in}}%
\pgfpathlineto{\pgfqpoint{1.291364in}{0.716603in}}%
\pgfpathlineto{\pgfqpoint{1.300455in}{0.919866in}}%
\pgfpathlineto{\pgfqpoint{1.309545in}{1.157873in}}%
\pgfpathlineto{\pgfqpoint{1.318636in}{1.409788in}}%
\pgfpathlineto{\pgfqpoint{1.327727in}{1.653559in}}%
\pgfpathlineto{\pgfqpoint{1.336818in}{1.867844in}}%
\pgfpathlineto{\pgfqpoint{1.345909in}{2.033885in}}%
\pgfpathlineto{\pgfqpoint{1.355000in}{2.137146in}}%
\pgfpathlineto{\pgfqpoint{1.364091in}{2.168587in}}%
\pgfpathlineto{\pgfqpoint{1.373182in}{2.125456in}}%
\pgfpathlineto{\pgfqpoint{1.382273in}{2.011529in}}%
\pgfpathlineto{\pgfqpoint{1.391364in}{1.836778in}}%
\pgfpathlineto{\pgfqpoint{1.400455in}{1.616503in}}%
\pgfpathlineto{\pgfqpoint{1.409545in}{1.369986in}}%
\pgfpathlineto{\pgfqpoint{1.418636in}{1.118809in}}%
\pgfpathlineto{\pgfqpoint{1.427727in}{0.884961in}}%
\pgfpathlineto{\pgfqpoint{1.436818in}{0.688912in}}%
\pgfpathlineto{\pgfqpoint{1.445909in}{0.547826in}}%
\pgfpathlineto{\pgfqpoint{1.455000in}{0.474054in}}%
\pgfusepath{stroke}%
\end{pgfscope}%
\begin{pgfscope}%
\pgfpathrectangle{\pgfqpoint{0.510000in}{0.379500in}}{\pgfqpoint{0.990000in}{1.874500in}}%
\pgfusepath{clip}%
\pgfsetrectcap%
\pgfsetroundjoin%
\pgfsetlinewidth{1.505625pt}%
\definecolor{currentstroke}{rgb}{0.000000,0.000000,1.000000}%
\pgfsetstrokecolor{currentstroke}%
\pgfsetdash{}{0pt}%
\pgfpathmoveto{\pgfqpoint{0.555000in}{1.276811in}}%
\pgfpathlineto{\pgfqpoint{0.564091in}{1.236961in}}%
\pgfpathlineto{\pgfqpoint{0.573182in}{1.197285in}}%
\pgfpathlineto{\pgfqpoint{0.582273in}{1.157873in}}%
\pgfpathlineto{\pgfqpoint{0.591364in}{1.118809in}}%
\pgfpathlineto{\pgfqpoint{0.600455in}{1.080181in}}%
\pgfpathlineto{\pgfqpoint{0.609545in}{1.042073in}}%
\pgfpathlineto{\pgfqpoint{0.618636in}{1.004568in}}%
\pgfpathlineto{\pgfqpoint{0.627727in}{0.967750in}}%
\pgfpathlineto{\pgfqpoint{0.636818in}{0.931699in}}%
\pgfpathlineto{\pgfqpoint{0.645909in}{0.896494in}}%
\pgfpathlineto{\pgfqpoint{0.655000in}{0.862213in}}%
\pgfpathlineto{\pgfqpoint{0.664091in}{0.828931in}}%
\pgfpathlineto{\pgfqpoint{0.673182in}{0.796722in}}%
\pgfpathlineto{\pgfqpoint{0.682273in}{0.765656in}}%
\pgfpathlineto{\pgfqpoint{0.691364in}{0.735801in}}%
\pgfpathlineto{\pgfqpoint{0.700455in}{0.707223in}}%
\pgfpathlineto{\pgfqpoint{0.709545in}{0.679985in}}%
\pgfpathlineto{\pgfqpoint{0.718636in}{0.654147in}}%
\pgfpathlineto{\pgfqpoint{0.727727in}{0.629766in}}%
\pgfpathlineto{\pgfqpoint{0.736818in}{0.606894in}}%
\pgfpathlineto{\pgfqpoint{0.745909in}{0.585583in}}%
\pgfpathlineto{\pgfqpoint{0.755000in}{0.565879in}}%
\pgfpathlineto{\pgfqpoint{0.764091in}{0.547826in}}%
\pgfpathlineto{\pgfqpoint{0.773182in}{0.531463in}}%
\pgfpathlineto{\pgfqpoint{0.782273in}{0.516827in}}%
\pgfpathlineto{\pgfqpoint{0.791364in}{0.503948in}}%
\pgfpathlineto{\pgfqpoint{0.800455in}{0.492857in}}%
\pgfpathlineto{\pgfqpoint{0.809545in}{0.483576in}}%
\pgfpathlineto{\pgfqpoint{0.818636in}{0.476127in}}%
\pgfpathlineto{\pgfqpoint{0.827727in}{0.470526in}}%
\pgfpathlineto{\pgfqpoint{0.836818in}{0.466785in}}%
\pgfpathlineto{\pgfqpoint{0.845909in}{0.464913in}}%
\pgfpathlineto{\pgfqpoint{0.855000in}{0.464913in}}%
\pgfpathlineto{\pgfqpoint{0.864091in}{0.466785in}}%
\pgfpathlineto{\pgfqpoint{0.873182in}{0.470526in}}%
\pgfpathlineto{\pgfqpoint{0.882273in}{0.476127in}}%
\pgfpathlineto{\pgfqpoint{0.891364in}{0.483576in}}%
\pgfpathlineto{\pgfqpoint{0.900455in}{0.492857in}}%
\pgfpathlineto{\pgfqpoint{0.909545in}{0.503948in}}%
\pgfpathlineto{\pgfqpoint{0.918636in}{0.516827in}}%
\pgfpathlineto{\pgfqpoint{0.927727in}{0.531463in}}%
\pgfpathlineto{\pgfqpoint{0.936818in}{0.547826in}}%
\pgfpathlineto{\pgfqpoint{0.945909in}{0.565879in}}%
\pgfpathlineto{\pgfqpoint{0.955000in}{0.585583in}}%
\pgfpathlineto{\pgfqpoint{0.964091in}{0.606894in}}%
\pgfpathlineto{\pgfqpoint{0.973182in}{0.629766in}}%
\pgfpathlineto{\pgfqpoint{0.982273in}{0.654147in}}%
\pgfpathlineto{\pgfqpoint{0.991364in}{0.679985in}}%
\pgfpathlineto{\pgfqpoint{1.000455in}{0.707223in}}%
\pgfpathlineto{\pgfqpoint{1.009545in}{0.735801in}}%
\pgfpathlineto{\pgfqpoint{1.018636in}{0.765656in}}%
\pgfpathlineto{\pgfqpoint{1.027727in}{0.796722in}}%
\pgfpathlineto{\pgfqpoint{1.036818in}{0.828931in}}%
\pgfpathlineto{\pgfqpoint{1.045909in}{0.862213in}}%
\pgfpathlineto{\pgfqpoint{1.055000in}{0.896494in}}%
\pgfpathlineto{\pgfqpoint{1.064091in}{0.931699in}}%
\pgfpathlineto{\pgfqpoint{1.073182in}{0.967750in}}%
\pgfpathlineto{\pgfqpoint{1.082273in}{1.004568in}}%
\pgfpathlineto{\pgfqpoint{1.091364in}{1.042073in}}%
\pgfpathlineto{\pgfqpoint{1.100455in}{1.080181in}}%
\pgfpathlineto{\pgfqpoint{1.109545in}{1.118809in}}%
\pgfpathlineto{\pgfqpoint{1.118636in}{1.157873in}}%
\pgfpathlineto{\pgfqpoint{1.127727in}{1.197285in}}%
\pgfpathlineto{\pgfqpoint{1.136818in}{1.236961in}}%
\pgfpathlineto{\pgfqpoint{1.145909in}{1.276811in}}%
\pgfpathlineto{\pgfqpoint{1.155000in}{1.316750in}}%
\pgfpathlineto{\pgfqpoint{1.164091in}{1.356689in}}%
\pgfpathlineto{\pgfqpoint{1.173182in}{1.396539in}}%
\pgfpathlineto{\pgfqpoint{1.182273in}{1.436215in}}%
\pgfpathlineto{\pgfqpoint{1.191364in}{1.475627in}}%
\pgfpathlineto{\pgfqpoint{1.200455in}{1.514691in}}%
\pgfpathlineto{\pgfqpoint{1.209545in}{1.553319in}}%
\pgfpathlineto{\pgfqpoint{1.218636in}{1.591427in}}%
\pgfpathlineto{\pgfqpoint{1.227727in}{1.628932in}}%
\pgfpathlineto{\pgfqpoint{1.236818in}{1.665750in}}%
\pgfpathlineto{\pgfqpoint{1.245909in}{1.701801in}}%
\pgfpathlineto{\pgfqpoint{1.255000in}{1.737006in}}%
\pgfpathlineto{\pgfqpoint{1.264091in}{1.771287in}}%
\pgfpathlineto{\pgfqpoint{1.273182in}{1.804569in}}%
\pgfpathlineto{\pgfqpoint{1.282273in}{1.836778in}}%
\pgfpathlineto{\pgfqpoint{1.291364in}{1.867844in}}%
\pgfpathlineto{\pgfqpoint{1.300455in}{1.897699in}}%
\pgfpathlineto{\pgfqpoint{1.309545in}{1.926277in}}%
\pgfpathlineto{\pgfqpoint{1.318636in}{1.953515in}}%
\pgfpathlineto{\pgfqpoint{1.327727in}{1.979353in}}%
\pgfpathlineto{\pgfqpoint{1.336818in}{2.003734in}}%
\pgfpathlineto{\pgfqpoint{1.345909in}{2.026606in}}%
\pgfpathlineto{\pgfqpoint{1.355000in}{2.047917in}}%
\pgfpathlineto{\pgfqpoint{1.364091in}{2.067621in}}%
\pgfpathlineto{\pgfqpoint{1.373182in}{2.085674in}}%
\pgfpathlineto{\pgfqpoint{1.382273in}{2.102037in}}%
\pgfpathlineto{\pgfqpoint{1.391364in}{2.116673in}}%
\pgfpathlineto{\pgfqpoint{1.400455in}{2.129552in}}%
\pgfpathlineto{\pgfqpoint{1.409545in}{2.140643in}}%
\pgfpathlineto{\pgfqpoint{1.418636in}{2.149924in}}%
\pgfpathlineto{\pgfqpoint{1.427727in}{2.157373in}}%
\pgfpathlineto{\pgfqpoint{1.436818in}{2.162974in}}%
\pgfpathlineto{\pgfqpoint{1.445909in}{2.166715in}}%
\pgfpathlineto{\pgfqpoint{1.455000in}{2.168587in}}%
\pgfusepath{stroke}%
\end{pgfscope}%
\begin{pgfscope}%
\pgfpathrectangle{\pgfqpoint{0.510000in}{0.379500in}}{\pgfqpoint{0.990000in}{1.874500in}}%
\pgfusepath{clip}%
\pgfsetrectcap%
\pgfsetroundjoin%
\pgfsetlinewidth{1.505625pt}%
\definecolor{currentstroke}{rgb}{1.000000,0.000000,0.000000}%
\pgfsetstrokecolor{currentstroke}%
\pgfsetdash{}{0pt}%
\pgfpathmoveto{\pgfqpoint{0.555000in}{1.330067in}}%
\pgfpathlineto{\pgfqpoint{0.564091in}{1.343381in}}%
\pgfpathlineto{\pgfqpoint{0.573182in}{1.356689in}}%
\pgfpathlineto{\pgfqpoint{0.582273in}{1.369986in}}%
\pgfpathlineto{\pgfqpoint{0.591364in}{1.383271in}}%
\pgfpathlineto{\pgfqpoint{0.600455in}{1.396539in}}%
\pgfpathlineto{\pgfqpoint{0.609545in}{1.409788in}}%
\pgfpathlineto{\pgfqpoint{0.618636in}{1.423014in}}%
\pgfpathlineto{\pgfqpoint{0.627727in}{1.436215in}}%
\pgfpathlineto{\pgfqpoint{0.636818in}{1.449386in}}%
\pgfpathlineto{\pgfqpoint{0.645909in}{1.462524in}}%
\pgfpathlineto{\pgfqpoint{0.655000in}{1.475627in}}%
\pgfpathlineto{\pgfqpoint{0.664091in}{1.488692in}}%
\pgfpathlineto{\pgfqpoint{0.673182in}{1.501714in}}%
\pgfpathlineto{\pgfqpoint{0.682273in}{1.514691in}}%
\pgfpathlineto{\pgfqpoint{0.691364in}{1.527619in}}%
\pgfpathlineto{\pgfqpoint{0.700455in}{1.540497in}}%
\pgfpathlineto{\pgfqpoint{0.709545in}{1.553319in}}%
\pgfpathlineto{\pgfqpoint{0.718636in}{1.566084in}}%
\pgfpathlineto{\pgfqpoint{0.727727in}{1.578788in}}%
\pgfpathlineto{\pgfqpoint{0.736818in}{1.591427in}}%
\pgfpathlineto{\pgfqpoint{0.745909in}{1.604000in}}%
\pgfpathlineto{\pgfqpoint{0.755000in}{1.616503in}}%
\pgfpathlineto{\pgfqpoint{0.764091in}{1.628932in}}%
\pgfpathlineto{\pgfqpoint{0.773182in}{1.641285in}}%
\pgfpathlineto{\pgfqpoint{0.782273in}{1.653559in}}%
\pgfpathlineto{\pgfqpoint{0.791364in}{1.665750in}}%
\pgfpathlineto{\pgfqpoint{0.800455in}{1.677856in}}%
\pgfpathlineto{\pgfqpoint{0.809545in}{1.689875in}}%
\pgfpathlineto{\pgfqpoint{0.818636in}{1.701801in}}%
\pgfpathlineto{\pgfqpoint{0.827727in}{1.713634in}}%
\pgfpathlineto{\pgfqpoint{0.836818in}{1.725370in}}%
\pgfpathlineto{\pgfqpoint{0.845909in}{1.737006in}}%
\pgfpathlineto{\pgfqpoint{0.855000in}{1.748539in}}%
\pgfpathlineto{\pgfqpoint{0.864091in}{1.759967in}}%
\pgfpathlineto{\pgfqpoint{0.873182in}{1.771287in}}%
\pgfpathlineto{\pgfqpoint{0.882273in}{1.782495in}}%
\pgfpathlineto{\pgfqpoint{0.891364in}{1.793590in}}%
\pgfpathlineto{\pgfqpoint{0.900455in}{1.804569in}}%
\pgfpathlineto{\pgfqpoint{0.909545in}{1.815428in}}%
\pgfpathlineto{\pgfqpoint{0.918636in}{1.826165in}}%
\pgfpathlineto{\pgfqpoint{0.927727in}{1.836778in}}%
\pgfpathlineto{\pgfqpoint{0.936818in}{1.847264in}}%
\pgfpathlineto{\pgfqpoint{0.945909in}{1.857620in}}%
\pgfpathlineto{\pgfqpoint{0.955000in}{1.867844in}}%
\pgfpathlineto{\pgfqpoint{0.964091in}{1.877934in}}%
\pgfpathlineto{\pgfqpoint{0.973182in}{1.887886in}}%
\pgfpathlineto{\pgfqpoint{0.982273in}{1.897699in}}%
\pgfpathlineto{\pgfqpoint{0.991364in}{1.907370in}}%
\pgfpathlineto{\pgfqpoint{1.000455in}{1.916897in}}%
\pgfpathlineto{\pgfqpoint{1.009545in}{1.926277in}}%
\pgfpathlineto{\pgfqpoint{1.018636in}{1.935508in}}%
\pgfpathlineto{\pgfqpoint{1.027727in}{1.944588in}}%
\pgfpathlineto{\pgfqpoint{1.036818in}{1.953515in}}%
\pgfpathlineto{\pgfqpoint{1.045909in}{1.962286in}}%
\pgfpathlineto{\pgfqpoint{1.055000in}{1.970899in}}%
\pgfpathlineto{\pgfqpoint{1.064091in}{1.979353in}}%
\pgfpathlineto{\pgfqpoint{1.073182in}{1.987645in}}%
\pgfpathlineto{\pgfqpoint{1.082273in}{1.995772in}}%
\pgfpathlineto{\pgfqpoint{1.091364in}{2.003734in}}%
\pgfpathlineto{\pgfqpoint{1.100455in}{2.011529in}}%
\pgfpathlineto{\pgfqpoint{1.109545in}{2.019153in}}%
\pgfpathlineto{\pgfqpoint{1.118636in}{2.026606in}}%
\pgfpathlineto{\pgfqpoint{1.127727in}{2.033885in}}%
\pgfpathlineto{\pgfqpoint{1.136818in}{2.040989in}}%
\pgfpathlineto{\pgfqpoint{1.145909in}{2.047917in}}%
\pgfpathlineto{\pgfqpoint{1.155000in}{2.054666in}}%
\pgfpathlineto{\pgfqpoint{1.164091in}{2.061234in}}%
\pgfpathlineto{\pgfqpoint{1.173182in}{2.067621in}}%
\pgfpathlineto{\pgfqpoint{1.182273in}{2.073824in}}%
\pgfpathlineto{\pgfqpoint{1.191364in}{2.079842in}}%
\pgfpathlineto{\pgfqpoint{1.200455in}{2.085674in}}%
\pgfpathlineto{\pgfqpoint{1.209545in}{2.091318in}}%
\pgfpathlineto{\pgfqpoint{1.218636in}{2.096773in}}%
\pgfpathlineto{\pgfqpoint{1.227727in}{2.102037in}}%
\pgfpathlineto{\pgfqpoint{1.236818in}{2.107109in}}%
\pgfpathlineto{\pgfqpoint{1.245909in}{2.111988in}}%
\pgfpathlineto{\pgfqpoint{1.255000in}{2.116673in}}%
\pgfpathlineto{\pgfqpoint{1.264091in}{2.121163in}}%
\pgfpathlineto{\pgfqpoint{1.273182in}{2.125456in}}%
\pgfpathlineto{\pgfqpoint{1.282273in}{2.129552in}}%
\pgfpathlineto{\pgfqpoint{1.291364in}{2.133449in}}%
\pgfpathlineto{\pgfqpoint{1.300455in}{2.137146in}}%
\pgfpathlineto{\pgfqpoint{1.309545in}{2.140643in}}%
\pgfpathlineto{\pgfqpoint{1.318636in}{2.143939in}}%
\pgfpathlineto{\pgfqpoint{1.327727in}{2.147033in}}%
\pgfpathlineto{\pgfqpoint{1.336818in}{2.149924in}}%
\pgfpathlineto{\pgfqpoint{1.345909in}{2.152611in}}%
\pgfpathlineto{\pgfqpoint{1.355000in}{2.155094in}}%
\pgfpathlineto{\pgfqpoint{1.364091in}{2.157373in}}%
\pgfpathlineto{\pgfqpoint{1.373182in}{2.159446in}}%
\pgfpathlineto{\pgfqpoint{1.382273in}{2.161313in}}%
\pgfpathlineto{\pgfqpoint{1.391364in}{2.162974in}}%
\pgfpathlineto{\pgfqpoint{1.400455in}{2.164428in}}%
\pgfpathlineto{\pgfqpoint{1.409545in}{2.165675in}}%
\pgfpathlineto{\pgfqpoint{1.418636in}{2.166715in}}%
\pgfpathlineto{\pgfqpoint{1.427727in}{2.167547in}}%
\pgfpathlineto{\pgfqpoint{1.436818in}{2.168171in}}%
\pgfpathlineto{\pgfqpoint{1.445909in}{2.168587in}}%
\pgfpathlineto{\pgfqpoint{1.455000in}{2.168795in}}%
\pgfusepath{stroke}%
\end{pgfscope}%
\begin{pgfscope}%
\pgfsetrectcap%
\pgfsetmiterjoin%
\pgfsetlinewidth{0.803000pt}%
\definecolor{currentstroke}{rgb}{0.000000,0.000000,0.000000}%
\pgfsetstrokecolor{currentstroke}%
\pgfsetdash{}{0pt}%
\pgfpathmoveto{\pgfqpoint{0.510000in}{0.379500in}}%
\pgfpathlineto{\pgfqpoint{0.510000in}{2.254000in}}%
\pgfusepath{stroke}%
\end{pgfscope}%
\begin{pgfscope}%
\pgfsetrectcap%
\pgfsetmiterjoin%
\pgfsetlinewidth{0.803000pt}%
\definecolor{currentstroke}{rgb}{0.000000,0.000000,0.000000}%
\pgfsetstrokecolor{currentstroke}%
\pgfsetdash{}{0pt}%
\pgfpathmoveto{\pgfqpoint{1.500000in}{0.379500in}}%
\pgfpathlineto{\pgfqpoint{1.500000in}{2.254000in}}%
\pgfusepath{stroke}%
\end{pgfscope}%
\begin{pgfscope}%
\pgfsetrectcap%
\pgfsetmiterjoin%
\pgfsetlinewidth{0.803000pt}%
\definecolor{currentstroke}{rgb}{0.000000,0.000000,0.000000}%
\pgfsetstrokecolor{currentstroke}%
\pgfsetdash{}{0pt}%
\pgfpathmoveto{\pgfqpoint{0.510000in}{0.379500in}}%
\pgfpathlineto{\pgfqpoint{1.500000in}{0.379500in}}%
\pgfusepath{stroke}%
\end{pgfscope}%
\begin{pgfscope}%
\pgfsetrectcap%
\pgfsetmiterjoin%
\pgfsetlinewidth{0.803000pt}%
\definecolor{currentstroke}{rgb}{0.000000,0.000000,0.000000}%
\pgfsetstrokecolor{currentstroke}%
\pgfsetdash{}{0pt}%
\pgfpathmoveto{\pgfqpoint{0.510000in}{2.254000in}}%
\pgfpathlineto{\pgfqpoint{1.500000in}{2.254000in}}%
\pgfusepath{stroke}%
\end{pgfscope}%
\end{pgfpicture}%
\makeatother%
\endgroup%

%% file: figures/acc_type_mean_only.pgf
\begingroup%
\makeatletter%
\begin{pgfpicture}%
\pgfpathrectangle{\pgfpointorigin}{\pgfqpoint{3.300000in}{3.200000in}}%
\pgfusepath{use as bounding box, clip}%
\begin{pgfscope}%
\pgfsetbuttcap%
\pgfsetmiterjoin%
\definecolor{currentfill}{rgb}{1.000000,1.000000,1.000000}%
\pgfsetfillcolor{currentfill}%
\pgfsetlinewidth{0.000000pt}%
\definecolor{currentstroke}{rgb}{1.000000,1.000000,1.000000}%
\pgfsetstrokecolor{currentstroke}%
\pgfsetdash{}{0pt}%
\pgfpathmoveto{\pgfqpoint{0.000000in}{0.000000in}}%
\pgfpathlineto{\pgfqpoint{3.300000in}{0.000000in}}%
\pgfpathlineto{\pgfqpoint{3.300000in}{3.200000in}}%
\pgfpathlineto{\pgfqpoint{0.000000in}{3.200000in}}%
\pgfpathlineto{\pgfqpoint{0.000000in}{0.000000in}}%
\pgfpathclose%
\pgfusepath{fill}%
\end{pgfscope}%
\begin{pgfscope}%
\pgfsetbuttcap%
\pgfsetmiterjoin%
\definecolor{currentfill}{rgb}{1.000000,1.000000,1.000000}%
\pgfsetfillcolor{currentfill}%
\pgfsetlinewidth{0.000000pt}%
\definecolor{currentstroke}{rgb}{0.000000,0.000000,0.000000}%
\pgfsetstrokecolor{currentstroke}%
\pgfsetstrokeopacity{0.000000}%
\pgfsetdash{}{0pt}%
\pgfpathmoveto{\pgfqpoint{0.594000in}{0.480000in}}%
\pgfpathlineto{\pgfqpoint{2.706000in}{0.480000in}}%
\pgfpathlineto{\pgfqpoint{2.706000in}{2.880000in}}%
\pgfpathlineto{\pgfqpoint{0.594000in}{2.880000in}}%
\pgfpathlineto{\pgfqpoint{0.594000in}{0.480000in}}%
\pgfpathclose%
\pgfusepath{fill}%
\end{pgfscope}%
\begin{pgfscope}%
\pgfpathrectangle{\pgfqpoint{0.594000in}{0.480000in}}{\pgfqpoint{2.112000in}{2.400000in}}%
\pgfusepath{clip}%
\pgfsetrectcap%
\pgfsetroundjoin%
\pgfsetlinewidth{0.803000pt}%
\definecolor{currentstroke}{rgb}{0.690196,0.690196,0.690196}%
\pgfsetstrokecolor{currentstroke}%
\pgfsetdash{}{0pt}%
\pgfpathmoveto{\pgfqpoint{0.594000in}{0.480000in}}%
\pgfpathlineto{\pgfqpoint{0.594000in}{2.880000in}}%
\pgfusepath{stroke}%
\end{pgfscope}%
\begin{pgfscope}%
\pgfsetbuttcap%
\pgfsetroundjoin%
\definecolor{currentfill}{rgb}{0.000000,0.000000,0.000000}%
\pgfsetfillcolor{currentfill}%
\pgfsetlinewidth{0.803000pt}%
\definecolor{currentstroke}{rgb}{0.000000,0.000000,0.000000}%
\pgfsetstrokecolor{currentstroke}%
\pgfsetdash{}{0pt}%
\pgfsys@defobject{currentmarker}{\pgfqpoint{0.000000in}{-0.048611in}}{\pgfqpoint{0.000000in}{0.000000in}}{%
\pgfpathmoveto{\pgfqpoint{0.000000in}{0.000000in}}%
\pgfpathlineto{\pgfqpoint{0.000000in}{-0.048611in}}%
\pgfusepath{stroke,fill}%
}%
\begin{pgfscope}%
\pgfsys@transformshift{0.594000in}{0.480000in}%
\pgfsys@useobject{currentmarker}{}%
\end{pgfscope}%
\end{pgfscope}%
\begin{pgfscope}%
\definecolor{textcolor}{rgb}{0.000000,0.000000,0.000000}%
\pgfsetstrokecolor{textcolor}%
\pgfsetfillcolor{textcolor}%
\pgftext[x=0.594000in,y=0.382778in,,top]{\color{textcolor}{\rmfamily\fontsize{9.000000}{10.800000}\selectfont\catcode`\^=\active\def^{\ifmmode\sp\else\^{}\fi}\catcode`\%=\active\def
\end{pgfscope}%
\begin{pgfscope}%
\pgfpathrectangle{\pgfqpoint{0.594000in}{0.480000in}}{\pgfqpoint{2.112000in}{2.400000in}}%
\pgfusepath{clip}%
\pgfsetrectcap%
\pgfsetroundjoin%
\pgfsetlinewidth{0.803000pt}%
\definecolor{currentstroke}{rgb}{0.690196,0.690196,0.690196}%
\pgfsetstrokecolor{currentstroke}%
\pgfsetdash{}{0pt}%
\pgfpathmoveto{\pgfqpoint{1.016400in}{0.480000in}}%
\pgfpathlineto{\pgfqpoint{1.016400in}{2.880000in}}%
\pgfusepath{stroke}%
\end{pgfscope}%
\begin{pgfscope}%
\pgfsetbuttcap%
\pgfsetroundjoin%
\definecolor{currentfill}{rgb}{0.000000,0.000000,0.000000}%
\pgfsetfillcolor{currentfill}%
\pgfsetlinewidth{0.803000pt}%
\definecolor{currentstroke}{rgb}{0.000000,0.000000,0.000000}%
\pgfsetstrokecolor{currentstroke}%
\pgfsetdash{}{0pt}%
\pgfsys@defobject{currentmarker}{\pgfqpoint{0.000000in}{-0.048611in}}{\pgfqpoint{0.000000in}{0.000000in}}{%
\pgfpathmoveto{\pgfqpoint{0.000000in}{0.000000in}}%
\pgfpathlineto{\pgfqpoint{0.000000in}{-0.048611in}}%
\pgfusepath{stroke,fill}%
}%
\begin{pgfscope}%
\pgfsys@transformshift{1.016400in}{0.480000in}%
\pgfsys@useobject{currentmarker}{}%
\end{pgfscope}%
\end{pgfscope}%
\begin{pgfscope}%
\definecolor{textcolor}{rgb}{0.000000,0.000000,0.000000}%
\pgfsetstrokecolor{textcolor}%
\pgfsetfillcolor{textcolor}%
\pgftext[x=1.016400in,y=0.382778in,,top]{\color{textcolor}{\rmfamily\fontsize{9.000000}{10.800000}\selectfont\catcode`\^=\active\def^{\ifmmode\sp\else\^{}\fi}\catcode`\%=\active\def
\end{pgfscope}%
\begin{pgfscope}%
\pgfpathrectangle{\pgfqpoint{0.594000in}{0.480000in}}{\pgfqpoint{2.112000in}{2.400000in}}%
\pgfusepath{clip}%
\pgfsetrectcap%
\pgfsetroundjoin%
\pgfsetlinewidth{0.803000pt}%
\definecolor{currentstroke}{rgb}{0.690196,0.690196,0.690196}%
\pgfsetstrokecolor{currentstroke}%
\pgfsetdash{}{0pt}%
\pgfpathmoveto{\pgfqpoint{1.438800in}{0.480000in}}%
\pgfpathlineto{\pgfqpoint{1.438800in}{2.880000in}}%
\pgfusepath{stroke}%
\end{pgfscope}%
\begin{pgfscope}%
\pgfsetbuttcap%
\pgfsetroundjoin%
\definecolor{currentfill}{rgb}{0.000000,0.000000,0.000000}%
\pgfsetfillcolor{currentfill}%
\pgfsetlinewidth{0.803000pt}%
\definecolor{currentstroke}{rgb}{0.000000,0.000000,0.000000}%
\pgfsetstrokecolor{currentstroke}%
\pgfsetdash{}{0pt}%
\pgfsys@defobject{currentmarker}{\pgfqpoint{0.000000in}{-0.048611in}}{\pgfqpoint{0.000000in}{0.000000in}}{%
\pgfpathmoveto{\pgfqpoint{0.000000in}{0.000000in}}%
\pgfpathlineto{\pgfqpoint{0.000000in}{-0.048611in}}%
\pgfusepath{stroke,fill}%
}%
\begin{pgfscope}%
\pgfsys@transformshift{1.438800in}{0.480000in}%
\pgfsys@useobject{currentmarker}{}%
\end{pgfscope}%
\end{pgfscope}%
\begin{pgfscope}%
\definecolor{textcolor}{rgb}{0.000000,0.000000,0.000000}%
\pgfsetstrokecolor{textcolor}%
\pgfsetfillcolor{textcolor}%
\pgftext[x=1.438800in,y=0.382778in,,top]{\color{textcolor}{\rmfamily\fontsize{9.000000}{10.800000}\selectfont\catcode`\^=\active\def^{\ifmmode\sp\else\^{}\fi}\catcode`\%=\active\def
\end{pgfscope}%
\begin{pgfscope}%
\pgfpathrectangle{\pgfqpoint{0.594000in}{0.480000in}}{\pgfqpoint{2.112000in}{2.400000in}}%
\pgfusepath{clip}%
\pgfsetrectcap%
\pgfsetroundjoin%
\pgfsetlinewidth{0.803000pt}%
\definecolor{currentstroke}{rgb}{0.690196,0.690196,0.690196}%
\pgfsetstrokecolor{currentstroke}%
\pgfsetdash{}{0pt}%
\pgfpathmoveto{\pgfqpoint{1.861200in}{0.480000in}}%
\pgfpathlineto{\pgfqpoint{1.861200in}{2.880000in}}%
\pgfusepath{stroke}%
\end{pgfscope}%
\begin{pgfscope}%
\pgfsetbuttcap%
\pgfsetroundjoin%
\definecolor{currentfill}{rgb}{0.000000,0.000000,0.000000}%
\pgfsetfillcolor{currentfill}%
\pgfsetlinewidth{0.803000pt}%
\definecolor{currentstroke}{rgb}{0.000000,0.000000,0.000000}%
\pgfsetstrokecolor{currentstroke}%
\pgfsetdash{}{0pt}%
\pgfsys@defobject{currentmarker}{\pgfqpoint{0.000000in}{-0.048611in}}{\pgfqpoint{0.000000in}{0.000000in}}{%
\pgfpathmoveto{\pgfqpoint{0.000000in}{0.000000in}}%
\pgfpathlineto{\pgfqpoint{0.000000in}{-0.048611in}}%
\pgfusepath{stroke,fill}%
}%
\begin{pgfscope}%
\pgfsys@transformshift{1.861200in}{0.480000in}%
\pgfsys@useobject{currentmarker}{}%
\end{pgfscope}%
\end{pgfscope}%
\begin{pgfscope}%
\definecolor{textcolor}{rgb}{0.000000,0.000000,0.000000}%
\pgfsetstrokecolor{textcolor}%
\pgfsetfillcolor{textcolor}%
\pgftext[x=1.861200in,y=0.382778in,,top]{\color{textcolor}{\rmfamily\fontsize{9.000000}{10.800000}\selectfont\catcode`\^=\active\def^{\ifmmode\sp\else\^{}\fi}\catcode`\%=\active\def
\end{pgfscope}%
\begin{pgfscope}%
\pgfpathrectangle{\pgfqpoint{0.594000in}{0.480000in}}{\pgfqpoint{2.112000in}{2.400000in}}%
\pgfusepath{clip}%
\pgfsetrectcap%
\pgfsetroundjoin%
\pgfsetlinewidth{0.803000pt}%
\definecolor{currentstroke}{rgb}{0.690196,0.690196,0.690196}%
\pgfsetstrokecolor{currentstroke}%
\pgfsetdash{}{0pt}%
\pgfpathmoveto{\pgfqpoint{2.283600in}{0.480000in}}%
\pgfpathlineto{\pgfqpoint{2.283600in}{2.880000in}}%
\pgfusepath{stroke}%
\end{pgfscope}%
\begin{pgfscope}%
\pgfsetbuttcap%
\pgfsetroundjoin%
\definecolor{currentfill}{rgb}{0.000000,0.000000,0.000000}%
\pgfsetfillcolor{currentfill}%
\pgfsetlinewidth{0.803000pt}%
\definecolor{currentstroke}{rgb}{0.000000,0.000000,0.000000}%
\pgfsetstrokecolor{currentstroke}%
\pgfsetdash{}{0pt}%
\pgfsys@defobject{currentmarker}{\pgfqpoint{0.000000in}{-0.048611in}}{\pgfqpoint{0.000000in}{0.000000in}}{%
\pgfpathmoveto{\pgfqpoint{0.000000in}{0.000000in}}%
\pgfpathlineto{\pgfqpoint{0.000000in}{-0.048611in}}%
\pgfusepath{stroke,fill}%
}%
\begin{pgfscope}%
\pgfsys@transformshift{2.283600in}{0.480000in}%
\pgfsys@useobject{currentmarker}{}%
\end{pgfscope}%
\end{pgfscope}%
\begin{pgfscope}%
\definecolor{textcolor}{rgb}{0.000000,0.000000,0.000000}%
\pgfsetstrokecolor{textcolor}%
\pgfsetfillcolor{textcolor}%
\pgftext[x=2.283600in,y=0.382778in,,top]{\color{textcolor}{\rmfamily\fontsize{9.000000}{10.800000}\selectfont\catcode`\^=\active\def^{\ifmmode\sp\else\^{}\fi}\catcode`\%=\active\def
\end{pgfscope}%
\begin{pgfscope}%
\pgfpathrectangle{\pgfqpoint{0.594000in}{0.480000in}}{\pgfqpoint{2.112000in}{2.400000in}}%
\pgfusepath{clip}%
\pgfsetrectcap%
\pgfsetroundjoin%
\pgfsetlinewidth{0.803000pt}%
\definecolor{currentstroke}{rgb}{0.690196,0.690196,0.690196}%
\pgfsetstrokecolor{currentstroke}%
\pgfsetdash{}{0pt}%
\pgfpathmoveto{\pgfqpoint{2.706000in}{0.480000in}}%
\pgfpathlineto{\pgfqpoint{2.706000in}{2.880000in}}%
\pgfusepath{stroke}%
\end{pgfscope}%
\begin{pgfscope}%
\pgfsetbuttcap%
\pgfsetroundjoin%
\definecolor{currentfill}{rgb}{0.000000,0.000000,0.000000}%
\pgfsetfillcolor{currentfill}%
\pgfsetlinewidth{0.803000pt}%
\definecolor{currentstroke}{rgb}{0.000000,0.000000,0.000000}%
\pgfsetstrokecolor{currentstroke}%
\pgfsetdash{}{0pt}%
\pgfsys@defobject{currentmarker}{\pgfqpoint{0.000000in}{-0.048611in}}{\pgfqpoint{0.000000in}{0.000000in}}{%
\pgfpathmoveto{\pgfqpoint{0.000000in}{0.000000in}}%
\pgfpathlineto{\pgfqpoint{0.000000in}{-0.048611in}}%
\pgfusepath{stroke,fill}%
}%
\begin{pgfscope}%
\pgfsys@transformshift{2.706000in}{0.480000in}%
\pgfsys@useobject{currentmarker}{}%
\end{pgfscope}%
\end{pgfscope}%
\begin{pgfscope}%
\definecolor{textcolor}{rgb}{0.000000,0.000000,0.000000}%
\pgfsetstrokecolor{textcolor}%
\pgfsetfillcolor{textcolor}%
\pgftext[x=2.706000in,y=0.382778in,,top]{\color{textcolor}{\rmfamily\fontsize{9.000000}{10.800000}\selectfont\catcode`\^=\active\def^{\ifmmode\sp\else\^{}\fi}\catcode`\%=\active\def
\end{pgfscope}%
\begin{pgfscope}%
\definecolor{textcolor}{rgb}{0.000000,0.000000,0.000000}%
\pgfsetstrokecolor{textcolor}%
\pgfsetfillcolor{textcolor}%
\pgftext[x=1.650000in,y=0.216111in,,top]{\color{textcolor}{\rmfamily\fontsize{10.000000}{12.000000}\selectfont\catcode`\^=\active\def^{\ifmmode\sp\else\^{}\fi}\catcode`\%=\active\def
\end{pgfscope}%
\begin{pgfscope}%
\pgfpathrectangle{\pgfqpoint{0.594000in}{0.480000in}}{\pgfqpoint{2.112000in}{2.400000in}}%
\pgfusepath{clip}%
\pgfsetrectcap%
\pgfsetroundjoin%
\pgfsetlinewidth{0.803000pt}%
\definecolor{currentstroke}{rgb}{0.690196,0.690196,0.690196}%
\pgfsetstrokecolor{currentstroke}%
\pgfsetdash{}{0pt}%
\pgfpathmoveto{\pgfqpoint{0.594000in}{0.589091in}}%
\pgfpathlineto{\pgfqpoint{2.706000in}{0.589091in}}%
\pgfusepath{stroke}%
\end{pgfscope}%
\begin{pgfscope}%
\pgfsetbuttcap%
\pgfsetroundjoin%
\definecolor{currentfill}{rgb}{0.000000,0.000000,0.000000}%
\pgfsetfillcolor{currentfill}%
\pgfsetlinewidth{0.803000pt}%
\definecolor{currentstroke}{rgb}{0.000000,0.000000,0.000000}%
\pgfsetstrokecolor{currentstroke}%
\pgfsetdash{}{0pt}%
\pgfsys@defobject{currentmarker}{\pgfqpoint{-0.048611in}{0.000000in}}{\pgfqpoint{-0.000000in}{0.000000in}}{%
\pgfpathmoveto{\pgfqpoint{-0.000000in}{0.000000in}}%
\pgfpathlineto{\pgfqpoint{-0.048611in}{0.000000in}}%
\pgfusepath{stroke,fill}%
}%
\begin{pgfscope}%
\pgfsys@transformshift{0.594000in}{0.589091in}%
\pgfsys@useobject{currentmarker}{}%
\end{pgfscope}%
\end{pgfscope}%
\begin{pgfscope}%
\definecolor{textcolor}{rgb}{0.000000,0.000000,0.000000}%
\pgfsetstrokecolor{textcolor}%
\pgfsetfillcolor{textcolor}%
\pgftext[x=0.268384in, y=0.545688in, left, base]{\color{textcolor}{\rmfamily\fontsize{9.000000}{10.800000}\selectfont\catcode`\^=\active\def^{\ifmmode\sp\else\^{}\fi}\catcode`\%=\active\def
\end{pgfscope}%
\begin{pgfscope}%
\pgfpathrectangle{\pgfqpoint{0.594000in}{0.480000in}}{\pgfqpoint{2.112000in}{2.400000in}}%
\pgfusepath{clip}%
\pgfsetrectcap%
\pgfsetroundjoin%
\pgfsetlinewidth{0.803000pt}%
\definecolor{currentstroke}{rgb}{0.690196,0.690196,0.690196}%
\pgfsetstrokecolor{currentstroke}%
\pgfsetdash{}{0pt}%
\pgfpathmoveto{\pgfqpoint{0.594000in}{0.856364in}}%
\pgfpathlineto{\pgfqpoint{2.706000in}{0.856364in}}%
\pgfusepath{stroke}%
\end{pgfscope}%
\begin{pgfscope}%
\pgfsetbuttcap%
\pgfsetroundjoin%
\definecolor{currentfill}{rgb}{0.000000,0.000000,0.000000}%
\pgfsetfillcolor{currentfill}%
\pgfsetlinewidth{0.803000pt}%
\definecolor{currentstroke}{rgb}{0.000000,0.000000,0.000000}%
\pgfsetstrokecolor{currentstroke}%
\pgfsetdash{}{0pt}%
\pgfsys@defobject{currentmarker}{\pgfqpoint{-0.048611in}{0.000000in}}{\pgfqpoint{-0.000000in}{0.000000in}}{%
\pgfpathmoveto{\pgfqpoint{-0.000000in}{0.000000in}}%
\pgfpathlineto{\pgfqpoint{-0.048611in}{0.000000in}}%
\pgfusepath{stroke,fill}%
}%
\begin{pgfscope}%
\pgfsys@transformshift{0.594000in}{0.856364in}%
\pgfsys@useobject{currentmarker}{}%
\end{pgfscope}%
\end{pgfscope}%
\begin{pgfscope}%
\definecolor{textcolor}{rgb}{0.000000,0.000000,0.000000}%
\pgfsetstrokecolor{textcolor}%
\pgfsetfillcolor{textcolor}%
\pgftext[x=0.268384in, y=0.812961in, left, base]{\color{textcolor}{\rmfamily\fontsize{9.000000}{10.800000}\selectfont\catcode`\^=\active\def^{\ifmmode\sp\else\^{}\fi}\catcode`\%=\active\def
\end{pgfscope}%
\begin{pgfscope}%
\pgfpathrectangle{\pgfqpoint{0.594000in}{0.480000in}}{\pgfqpoint{2.112000in}{2.400000in}}%
\pgfusepath{clip}%
\pgfsetrectcap%
\pgfsetroundjoin%
\pgfsetlinewidth{0.803000pt}%
\definecolor{currentstroke}{rgb}{0.690196,0.690196,0.690196}%
\pgfsetstrokecolor{currentstroke}%
\pgfsetdash{}{0pt}%
\pgfpathmoveto{\pgfqpoint{0.594000in}{1.123636in}}%
\pgfpathlineto{\pgfqpoint{2.706000in}{1.123636in}}%
\pgfusepath{stroke}%
\end{pgfscope}%
\begin{pgfscope}%
\pgfsetbuttcap%
\pgfsetroundjoin%
\definecolor{currentfill}{rgb}{0.000000,0.000000,0.000000}%
\pgfsetfillcolor{currentfill}%
\pgfsetlinewidth{0.803000pt}%
\definecolor{currentstroke}{rgb}{0.000000,0.000000,0.000000}%
\pgfsetstrokecolor{currentstroke}%
\pgfsetdash{}{0pt}%
\pgfsys@defobject{currentmarker}{\pgfqpoint{-0.048611in}{0.000000in}}{\pgfqpoint{-0.000000in}{0.000000in}}{%
\pgfpathmoveto{\pgfqpoint{-0.000000in}{0.000000in}}%
\pgfpathlineto{\pgfqpoint{-0.048611in}{0.000000in}}%
\pgfusepath{stroke,fill}%
}%
\begin{pgfscope}%
\pgfsys@transformshift{0.594000in}{1.123636in}%
\pgfsys@useobject{currentmarker}{}%
\end{pgfscope}%
\end{pgfscope}%
\begin{pgfscope}%
\definecolor{textcolor}{rgb}{0.000000,0.000000,0.000000}%
\pgfsetstrokecolor{textcolor}%
\pgfsetfillcolor{textcolor}%
\pgftext[x=0.268384in, y=1.080234in, left, base]{\color{textcolor}{\rmfamily\fontsize{9.000000}{10.800000}\selectfont\catcode`\^=\active\def^{\ifmmode\sp\else\^{}\fi}\catcode`\%=\active\def
\end{pgfscope}%
\begin{pgfscope}%
\pgfpathrectangle{\pgfqpoint{0.594000in}{0.480000in}}{\pgfqpoint{2.112000in}{2.400000in}}%
\pgfusepath{clip}%
\pgfsetrectcap%
\pgfsetroundjoin%
\pgfsetlinewidth{0.803000pt}%
\definecolor{currentstroke}{rgb}{0.690196,0.690196,0.690196}%
\pgfsetstrokecolor{currentstroke}%
\pgfsetdash{}{0pt}%
\pgfpathmoveto{\pgfqpoint{0.594000in}{1.390909in}}%
\pgfpathlineto{\pgfqpoint{2.706000in}{1.390909in}}%
\pgfusepath{stroke}%
\end{pgfscope}%
\begin{pgfscope}%
\pgfsetbuttcap%
\pgfsetroundjoin%
\definecolor{currentfill}{rgb}{0.000000,0.000000,0.000000}%
\pgfsetfillcolor{currentfill}%
\pgfsetlinewidth{0.803000pt}%
\definecolor{currentstroke}{rgb}{0.000000,0.000000,0.000000}%
\pgfsetstrokecolor{currentstroke}%
\pgfsetdash{}{0pt}%
\pgfsys@defobject{currentmarker}{\pgfqpoint{-0.048611in}{0.000000in}}{\pgfqpoint{-0.000000in}{0.000000in}}{%
\pgfpathmoveto{\pgfqpoint{-0.000000in}{0.000000in}}%
\pgfpathlineto{\pgfqpoint{-0.048611in}{0.000000in}}%
\pgfusepath{stroke,fill}%
}%
\begin{pgfscope}%
\pgfsys@transformshift{0.594000in}{1.390909in}%
\pgfsys@useobject{currentmarker}{}%
\end{pgfscope}%
\end{pgfscope}%
\begin{pgfscope}%
\definecolor{textcolor}{rgb}{0.000000,0.000000,0.000000}%
\pgfsetstrokecolor{textcolor}%
\pgfsetfillcolor{textcolor}%
\pgftext[x=0.268384in, y=1.347506in, left, base]{\color{textcolor}{\rmfamily\fontsize{9.000000}{10.800000}\selectfont\catcode`\^=\active\def^{\ifmmode\sp\else\^{}\fi}\catcode`\%=\active\def
\end{pgfscope}%
\begin{pgfscope}%
\pgfpathrectangle{\pgfqpoint{0.594000in}{0.480000in}}{\pgfqpoint{2.112000in}{2.400000in}}%
\pgfusepath{clip}%
\pgfsetrectcap%
\pgfsetroundjoin%
\pgfsetlinewidth{0.803000pt}%
\definecolor{currentstroke}{rgb}{0.690196,0.690196,0.690196}%
\pgfsetstrokecolor{currentstroke}%
\pgfsetdash{}{0pt}%
\pgfpathmoveto{\pgfqpoint{0.594000in}{1.658182in}}%
\pgfpathlineto{\pgfqpoint{2.706000in}{1.658182in}}%
\pgfusepath{stroke}%
\end{pgfscope}%
\begin{pgfscope}%
\pgfsetbuttcap%
\pgfsetroundjoin%
\definecolor{currentfill}{rgb}{0.000000,0.000000,0.000000}%
\pgfsetfillcolor{currentfill}%
\pgfsetlinewidth{0.803000pt}%
\definecolor{currentstroke}{rgb}{0.000000,0.000000,0.000000}%
\pgfsetstrokecolor{currentstroke}%
\pgfsetdash{}{0pt}%
\pgfsys@defobject{currentmarker}{\pgfqpoint{-0.048611in}{0.000000in}}{\pgfqpoint{-0.000000in}{0.000000in}}{%
\pgfpathmoveto{\pgfqpoint{-0.000000in}{0.000000in}}%
\pgfpathlineto{\pgfqpoint{-0.048611in}{0.000000in}}%
\pgfusepath{stroke,fill}%
}%
\begin{pgfscope}%
\pgfsys@transformshift{0.594000in}{1.658182in}%
\pgfsys@useobject{currentmarker}{}%
\end{pgfscope}%
\end{pgfscope}%
\begin{pgfscope}%
\definecolor{textcolor}{rgb}{0.000000,0.000000,0.000000}%
\pgfsetstrokecolor{textcolor}%
\pgfsetfillcolor{textcolor}%
\pgftext[x=0.268384in, y=1.614779in, left, base]{\color{textcolor}{\rmfamily\fontsize{9.000000}{10.800000}\selectfont\catcode`\^=\active\def^{\ifmmode\sp\else\^{}\fi}\catcode`\%=\active\def
\end{pgfscope}%
\begin{pgfscope}%
\pgfpathrectangle{\pgfqpoint{0.594000in}{0.480000in}}{\pgfqpoint{2.112000in}{2.400000in}}%
\pgfusepath{clip}%
\pgfsetrectcap%
\pgfsetroundjoin%
\pgfsetlinewidth{0.803000pt}%
\definecolor{currentstroke}{rgb}{0.690196,0.690196,0.690196}%
\pgfsetstrokecolor{currentstroke}%
\pgfsetdash{}{0pt}%
\pgfpathmoveto{\pgfqpoint{0.594000in}{1.925455in}}%
\pgfpathlineto{\pgfqpoint{2.706000in}{1.925455in}}%
\pgfusepath{stroke}%
\end{pgfscope}%
\begin{pgfscope}%
\pgfsetbuttcap%
\pgfsetroundjoin%
\definecolor{currentfill}{rgb}{0.000000,0.000000,0.000000}%
\pgfsetfillcolor{currentfill}%
\pgfsetlinewidth{0.803000pt}%
\definecolor{currentstroke}{rgb}{0.000000,0.000000,0.000000}%
\pgfsetstrokecolor{currentstroke}%
\pgfsetdash{}{0pt}%
\pgfsys@defobject{currentmarker}{\pgfqpoint{-0.048611in}{0.000000in}}{\pgfqpoint{-0.000000in}{0.000000in}}{%
\pgfpathmoveto{\pgfqpoint{-0.000000in}{0.000000in}}%
\pgfpathlineto{\pgfqpoint{-0.048611in}{0.000000in}}%
\pgfusepath{stroke,fill}%
}%
\begin{pgfscope}%
\pgfsys@transformshift{0.594000in}{1.925455in}%
\pgfsys@useobject{currentmarker}{}%
\end{pgfscope}%
\end{pgfscope}%
\begin{pgfscope}%
\definecolor{textcolor}{rgb}{0.000000,0.000000,0.000000}%
\pgfsetstrokecolor{textcolor}%
\pgfsetfillcolor{textcolor}%
\pgftext[x=0.268384in, y=1.882052in, left, base]{\color{textcolor}{\rmfamily\fontsize{9.000000}{10.800000}\selectfont\catcode`\^=\active\def^{\ifmmode\sp\else\^{}\fi}\catcode`\%=\active\def
\end{pgfscope}%
\begin{pgfscope}%
\pgfpathrectangle{\pgfqpoint{0.594000in}{0.480000in}}{\pgfqpoint{2.112000in}{2.400000in}}%
\pgfusepath{clip}%
\pgfsetrectcap%
\pgfsetroundjoin%
\pgfsetlinewidth{0.803000pt}%
\definecolor{currentstroke}{rgb}{0.690196,0.690196,0.690196}%
\pgfsetstrokecolor{currentstroke}%
\pgfsetdash{}{0pt}%
\pgfpathmoveto{\pgfqpoint{0.594000in}{2.192727in}}%
\pgfpathlineto{\pgfqpoint{2.706000in}{2.192727in}}%
\pgfusepath{stroke}%
\end{pgfscope}%
\begin{pgfscope}%
\pgfsetbuttcap%
\pgfsetroundjoin%
\definecolor{currentfill}{rgb}{0.000000,0.000000,0.000000}%
\pgfsetfillcolor{currentfill}%
\pgfsetlinewidth{0.803000pt}%
\definecolor{currentstroke}{rgb}{0.000000,0.000000,0.000000}%
\pgfsetstrokecolor{currentstroke}%
\pgfsetdash{}{0pt}%
\pgfsys@defobject{currentmarker}{\pgfqpoint{-0.048611in}{0.000000in}}{\pgfqpoint{-0.000000in}{0.000000in}}{%
\pgfpathmoveto{\pgfqpoint{-0.000000in}{0.000000in}}%
\pgfpathlineto{\pgfqpoint{-0.048611in}{0.000000in}}%
\pgfusepath{stroke,fill}%
}%
\begin{pgfscope}%
\pgfsys@transformshift{0.594000in}{2.192727in}%
\pgfsys@useobject{currentmarker}{}%
\end{pgfscope}%
\end{pgfscope}%
\begin{pgfscope}%
\definecolor{textcolor}{rgb}{0.000000,0.000000,0.000000}%
\pgfsetstrokecolor{textcolor}%
\pgfsetfillcolor{textcolor}%
\pgftext[x=0.268384in, y=2.149324in, left, base]{\color{textcolor}{\rmfamily\fontsize{9.000000}{10.800000}\selectfont\catcode`\^=\active\def^{\ifmmode\sp\else\^{}\fi}\catcode`\%=\active\def
\end{pgfscope}%
\begin{pgfscope}%
\pgfpathrectangle{\pgfqpoint{0.594000in}{0.480000in}}{\pgfqpoint{2.112000in}{2.400000in}}%
\pgfusepath{clip}%
\pgfsetrectcap%
\pgfsetroundjoin%
\pgfsetlinewidth{0.803000pt}%
\definecolor{currentstroke}{rgb}{0.690196,0.690196,0.690196}%
\pgfsetstrokecolor{currentstroke}%
\pgfsetdash{}{0pt}%
\pgfpathmoveto{\pgfqpoint{0.594000in}{2.460000in}}%
\pgfpathlineto{\pgfqpoint{2.706000in}{2.460000in}}%
\pgfusepath{stroke}%
\end{pgfscope}%
\begin{pgfscope}%
\pgfsetbuttcap%
\pgfsetroundjoin%
\definecolor{currentfill}{rgb}{0.000000,0.000000,0.000000}%
\pgfsetfillcolor{currentfill}%
\pgfsetlinewidth{0.803000pt}%
\definecolor{currentstroke}{rgb}{0.000000,0.000000,0.000000}%
\pgfsetstrokecolor{currentstroke}%
\pgfsetdash{}{0pt}%
\pgfsys@defobject{currentmarker}{\pgfqpoint{-0.048611in}{0.000000in}}{\pgfqpoint{-0.000000in}{0.000000in}}{%
\pgfpathmoveto{\pgfqpoint{-0.000000in}{0.000000in}}%
\pgfpathlineto{\pgfqpoint{-0.048611in}{0.000000in}}%
\pgfusepath{stroke,fill}%
}%
\begin{pgfscope}%
\pgfsys@transformshift{0.594000in}{2.460000in}%
\pgfsys@useobject{currentmarker}{}%
\end{pgfscope}%
\end{pgfscope}%
\begin{pgfscope}%
\definecolor{textcolor}{rgb}{0.000000,0.000000,0.000000}%
\pgfsetstrokecolor{textcolor}%
\pgfsetfillcolor{textcolor}%
\pgftext[x=0.268384in, y=2.416597in, left, base]{\color{textcolor}{\rmfamily\fontsize{9.000000}{10.800000}\selectfont\catcode`\^=\active\def^{\ifmmode\sp\else\^{}\fi}\catcode`\%=\active\def
\end{pgfscope}%
\begin{pgfscope}%
\pgfpathrectangle{\pgfqpoint{0.594000in}{0.480000in}}{\pgfqpoint{2.112000in}{2.400000in}}%
\pgfusepath{clip}%
\pgfsetrectcap%
\pgfsetroundjoin%
\pgfsetlinewidth{0.803000pt}%
\definecolor{currentstroke}{rgb}{0.690196,0.690196,0.690196}%
\pgfsetstrokecolor{currentstroke}%
\pgfsetdash{}{0pt}%
\pgfpathmoveto{\pgfqpoint{0.594000in}{2.727273in}}%
\pgfpathlineto{\pgfqpoint{2.706000in}{2.727273in}}%
\pgfusepath{stroke}%
\end{pgfscope}%
\begin{pgfscope}%
\pgfsetbuttcap%
\pgfsetroundjoin%
\definecolor{currentfill}{rgb}{0.000000,0.000000,0.000000}%
\pgfsetfillcolor{currentfill}%
\pgfsetlinewidth{0.803000pt}%
\definecolor{currentstroke}{rgb}{0.000000,0.000000,0.000000}%
\pgfsetstrokecolor{currentstroke}%
\pgfsetdash{}{0pt}%
\pgfsys@defobject{currentmarker}{\pgfqpoint{-0.048611in}{0.000000in}}{\pgfqpoint{-0.000000in}{0.000000in}}{%
\pgfpathmoveto{\pgfqpoint{-0.000000in}{0.000000in}}%
\pgfpathlineto{\pgfqpoint{-0.048611in}{0.000000in}}%
\pgfusepath{stroke,fill}%
}%
\begin{pgfscope}%
\pgfsys@transformshift{0.594000in}{2.727273in}%
\pgfsys@useobject{currentmarker}{}%
\end{pgfscope}%
\end{pgfscope}%
\begin{pgfscope}%
\definecolor{textcolor}{rgb}{0.000000,0.000000,0.000000}%
\pgfsetstrokecolor{textcolor}%
\pgfsetfillcolor{textcolor}%
\pgftext[x=0.268384in, y=2.683870in, left, base]{\color{textcolor}{\rmfamily\fontsize{9.000000}{10.800000}\selectfont\catcode`\^=\active\def^{\ifmmode\sp\else\^{}\fi}\catcode`\%=\active\def
\end{pgfscope}%
\begin{pgfscope}%
\definecolor{textcolor}{rgb}{0.000000,0.000000,0.000000}%
\pgfsetstrokecolor{textcolor}%
\pgfsetfillcolor{textcolor}%
\pgftext[x=0.212829in,y=1.680000in,,bottom,rotate=90.000000]{\color{textcolor}{\rmfamily\fontsize{10.000000}{12.000000}\selectfont\catcode`\^=\active\def^{\ifmmode\sp\else\^{}\fi}\catcode`\%=\active\def
\end{pgfscope}%
\begin{pgfscope}%
\pgfpathrectangle{\pgfqpoint{0.594000in}{0.480000in}}{\pgfqpoint{2.112000in}{2.400000in}}%
\pgfusepath{clip}%
\pgfsetrectcap%
\pgfsetroundjoin%
\pgfsetlinewidth{1.505625pt}%
\definecolor{currentstroke}{rgb}{0.000000,0.000000,0.000000}%
\pgfsetstrokecolor{currentstroke}%
\pgfsetstrokeopacity{0.300000}%
\pgfsetdash{}{0pt}%
\pgfpathmoveto{\pgfqpoint{0.594000in}{0.589091in}}%
\pgfpathlineto{\pgfqpoint{0.613853in}{0.826677in}}%
\pgfpathlineto{\pgfqpoint{0.633706in}{1.038391in}}%
\pgfpathlineto{\pgfqpoint{0.653558in}{1.227051in}}%
\pgfpathlineto{\pgfqpoint{0.673411in}{1.395167in}}%
\pgfpathlineto{\pgfqpoint{0.693264in}{1.544976in}}%
\pgfpathlineto{\pgfqpoint{0.713117in}{1.678472in}}%
\pgfpathlineto{\pgfqpoint{0.733392in}{1.799817in}}%
\pgfpathlineto{\pgfqpoint{0.753667in}{1.907682in}}%
\pgfpathlineto{\pgfqpoint{0.773942in}{2.003567in}}%
\pgfpathlineto{\pgfqpoint{0.794218in}{2.088800in}}%
\pgfpathlineto{\pgfqpoint{0.814493in}{2.164567in}}%
\pgfpathlineto{\pgfqpoint{0.834768in}{2.231917in}}%
\pgfpathlineto{\pgfqpoint{0.855043in}{2.291787in}}%
\pgfpathlineto{\pgfqpoint{0.875741in}{2.346050in}}%
\pgfpathlineto{\pgfqpoint{0.896438in}{2.394167in}}%
\pgfpathlineto{\pgfqpoint{0.917558in}{2.437653in}}%
\pgfpathlineto{\pgfqpoint{0.938678in}{2.476120in}}%
\pgfpathlineto{\pgfqpoint{0.960221in}{2.510785in}}%
\pgfpathlineto{\pgfqpoint{0.982186in}{2.541937in}}%
\pgfpathlineto{\pgfqpoint{1.004573in}{2.569851in}}%
\pgfpathlineto{\pgfqpoint{1.027382in}{2.594795in}}%
\pgfpathlineto{\pgfqpoint{1.051037in}{2.617400in}}%
\pgfpathlineto{\pgfqpoint{1.075536in}{2.637758in}}%
\pgfpathlineto{\pgfqpoint{1.101302in}{2.656263in}}%
\pgfpathlineto{\pgfqpoint{1.128336in}{2.672920in}}%
\pgfpathlineto{\pgfqpoint{1.157059in}{2.687974in}}%
\pgfpathlineto{\pgfqpoint{1.187472in}{2.701401in}}%
\pgfpathlineto{\pgfqpoint{1.220419in}{2.713506in}}%
\pgfpathlineto{\pgfqpoint{1.256323in}{2.724309in}}%
\pgfpathlineto{\pgfqpoint{1.296029in}{2.733905in}}%
\pgfpathlineto{\pgfqpoint{1.340381in}{2.742308in}}%
\pgfpathlineto{\pgfqpoint{1.391069in}{2.749601in}}%
\pgfpathlineto{\pgfqpoint{1.450205in}{2.755794in}}%
\pgfpathlineto{\pgfqpoint{1.520746in}{2.760875in}}%
\pgfpathlineto{\pgfqpoint{1.608605in}{2.764885in}}%
\pgfpathlineto{\pgfqpoint{1.725187in}{2.767848in}}%
\pgfpathlineto{\pgfqpoint{1.896259in}{2.769776in}}%
\pgfpathlineto{\pgfqpoint{2.209680in}{2.770725in}}%
\pgfpathlineto{\pgfqpoint{2.706422in}{2.770899in}}%
\pgfpathlineto{\pgfqpoint{2.706422in}{2.770899in}}%
\pgfusepath{stroke}%
\end{pgfscope}%
\begin{pgfscope}%
\pgfpathrectangle{\pgfqpoint{0.594000in}{0.480000in}}{\pgfqpoint{2.112000in}{2.400000in}}%
\pgfusepath{clip}%
\pgfsetrectcap%
\pgfsetroundjoin%
\pgfsetlinewidth{1.505625pt}%
\definecolor{currentstroke}{rgb}{0.000000,0.000000,0.000000}%
\pgfsetstrokecolor{currentstroke}%
\pgfsetdash{}{0pt}%
\pgfpathmoveto{\pgfqpoint{0.594000in}{0.589091in}}%
\pgfpathlineto{\pgfqpoint{0.613853in}{0.826677in}}%
\pgfpathlineto{\pgfqpoint{0.633706in}{1.038391in}}%
\pgfpathlineto{\pgfqpoint{0.653558in}{1.227051in}}%
\pgfpathlineto{\pgfqpoint{0.673411in}{1.395167in}}%
\pgfpathlineto{\pgfqpoint{0.693264in}{1.544976in}}%
\pgfpathlineto{\pgfqpoint{0.713117in}{1.678472in}}%
\pgfpathlineto{\pgfqpoint{0.733392in}{1.799817in}}%
\pgfpathlineto{\pgfqpoint{0.753667in}{1.907682in}}%
\pgfpathlineto{\pgfqpoint{0.773942in}{2.003567in}}%
\pgfpathlineto{\pgfqpoint{0.794218in}{2.088800in}}%
\pgfpathlineto{\pgfqpoint{0.814493in}{2.164567in}}%
\pgfpathlineto{\pgfqpoint{0.834768in}{2.231917in}}%
\pgfpathlineto{\pgfqpoint{0.855043in}{2.291787in}}%
\pgfpathlineto{\pgfqpoint{0.875741in}{2.346050in}}%
\pgfpathlineto{\pgfqpoint{0.896438in}{2.394167in}}%
\pgfpathlineto{\pgfqpoint{0.917558in}{2.437653in}}%
\pgfpathlineto{\pgfqpoint{0.938678in}{2.476120in}}%
\pgfpathlineto{\pgfqpoint{0.960221in}{2.510785in}}%
\pgfpathlineto{\pgfqpoint{0.982186in}{2.541937in}}%
\pgfpathlineto{\pgfqpoint{1.004573in}{2.569851in}}%
\pgfpathlineto{\pgfqpoint{1.027382in}{2.594795in}}%
\pgfpathlineto{\pgfqpoint{1.051037in}{2.617400in}}%
\pgfpathlineto{\pgfqpoint{1.075536in}{2.637758in}}%
\pgfpathlineto{\pgfqpoint{1.101302in}{2.656263in}}%
\pgfpathlineto{\pgfqpoint{1.128336in}{2.672920in}}%
\pgfpathlineto{\pgfqpoint{1.157059in}{2.687974in}}%
\pgfpathlineto{\pgfqpoint{1.187472in}{2.701401in}}%
\pgfpathlineto{\pgfqpoint{1.220419in}{2.713506in}}%
\pgfpathlineto{\pgfqpoint{1.256323in}{2.724309in}}%
\pgfpathlineto{\pgfqpoint{1.296029in}{2.733905in}}%
\pgfpathlineto{\pgfqpoint{1.340381in}{2.742308in}}%
\pgfpathlineto{\pgfqpoint{1.391069in}{2.749601in}}%
\pgfpathlineto{\pgfqpoint{1.450205in}{2.755794in}}%
\pgfpathlineto{\pgfqpoint{1.520746in}{2.760875in}}%
\pgfpathlineto{\pgfqpoint{1.608605in}{2.764885in}}%
\pgfpathlineto{\pgfqpoint{1.725187in}{2.767848in}}%
\pgfpathlineto{\pgfqpoint{1.896259in}{2.769776in}}%
\pgfpathlineto{\pgfqpoint{2.209680in}{2.770725in}}%
\pgfpathlineto{\pgfqpoint{2.706422in}{2.770899in}}%
\pgfpathlineto{\pgfqpoint{2.706422in}{2.770899in}}%
\pgfusepath{stroke}%
\end{pgfscope}%
\begin{pgfscope}%
\pgfpathrectangle{\pgfqpoint{0.594000in}{0.480000in}}{\pgfqpoint{2.112000in}{2.400000in}}%
\pgfusepath{clip}%
\pgfsetrectcap%
\pgfsetroundjoin%
\pgfsetlinewidth{1.505625pt}%
\definecolor{currentstroke}{rgb}{1.000000,0.000000,0.000000}%
\pgfsetstrokecolor{currentstroke}%
\pgfsetdash{}{0pt}%
\pgfpathmoveto{\pgfqpoint{0.594000in}{0.589091in}}%
\pgfpathlineto{\pgfqpoint{0.608362in}{0.924673in}}%
\pgfpathlineto{\pgfqpoint{0.622723in}{1.208640in}}%
\pgfpathlineto{\pgfqpoint{0.637085in}{1.448930in}}%
\pgfpathlineto{\pgfqpoint{0.651446in}{1.652261in}}%
\pgfpathlineto{\pgfqpoint{0.665808in}{1.824319in}}%
\pgfpathlineto{\pgfqpoint{0.680170in}{1.969912in}}%
\pgfpathlineto{\pgfqpoint{0.694531in}{2.093112in}}%
\pgfpathlineto{\pgfqpoint{0.708893in}{2.197363in}}%
\pgfpathlineto{\pgfqpoint{0.723254in}{2.285579in}}%
\pgfpathlineto{\pgfqpoint{0.737616in}{2.360227in}}%
\pgfpathlineto{\pgfqpoint{0.751978in}{2.423394in}}%
\pgfpathlineto{\pgfqpoint{0.766339in}{2.476844in}}%
\pgfpathlineto{\pgfqpoint{0.780701in}{2.522074in}}%
\pgfpathlineto{\pgfqpoint{0.795485in}{2.561379in}}%
\pgfpathlineto{\pgfqpoint{0.810269in}{2.594475in}}%
\pgfpathlineto{\pgfqpoint{0.825475in}{2.623072in}}%
\pgfpathlineto{\pgfqpoint{0.840682in}{2.647033in}}%
\pgfpathlineto{\pgfqpoint{0.856310in}{2.667620in}}%
\pgfpathlineto{\pgfqpoint{0.872784in}{2.685627in}}%
\pgfpathlineto{\pgfqpoint{0.889680in}{2.700840in}}%
\pgfpathlineto{\pgfqpoint{0.907843in}{2.714181in}}%
\pgfpathlineto{\pgfqpoint{0.927274in}{2.725654in}}%
\pgfpathlineto{\pgfqpoint{0.948394in}{2.735509in}}%
\pgfpathlineto{\pgfqpoint{0.972048in}{2.744022in}}%
\pgfpathlineto{\pgfqpoint{0.998659in}{2.751178in}}%
\pgfpathlineto{\pgfqpoint{1.029494in}{2.757124in}}%
\pgfpathlineto{\pgfqpoint{1.066666in}{2.761962in}}%
\pgfpathlineto{\pgfqpoint{1.113552in}{2.765722in}}%
\pgfpathlineto{\pgfqpoint{1.176912in}{2.768427in}}%
\pgfpathlineto{\pgfqpoint{1.273642in}{2.770103in}}%
\pgfpathlineto{\pgfqpoint{1.469635in}{2.770827in}}%
\pgfpathlineto{\pgfqpoint{2.706422in}{2.770909in}}%
\pgfpathlineto{\pgfqpoint{2.706422in}{2.770909in}}%
\pgfusepath{stroke}%
\end{pgfscope}%
\begin{pgfscope}%
\pgfpathrectangle{\pgfqpoint{0.594000in}{0.480000in}}{\pgfqpoint{2.112000in}{2.400000in}}%
\pgfusepath{clip}%
\pgfsetrectcap%
\pgfsetroundjoin%
\pgfsetlinewidth{1.505625pt}%
\definecolor{currentstroke}{rgb}{0.000000,0.000000,1.000000}%
\pgfsetstrokecolor{currentstroke}%
\pgfsetdash{}{0pt}%
\pgfpathmoveto{\pgfqpoint{0.594000in}{0.589091in}}%
\pgfpathlineto{\pgfqpoint{0.605827in}{0.724816in}}%
\pgfpathlineto{\pgfqpoint{0.617654in}{0.835212in}}%
\pgfpathlineto{\pgfqpoint{0.629482in}{0.925006in}}%
\pgfpathlineto{\pgfqpoint{0.641309in}{0.998042in}}%
\pgfpathlineto{\pgfqpoint{0.653136in}{1.057448in}}%
\pgfpathlineto{\pgfqpoint{0.664963in}{1.105768in}}%
\pgfpathlineto{\pgfqpoint{0.676790in}{1.145070in}}%
\pgfpathlineto{\pgfqpoint{0.688618in}{1.177037in}}%
\pgfpathlineto{\pgfqpoint{0.700867in}{1.203871in}}%
\pgfpathlineto{\pgfqpoint{0.713117in}{1.225538in}}%
\pgfpathlineto{\pgfqpoint{0.725789in}{1.243570in}}%
\pgfpathlineto{\pgfqpoint{0.738883in}{1.258451in}}%
\pgfpathlineto{\pgfqpoint{0.752822in}{1.270964in}}%
\pgfpathlineto{\pgfqpoint{0.767606in}{1.281295in}}%
\pgfpathlineto{\pgfqpoint{0.784080in}{1.290063in}}%
\pgfpathlineto{\pgfqpoint{0.802666in}{1.297353in}}%
\pgfpathlineto{\pgfqpoint{0.824208in}{1.303314in}}%
\pgfpathlineto{\pgfqpoint{0.849974in}{1.308043in}}%
\pgfpathlineto{\pgfqpoint{0.882499in}{1.311649in}}%
\pgfpathlineto{\pgfqpoint{0.926851in}{1.314191in}}%
\pgfpathlineto{\pgfqpoint{0.996547in}{1.315720in}}%
\pgfpathlineto{\pgfqpoint{1.149034in}{1.316319in}}%
\pgfpathlineto{\pgfqpoint{2.629968in}{1.316364in}}%
\pgfpathlineto{\pgfqpoint{2.706422in}{1.316364in}}%
\pgfpathlineto{\pgfqpoint{2.706422in}{1.316364in}}%
\pgfusepath{stroke}%
\end{pgfscope}%
\begin{pgfscope}%
\pgfsetrectcap%
\pgfsetmiterjoin%
\pgfsetlinewidth{0.803000pt}%
\definecolor{currentstroke}{rgb}{0.000000,0.000000,0.000000}%
\pgfsetstrokecolor{currentstroke}%
\pgfsetdash{}{0pt}%
\pgfpathmoveto{\pgfqpoint{0.594000in}{0.480000in}}%
\pgfpathlineto{\pgfqpoint{0.594000in}{2.880000in}}%
\pgfusepath{stroke}%
\end{pgfscope}%
\begin{pgfscope}%
\pgfsetrectcap%
\pgfsetmiterjoin%
\pgfsetlinewidth{0.803000pt}%
\definecolor{currentstroke}{rgb}{0.000000,0.000000,0.000000}%
\pgfsetstrokecolor{currentstroke}%
\pgfsetdash{}{0pt}%
\pgfpathmoveto{\pgfqpoint{2.706000in}{0.480000in}}%
\pgfpathlineto{\pgfqpoint{2.706000in}{2.880000in}}%
\pgfusepath{stroke}%
\end{pgfscope}%
\begin{pgfscope}%
\pgfsetrectcap%
\pgfsetmiterjoin%
\pgfsetlinewidth{0.803000pt}%
\definecolor{currentstroke}{rgb}{0.000000,0.000000,0.000000}%
\pgfsetstrokecolor{currentstroke}%
\pgfsetdash{}{0pt}%
\pgfpathmoveto{\pgfqpoint{0.594000in}{0.480000in}}%
\pgfpathlineto{\pgfqpoint{2.706000in}{0.480000in}}%
\pgfusepath{stroke}%
\end{pgfscope}%
\begin{pgfscope}%
\pgfsetrectcap%
\pgfsetmiterjoin%
\pgfsetlinewidth{0.803000pt}%
\definecolor{currentstroke}{rgb}{0.000000,0.000000,0.000000}%
\pgfsetstrokecolor{currentstroke}%
\pgfsetdash{}{0pt}%
\pgfpathmoveto{\pgfqpoint{0.594000in}{2.880000in}}%
\pgfpathlineto{\pgfqpoint{2.706000in}{2.880000in}}%
\pgfusepath{stroke}%
\end{pgfscope}%
\begin{pgfscope}%
\definecolor{textcolor}{rgb}{0.000000,0.000000,0.000000}%
\pgfsetstrokecolor{textcolor}%
\pgfsetfillcolor{textcolor}%
\pgftext[x=1.650000in,y=2.963333in,,base]{\color{textcolor}{\rmfamily\fontsize{12.000000}{14.400000}\selectfont\catcode`\^=\active\def^{\ifmmode\sp\else\^{}\fi}\catcode`\%=\active\def
\end{pgfscope}%
\begin{pgfscope}%
\pgfsetbuttcap%
\pgfsetmiterjoin%
\definecolor{currentfill}{rgb}{1.000000,1.000000,1.000000}%
\pgfsetfillcolor{currentfill}%
\pgfsetfillopacity{0.800000}%
\pgfsetlinewidth{1.003750pt}%
\definecolor{currentstroke}{rgb}{0.800000,0.800000,0.800000}%
\pgfsetstrokecolor{currentstroke}%
\pgfsetstrokeopacity{0.800000}%
\pgfsetdash{}{0pt}%
\pgfpathmoveto{\pgfqpoint{1.333630in}{0.542500in}}%
\pgfpathlineto{\pgfqpoint{2.618500in}{0.542500in}}%
\pgfpathquadraticcurveto{\pgfqpoint{2.643500in}{0.542500in}}{\pgfqpoint{2.643500in}{0.567500in}}%
\pgfpathlineto{\pgfqpoint{2.643500in}{1.077917in}}%
\pgfpathquadraticcurveto{\pgfqpoint{2.643500in}{1.102917in}}{\pgfqpoint{2.618500in}{1.102917in}}%
\pgfpathlineto{\pgfqpoint{1.333630in}{1.102917in}}%
\pgfpathquadraticcurveto{\pgfqpoint{1.308630in}{1.102917in}}{\pgfqpoint{1.308630in}{1.077917in}}%
\pgfpathlineto{\pgfqpoint{1.308630in}{0.567500in}}%
\pgfpathquadraticcurveto{\pgfqpoint{1.308630in}{0.542500in}}{\pgfqpoint{1.333630in}{0.542500in}}%
\pgfpathlineto{\pgfqpoint{1.333630in}{0.542500in}}%
\pgfpathclose%
\pgfusepath{stroke,fill}%
\end{pgfscope}%
\begin{pgfscope}%
\pgfsetrectcap%
\pgfsetroundjoin%
\pgfsetlinewidth{1.505625pt}%
\definecolor{currentstroke}{rgb}{0.000000,0.000000,0.000000}%
\pgfsetstrokecolor{currentstroke}%
\pgfsetdash{}{0pt}%
\pgfpathmoveto{\pgfqpoint{1.358630in}{1.009167in}}%
\pgfpathlineto{\pgfqpoint{1.483630in}{1.009167in}}%
\pgfpathlineto{\pgfqpoint{1.608630in}{1.009167in}}%
\pgfusepath{stroke}%
\end{pgfscope}%
\begin{pgfscope}%
\definecolor{textcolor}{rgb}{0.000000,0.000000,0.000000}%
\pgfsetstrokecolor{textcolor}%
\pgfsetfillcolor{textcolor}%
\pgftext[x=1.708630in,y=0.965417in,left,base]{\color{textcolor}{\rmfamily\fontsize{9.000000}{10.800000}\selectfont\catcode`\^=\active\def^{\ifmmode\sp\else\^{}\fi}\catcode`\%=\active\def
\end{pgfscope}%
\begin{pgfscope}%
\pgfsetrectcap%
\pgfsetroundjoin%
\pgfsetlinewidth{1.505625pt}%
\definecolor{currentstroke}{rgb}{1.000000,0.000000,0.000000}%
\pgfsetstrokecolor{currentstroke}%
\pgfsetdash{}{0pt}%
\pgfpathmoveto{\pgfqpoint{1.358630in}{0.834861in}}%
\pgfpathlineto{\pgfqpoint{1.483630in}{0.834861in}}%
\pgfpathlineto{\pgfqpoint{1.608630in}{0.834861in}}%
\pgfusepath{stroke}%
\end{pgfscope}%
\begin{pgfscope}%
\definecolor{textcolor}{rgb}{0.000000,0.000000,0.000000}%
\pgfsetstrokecolor{textcolor}%
\pgfsetfillcolor{textcolor}%
\pgftext[x=1.708630in,y=0.791111in,left,base]{\color{textcolor}{\rmfamily\fontsize{9.000000}{10.800000}\selectfont\catcode`\^=\active\def^{\ifmmode\sp\else\^{}\fi}\catcode`\%=\active\def
\end{pgfscope}%
\begin{pgfscope}%
\pgfsetrectcap%
\pgfsetroundjoin%
\pgfsetlinewidth{1.505625pt}%
\definecolor{currentstroke}{rgb}{0.000000,0.000000,1.000000}%
\pgfsetstrokecolor{currentstroke}%
\pgfsetdash{}{0pt}%
\pgfpathmoveto{\pgfqpoint{1.358630in}{0.660556in}}%
\pgfpathlineto{\pgfqpoint{1.483630in}{0.660556in}}%
\pgfpathlineto{\pgfqpoint{1.608630in}{0.660556in}}%
\pgfusepath{stroke}%
\end{pgfscope}%
\begin{pgfscope}%
\definecolor{textcolor}{rgb}{0.000000,0.000000,0.000000}%
\pgfsetstrokecolor{textcolor}%
\pgfsetfillcolor{textcolor}%
\pgftext[x=1.708630in,y=0.616806in,left,base]{\color{textcolor}{\rmfamily\fontsize{9.000000}{10.800000}\selectfont\catcode`\^=\active\def^{\ifmmode\sp\else\^{}\fi}\catcode`\%=\active\def
\end{pgfscope}%
\end{pgfpicture}%
\makeatother%
\endgroup%

%% file: figures/acc_type_with_mean.pgf
\begingroup%
\makeatletter%
\begin{pgfpicture}%
\pgfpathrectangle{\pgfpointorigin}{\pgfqpoint{3.300000in}{3.200000in}}%
\pgfusepath{use as bounding box, clip}%
\begin{pgfscope}%
\pgfsetbuttcap%
\pgfsetmiterjoin%
\definecolor{currentfill}{rgb}{1.000000,1.000000,1.000000}%
\pgfsetfillcolor{currentfill}%
\pgfsetlinewidth{0.000000pt}%
\definecolor{currentstroke}{rgb}{1.000000,1.000000,1.000000}%
\pgfsetstrokecolor{currentstroke}%
\pgfsetdash{}{0pt}%
\pgfpathmoveto{\pgfqpoint{0.000000in}{0.000000in}}%
\pgfpathlineto{\pgfqpoint{3.300000in}{0.000000in}}%
\pgfpathlineto{\pgfqpoint{3.300000in}{3.200000in}}%
\pgfpathlineto{\pgfqpoint{0.000000in}{3.200000in}}%
\pgfpathlineto{\pgfqpoint{0.000000in}{0.000000in}}%
\pgfpathclose%
\pgfusepath{fill}%
\end{pgfscope}%
\begin{pgfscope}%
\pgfsetbuttcap%
\pgfsetmiterjoin%
\definecolor{currentfill}{rgb}{1.000000,1.000000,1.000000}%
\pgfsetfillcolor{currentfill}%
\pgfsetlinewidth{0.000000pt}%
\definecolor{currentstroke}{rgb}{0.000000,0.000000,0.000000}%
\pgfsetstrokecolor{currentstroke}%
\pgfsetstrokeopacity{0.000000}%
\pgfsetdash{}{0pt}%
\pgfpathmoveto{\pgfqpoint{0.594000in}{0.480000in}}%
\pgfpathlineto{\pgfqpoint{2.706000in}{0.480000in}}%
\pgfpathlineto{\pgfqpoint{2.706000in}{2.880000in}}%
\pgfpathlineto{\pgfqpoint{0.594000in}{2.880000in}}%
\pgfpathlineto{\pgfqpoint{0.594000in}{0.480000in}}%
\pgfpathclose%
\pgfusepath{fill}%
\end{pgfscope}%
\begin{pgfscope}%
\pgfpathrectangle{\pgfqpoint{0.594000in}{0.480000in}}{\pgfqpoint{2.112000in}{2.400000in}}%
\pgfusepath{clip}%
\pgfsetrectcap%
\pgfsetroundjoin%
\pgfsetlinewidth{0.803000pt}%
\definecolor{currentstroke}{rgb}{0.690196,0.690196,0.690196}%
\pgfsetstrokecolor{currentstroke}%
\pgfsetdash{}{0pt}%
\pgfpathmoveto{\pgfqpoint{0.594000in}{0.480000in}}%
\pgfpathlineto{\pgfqpoint{0.594000in}{2.880000in}}%
\pgfusepath{stroke}%
\end{pgfscope}%
\begin{pgfscope}%
\pgfsetbuttcap%
\pgfsetroundjoin%
\definecolor{currentfill}{rgb}{0.000000,0.000000,0.000000}%
\pgfsetfillcolor{currentfill}%
\pgfsetlinewidth{0.803000pt}%
\definecolor{currentstroke}{rgb}{0.000000,0.000000,0.000000}%
\pgfsetstrokecolor{currentstroke}%
\pgfsetdash{}{0pt}%
\pgfsys@defobject{currentmarker}{\pgfqpoint{0.000000in}{-0.048611in}}{\pgfqpoint{0.000000in}{0.000000in}}{%
\pgfpathmoveto{\pgfqpoint{0.000000in}{0.000000in}}%
\pgfpathlineto{\pgfqpoint{0.000000in}{-0.048611in}}%
\pgfusepath{stroke,fill}%
}%
\begin{pgfscope}%
\pgfsys@transformshift{0.594000in}{0.480000in}%
\pgfsys@useobject{currentmarker}{}%
\end{pgfscope}%
\end{pgfscope}%
\begin{pgfscope}%
\definecolor{textcolor}{rgb}{0.000000,0.000000,0.000000}%
\pgfsetstrokecolor{textcolor}%
\pgfsetfillcolor{textcolor}%
\pgftext[x=0.594000in,y=0.382778in,,top]{\color{textcolor}{\rmfamily\fontsize{9.000000}{10.800000}\selectfont\catcode`\^=\active\def^{\ifmmode\sp\else\^{}\fi}\catcode`\%=\active\def
\end{pgfscope}%
\begin{pgfscope}%
\pgfpathrectangle{\pgfqpoint{0.594000in}{0.480000in}}{\pgfqpoint{2.112000in}{2.400000in}}%
\pgfusepath{clip}%
\pgfsetrectcap%
\pgfsetroundjoin%
\pgfsetlinewidth{0.803000pt}%
\definecolor{currentstroke}{rgb}{0.690196,0.690196,0.690196}%
\pgfsetstrokecolor{currentstroke}%
\pgfsetdash{}{0pt}%
\pgfpathmoveto{\pgfqpoint{1.016400in}{0.480000in}}%
\pgfpathlineto{\pgfqpoint{1.016400in}{2.880000in}}%
\pgfusepath{stroke}%
\end{pgfscope}%
\begin{pgfscope}%
\pgfsetbuttcap%
\pgfsetroundjoin%
\definecolor{currentfill}{rgb}{0.000000,0.000000,0.000000}%
\pgfsetfillcolor{currentfill}%
\pgfsetlinewidth{0.803000pt}%
\definecolor{currentstroke}{rgb}{0.000000,0.000000,0.000000}%
\pgfsetstrokecolor{currentstroke}%
\pgfsetdash{}{0pt}%
\pgfsys@defobject{currentmarker}{\pgfqpoint{0.000000in}{-0.048611in}}{\pgfqpoint{0.000000in}{0.000000in}}{%
\pgfpathmoveto{\pgfqpoint{0.000000in}{0.000000in}}%
\pgfpathlineto{\pgfqpoint{0.000000in}{-0.048611in}}%
\pgfusepath{stroke,fill}%
}%
\begin{pgfscope}%
\pgfsys@transformshift{1.016400in}{0.480000in}%
\pgfsys@useobject{currentmarker}{}%
\end{pgfscope}%
\end{pgfscope}%
\begin{pgfscope}%
\definecolor{textcolor}{rgb}{0.000000,0.000000,0.000000}%
\pgfsetstrokecolor{textcolor}%
\pgfsetfillcolor{textcolor}%
\pgftext[x=1.016400in,y=0.382778in,,top]{\color{textcolor}{\rmfamily\fontsize{9.000000}{10.800000}\selectfont\catcode`\^=\active\def^{\ifmmode\sp\else\^{}\fi}\catcode`\%=\active\def
\end{pgfscope}%
\begin{pgfscope}%
\pgfpathrectangle{\pgfqpoint{0.594000in}{0.480000in}}{\pgfqpoint{2.112000in}{2.400000in}}%
\pgfusepath{clip}%
\pgfsetrectcap%
\pgfsetroundjoin%
\pgfsetlinewidth{0.803000pt}%
\definecolor{currentstroke}{rgb}{0.690196,0.690196,0.690196}%
\pgfsetstrokecolor{currentstroke}%
\pgfsetdash{}{0pt}%
\pgfpathmoveto{\pgfqpoint{1.438800in}{0.480000in}}%
\pgfpathlineto{\pgfqpoint{1.438800in}{2.880000in}}%
\pgfusepath{stroke}%
\end{pgfscope}%
\begin{pgfscope}%
\pgfsetbuttcap%
\pgfsetroundjoin%
\definecolor{currentfill}{rgb}{0.000000,0.000000,0.000000}%
\pgfsetfillcolor{currentfill}%
\pgfsetlinewidth{0.803000pt}%
\definecolor{currentstroke}{rgb}{0.000000,0.000000,0.000000}%
\pgfsetstrokecolor{currentstroke}%
\pgfsetdash{}{0pt}%
\pgfsys@defobject{currentmarker}{\pgfqpoint{0.000000in}{-0.048611in}}{\pgfqpoint{0.000000in}{0.000000in}}{%
\pgfpathmoveto{\pgfqpoint{0.000000in}{0.000000in}}%
\pgfpathlineto{\pgfqpoint{0.000000in}{-0.048611in}}%
\pgfusepath{stroke,fill}%
}%
\begin{pgfscope}%
\pgfsys@transformshift{1.438800in}{0.480000in}%
\pgfsys@useobject{currentmarker}{}%
\end{pgfscope}%
\end{pgfscope}%
\begin{pgfscope}%
\definecolor{textcolor}{rgb}{0.000000,0.000000,0.000000}%
\pgfsetstrokecolor{textcolor}%
\pgfsetfillcolor{textcolor}%
\pgftext[x=1.438800in,y=0.382778in,,top]{\color{textcolor}{\rmfamily\fontsize{9.000000}{10.800000}\selectfont\catcode`\^=\active\def^{\ifmmode\sp\else\^{}\fi}\catcode`\%=\active\def
\end{pgfscope}%
\begin{pgfscope}%
\pgfpathrectangle{\pgfqpoint{0.594000in}{0.480000in}}{\pgfqpoint{2.112000in}{2.400000in}}%
\pgfusepath{clip}%
\pgfsetrectcap%
\pgfsetroundjoin%
\pgfsetlinewidth{0.803000pt}%
\definecolor{currentstroke}{rgb}{0.690196,0.690196,0.690196}%
\pgfsetstrokecolor{currentstroke}%
\pgfsetdash{}{0pt}%
\pgfpathmoveto{\pgfqpoint{1.861200in}{0.480000in}}%
\pgfpathlineto{\pgfqpoint{1.861200in}{2.880000in}}%
\pgfusepath{stroke}%
\end{pgfscope}%
\begin{pgfscope}%
\pgfsetbuttcap%
\pgfsetroundjoin%
\definecolor{currentfill}{rgb}{0.000000,0.000000,0.000000}%
\pgfsetfillcolor{currentfill}%
\pgfsetlinewidth{0.803000pt}%
\definecolor{currentstroke}{rgb}{0.000000,0.000000,0.000000}%
\pgfsetstrokecolor{currentstroke}%
\pgfsetdash{}{0pt}%
\pgfsys@defobject{currentmarker}{\pgfqpoint{0.000000in}{-0.048611in}}{\pgfqpoint{0.000000in}{0.000000in}}{%
\pgfpathmoveto{\pgfqpoint{0.000000in}{0.000000in}}%
\pgfpathlineto{\pgfqpoint{0.000000in}{-0.048611in}}%
\pgfusepath{stroke,fill}%
}%
\begin{pgfscope}%
\pgfsys@transformshift{1.861200in}{0.480000in}%
\pgfsys@useobject{currentmarker}{}%
\end{pgfscope}%
\end{pgfscope}%
\begin{pgfscope}%
\definecolor{textcolor}{rgb}{0.000000,0.000000,0.000000}%
\pgfsetstrokecolor{textcolor}%
\pgfsetfillcolor{textcolor}%
\pgftext[x=1.861200in,y=0.382778in,,top]{\color{textcolor}{\rmfamily\fontsize{9.000000}{10.800000}\selectfont\catcode`\^=\active\def^{\ifmmode\sp\else\^{}\fi}\catcode`\%=\active\def
\end{pgfscope}%
\begin{pgfscope}%
\pgfpathrectangle{\pgfqpoint{0.594000in}{0.480000in}}{\pgfqpoint{2.112000in}{2.400000in}}%
\pgfusepath{clip}%
\pgfsetrectcap%
\pgfsetroundjoin%
\pgfsetlinewidth{0.803000pt}%
\definecolor{currentstroke}{rgb}{0.690196,0.690196,0.690196}%
\pgfsetstrokecolor{currentstroke}%
\pgfsetdash{}{0pt}%
\pgfpathmoveto{\pgfqpoint{2.283600in}{0.480000in}}%
\pgfpathlineto{\pgfqpoint{2.283600in}{2.880000in}}%
\pgfusepath{stroke}%
\end{pgfscope}%
\begin{pgfscope}%
\pgfsetbuttcap%
\pgfsetroundjoin%
\definecolor{currentfill}{rgb}{0.000000,0.000000,0.000000}%
\pgfsetfillcolor{currentfill}%
\pgfsetlinewidth{0.803000pt}%
\definecolor{currentstroke}{rgb}{0.000000,0.000000,0.000000}%
\pgfsetstrokecolor{currentstroke}%
\pgfsetdash{}{0pt}%
\pgfsys@defobject{currentmarker}{\pgfqpoint{0.000000in}{-0.048611in}}{\pgfqpoint{0.000000in}{0.000000in}}{%
\pgfpathmoveto{\pgfqpoint{0.000000in}{0.000000in}}%
\pgfpathlineto{\pgfqpoint{0.000000in}{-0.048611in}}%
\pgfusepath{stroke,fill}%
}%
\begin{pgfscope}%
\pgfsys@transformshift{2.283600in}{0.480000in}%
\pgfsys@useobject{currentmarker}{}%
\end{pgfscope}%
\end{pgfscope}%
\begin{pgfscope}%
\definecolor{textcolor}{rgb}{0.000000,0.000000,0.000000}%
\pgfsetstrokecolor{textcolor}%
\pgfsetfillcolor{textcolor}%
\pgftext[x=2.283600in,y=0.382778in,,top]{\color{textcolor}{\rmfamily\fontsize{9.000000}{10.800000}\selectfont\catcode`\^=\active\def^{\ifmmode\sp\else\^{}\fi}\catcode`\%=\active\def
\end{pgfscope}%
\begin{pgfscope}%
\pgfpathrectangle{\pgfqpoint{0.594000in}{0.480000in}}{\pgfqpoint{2.112000in}{2.400000in}}%
\pgfusepath{clip}%
\pgfsetrectcap%
\pgfsetroundjoin%
\pgfsetlinewidth{0.803000pt}%
\definecolor{currentstroke}{rgb}{0.690196,0.690196,0.690196}%
\pgfsetstrokecolor{currentstroke}%
\pgfsetdash{}{0pt}%
\pgfpathmoveto{\pgfqpoint{2.706000in}{0.480000in}}%
\pgfpathlineto{\pgfqpoint{2.706000in}{2.880000in}}%
\pgfusepath{stroke}%
\end{pgfscope}%
\begin{pgfscope}%
\pgfsetbuttcap%
\pgfsetroundjoin%
\definecolor{currentfill}{rgb}{0.000000,0.000000,0.000000}%
\pgfsetfillcolor{currentfill}%
\pgfsetlinewidth{0.803000pt}%
\definecolor{currentstroke}{rgb}{0.000000,0.000000,0.000000}%
\pgfsetstrokecolor{currentstroke}%
\pgfsetdash{}{0pt}%
\pgfsys@defobject{currentmarker}{\pgfqpoint{0.000000in}{-0.048611in}}{\pgfqpoint{0.000000in}{0.000000in}}{%
\pgfpathmoveto{\pgfqpoint{0.000000in}{0.000000in}}%
\pgfpathlineto{\pgfqpoint{0.000000in}{-0.048611in}}%
\pgfusepath{stroke,fill}%
}%
\begin{pgfscope}%
\pgfsys@transformshift{2.706000in}{0.480000in}%
\pgfsys@useobject{currentmarker}{}%
\end{pgfscope}%
\end{pgfscope}%
\begin{pgfscope}%
\definecolor{textcolor}{rgb}{0.000000,0.000000,0.000000}%
\pgfsetstrokecolor{textcolor}%
\pgfsetfillcolor{textcolor}%
\pgftext[x=2.706000in,y=0.382778in,,top]{\color{textcolor}{\rmfamily\fontsize{9.000000}{10.800000}\selectfont\catcode`\^=\active\def^{\ifmmode\sp\else\^{}\fi}\catcode`\%=\active\def
\end{pgfscope}%
\begin{pgfscope}%
\definecolor{textcolor}{rgb}{0.000000,0.000000,0.000000}%
\pgfsetstrokecolor{textcolor}%
\pgfsetfillcolor{textcolor}%
\pgftext[x=1.650000in,y=0.216111in,,top]{\color{textcolor}{\rmfamily\fontsize{10.000000}{12.000000}\selectfont\catcode`\^=\active\def^{\ifmmode\sp\else\^{}\fi}\catcode`\%=\active\def
\end{pgfscope}%
\begin{pgfscope}%
\pgfpathrectangle{\pgfqpoint{0.594000in}{0.480000in}}{\pgfqpoint{2.112000in}{2.400000in}}%
\pgfusepath{clip}%
\pgfsetrectcap%
\pgfsetroundjoin%
\pgfsetlinewidth{0.803000pt}%
\definecolor{currentstroke}{rgb}{0.690196,0.690196,0.690196}%
\pgfsetstrokecolor{currentstroke}%
\pgfsetdash{}{0pt}%
\pgfpathmoveto{\pgfqpoint{0.594000in}{0.589091in}}%
\pgfpathlineto{\pgfqpoint{2.706000in}{0.589091in}}%
\pgfusepath{stroke}%
\end{pgfscope}%
\begin{pgfscope}%
\pgfsetbuttcap%
\pgfsetroundjoin%
\definecolor{currentfill}{rgb}{0.000000,0.000000,0.000000}%
\pgfsetfillcolor{currentfill}%
\pgfsetlinewidth{0.803000pt}%
\definecolor{currentstroke}{rgb}{0.000000,0.000000,0.000000}%
\pgfsetstrokecolor{currentstroke}%
\pgfsetdash{}{0pt}%
\pgfsys@defobject{currentmarker}{\pgfqpoint{-0.048611in}{0.000000in}}{\pgfqpoint{-0.000000in}{0.000000in}}{%
\pgfpathmoveto{\pgfqpoint{-0.000000in}{0.000000in}}%
\pgfpathlineto{\pgfqpoint{-0.048611in}{0.000000in}}%
\pgfusepath{stroke,fill}%
}%
\begin{pgfscope}%
\pgfsys@transformshift{0.594000in}{0.589091in}%
\pgfsys@useobject{currentmarker}{}%
\end{pgfscope}%
\end{pgfscope}%
\begin{pgfscope}%
\definecolor{textcolor}{rgb}{0.000000,0.000000,0.000000}%
\pgfsetstrokecolor{textcolor}%
\pgfsetfillcolor{textcolor}%
\pgftext[x=0.332620in, y=0.545688in, left, base]{\color{textcolor}{\rmfamily\fontsize{9.000000}{10.800000}\selectfont\catcode`\^=\active\def^{\ifmmode\sp\else\^{}\fi}\catcode`\%=\active\def
\end{pgfscope}%
\begin{pgfscope}%
\pgfpathrectangle{\pgfqpoint{0.594000in}{0.480000in}}{\pgfqpoint{2.112000in}{2.400000in}}%
\pgfusepath{clip}%
\pgfsetrectcap%
\pgfsetroundjoin%
\pgfsetlinewidth{0.803000pt}%
\definecolor{currentstroke}{rgb}{0.690196,0.690196,0.690196}%
\pgfsetstrokecolor{currentstroke}%
\pgfsetdash{}{0pt}%
\pgfpathmoveto{\pgfqpoint{0.594000in}{1.048270in}}%
\pgfpathlineto{\pgfqpoint{2.706000in}{1.048270in}}%
\pgfusepath{stroke}%
\end{pgfscope}%
\begin{pgfscope}%
\pgfsetbuttcap%
\pgfsetroundjoin%
\definecolor{currentfill}{rgb}{0.000000,0.000000,0.000000}%
\pgfsetfillcolor{currentfill}%
\pgfsetlinewidth{0.803000pt}%
\definecolor{currentstroke}{rgb}{0.000000,0.000000,0.000000}%
\pgfsetstrokecolor{currentstroke}%
\pgfsetdash{}{0pt}%
\pgfsys@defobject{currentmarker}{\pgfqpoint{-0.048611in}{0.000000in}}{\pgfqpoint{-0.000000in}{0.000000in}}{%
\pgfpathmoveto{\pgfqpoint{-0.000000in}{0.000000in}}%
\pgfpathlineto{\pgfqpoint{-0.048611in}{0.000000in}}%
\pgfusepath{stroke,fill}%
}%
\begin{pgfscope}%
\pgfsys@transformshift{0.594000in}{1.048270in}%
\pgfsys@useobject{currentmarker}{}%
\end{pgfscope}%
\end{pgfscope}%
\begin{pgfscope}%
\definecolor{textcolor}{rgb}{0.000000,0.000000,0.000000}%
\pgfsetstrokecolor{textcolor}%
\pgfsetfillcolor{textcolor}%
\pgftext[x=0.332620in, y=1.004867in, left, base]{\color{textcolor}{\rmfamily\fontsize{9.000000}{10.800000}\selectfont\catcode`\^=\active\def^{\ifmmode\sp\else\^{}\fi}\catcode`\%=\active\def
\end{pgfscope}%
\begin{pgfscope}%
\pgfpathrectangle{\pgfqpoint{0.594000in}{0.480000in}}{\pgfqpoint{2.112000in}{2.400000in}}%
\pgfusepath{clip}%
\pgfsetrectcap%
\pgfsetroundjoin%
\pgfsetlinewidth{0.803000pt}%
\definecolor{currentstroke}{rgb}{0.690196,0.690196,0.690196}%
\pgfsetstrokecolor{currentstroke}%
\pgfsetdash{}{0pt}%
\pgfpathmoveto{\pgfqpoint{0.594000in}{1.507449in}}%
\pgfpathlineto{\pgfqpoint{2.706000in}{1.507449in}}%
\pgfusepath{stroke}%
\end{pgfscope}%
\begin{pgfscope}%
\pgfsetbuttcap%
\pgfsetroundjoin%
\definecolor{currentfill}{rgb}{0.000000,0.000000,0.000000}%
\pgfsetfillcolor{currentfill}%
\pgfsetlinewidth{0.803000pt}%
\definecolor{currentstroke}{rgb}{0.000000,0.000000,0.000000}%
\pgfsetstrokecolor{currentstroke}%
\pgfsetdash{}{0pt}%
\pgfsys@defobject{currentmarker}{\pgfqpoint{-0.048611in}{0.000000in}}{\pgfqpoint{-0.000000in}{0.000000in}}{%
\pgfpathmoveto{\pgfqpoint{-0.000000in}{0.000000in}}%
\pgfpathlineto{\pgfqpoint{-0.048611in}{0.000000in}}%
\pgfusepath{stroke,fill}%
}%
\begin{pgfscope}%
\pgfsys@transformshift{0.594000in}{1.507449in}%
\pgfsys@useobject{currentmarker}{}%
\end{pgfscope}%
\end{pgfscope}%
\begin{pgfscope}%
\definecolor{textcolor}{rgb}{0.000000,0.000000,0.000000}%
\pgfsetstrokecolor{textcolor}%
\pgfsetfillcolor{textcolor}%
\pgftext[x=0.332620in, y=1.464046in, left, base]{\color{textcolor}{\rmfamily\fontsize{9.000000}{10.800000}\selectfont\catcode`\^=\active\def^{\ifmmode\sp\else\^{}\fi}\catcode`\%=\active\def
\end{pgfscope}%
\begin{pgfscope}%
\pgfpathrectangle{\pgfqpoint{0.594000in}{0.480000in}}{\pgfqpoint{2.112000in}{2.400000in}}%
\pgfusepath{clip}%
\pgfsetrectcap%
\pgfsetroundjoin%
\pgfsetlinewidth{0.803000pt}%
\definecolor{currentstroke}{rgb}{0.690196,0.690196,0.690196}%
\pgfsetstrokecolor{currentstroke}%
\pgfsetdash{}{0pt}%
\pgfpathmoveto{\pgfqpoint{0.594000in}{1.966628in}}%
\pgfpathlineto{\pgfqpoint{2.706000in}{1.966628in}}%
\pgfusepath{stroke}%
\end{pgfscope}%
\begin{pgfscope}%
\pgfsetbuttcap%
\pgfsetroundjoin%
\definecolor{currentfill}{rgb}{0.000000,0.000000,0.000000}%
\pgfsetfillcolor{currentfill}%
\pgfsetlinewidth{0.803000pt}%
\definecolor{currentstroke}{rgb}{0.000000,0.000000,0.000000}%
\pgfsetstrokecolor{currentstroke}%
\pgfsetdash{}{0pt}%
\pgfsys@defobject{currentmarker}{\pgfqpoint{-0.048611in}{0.000000in}}{\pgfqpoint{-0.000000in}{0.000000in}}{%
\pgfpathmoveto{\pgfqpoint{-0.000000in}{0.000000in}}%
\pgfpathlineto{\pgfqpoint{-0.048611in}{0.000000in}}%
\pgfusepath{stroke,fill}%
}%
\begin{pgfscope}%
\pgfsys@transformshift{0.594000in}{1.966628in}%
\pgfsys@useobject{currentmarker}{}%
\end{pgfscope}%
\end{pgfscope}%
\begin{pgfscope}%
\definecolor{textcolor}{rgb}{0.000000,0.000000,0.000000}%
\pgfsetstrokecolor{textcolor}%
\pgfsetfillcolor{textcolor}%
\pgftext[x=0.332620in, y=1.923225in, left, base]{\color{textcolor}{\rmfamily\fontsize{9.000000}{10.800000}\selectfont\catcode`\^=\active\def^{\ifmmode\sp\else\^{}\fi}\catcode`\%=\active\def
\end{pgfscope}%
\begin{pgfscope}%
\pgfpathrectangle{\pgfqpoint{0.594000in}{0.480000in}}{\pgfqpoint{2.112000in}{2.400000in}}%
\pgfusepath{clip}%
\pgfsetrectcap%
\pgfsetroundjoin%
\pgfsetlinewidth{0.803000pt}%
\definecolor{currentstroke}{rgb}{0.690196,0.690196,0.690196}%
\pgfsetstrokecolor{currentstroke}%
\pgfsetdash{}{0pt}%
\pgfpathmoveto{\pgfqpoint{0.594000in}{2.425806in}}%
\pgfpathlineto{\pgfqpoint{2.706000in}{2.425806in}}%
\pgfusepath{stroke}%
\end{pgfscope}%
\begin{pgfscope}%
\pgfsetbuttcap%
\pgfsetroundjoin%
\definecolor{currentfill}{rgb}{0.000000,0.000000,0.000000}%
\pgfsetfillcolor{currentfill}%
\pgfsetlinewidth{0.803000pt}%
\definecolor{currentstroke}{rgb}{0.000000,0.000000,0.000000}%
\pgfsetstrokecolor{currentstroke}%
\pgfsetdash{}{0pt}%
\pgfsys@defobject{currentmarker}{\pgfqpoint{-0.048611in}{0.000000in}}{\pgfqpoint{-0.000000in}{0.000000in}}{%
\pgfpathmoveto{\pgfqpoint{-0.000000in}{0.000000in}}%
\pgfpathlineto{\pgfqpoint{-0.048611in}{0.000000in}}%
\pgfusepath{stroke,fill}%
}%
\begin{pgfscope}%
\pgfsys@transformshift{0.594000in}{2.425806in}%
\pgfsys@useobject{currentmarker}{}%
\end{pgfscope}%
\end{pgfscope}%
\begin{pgfscope}%
\definecolor{textcolor}{rgb}{0.000000,0.000000,0.000000}%
\pgfsetstrokecolor{textcolor}%
\pgfsetfillcolor{textcolor}%
\pgftext[x=0.332620in, y=2.382404in, left, base]{\color{textcolor}{\rmfamily\fontsize{9.000000}{10.800000}\selectfont\catcode`\^=\active\def^{\ifmmode\sp\else\^{}\fi}\catcode`\%=\active\def
\end{pgfscope}%
\begin{pgfscope}%
\definecolor{textcolor}{rgb}{0.000000,0.000000,0.000000}%
\pgfsetstrokecolor{textcolor}%
\pgfsetfillcolor{textcolor}%
\pgftext[x=0.277064in,y=1.680000in,,bottom,rotate=90.000000]{\color{textcolor}{\rmfamily\fontsize{10.000000}{12.000000}\selectfont\catcode`\^=\active\def^{\ifmmode\sp\else\^{}\fi}\catcode`\%=\active\def
\end{pgfscope}%
\begin{pgfscope}%
\pgfpathrectangle{\pgfqpoint{0.594000in}{0.480000in}}{\pgfqpoint{2.112000in}{2.400000in}}%
\pgfusepath{clip}%
\pgfsetrectcap%
\pgfsetroundjoin%
\pgfsetlinewidth{1.505625pt}%
\definecolor{currentstroke}{rgb}{0.000000,0.000000,0.000000}%
\pgfsetstrokecolor{currentstroke}%
\pgfsetstrokeopacity{0.300000}%
\pgfsetdash{}{0pt}%
\pgfpathmoveto{\pgfqpoint{0.594000in}{0.589091in}}%
\pgfpathlineto{\pgfqpoint{0.597802in}{0.640794in}}%
\pgfpathlineto{\pgfqpoint{0.602870in}{0.743684in}}%
\pgfpathlineto{\pgfqpoint{0.610474in}{0.952319in}}%
\pgfpathlineto{\pgfqpoint{0.626525in}{1.406821in}}%
\pgfpathlineto{\pgfqpoint{0.632016in}{1.502313in}}%
\pgfpathlineto{\pgfqpoint{0.636240in}{1.541870in}}%
\pgfpathlineto{\pgfqpoint{0.638774in}{1.550814in}}%
\pgfpathlineto{\pgfqpoint{0.640042in}{1.551210in}}%
\pgfpathlineto{\pgfqpoint{0.641309in}{1.548982in}}%
\pgfpathlineto{\pgfqpoint{0.643421in}{1.539713in}}%
\pgfpathlineto{\pgfqpoint{0.646800in}{1.511832in}}%
\pgfpathlineto{\pgfqpoint{0.651869in}{1.446514in}}%
\pgfpathlineto{\pgfqpoint{0.669187in}{1.202014in}}%
\pgfpathlineto{\pgfqpoint{0.672566in}{1.184668in}}%
\pgfpathlineto{\pgfqpoint{0.674256in}{1.182458in}}%
\pgfpathlineto{\pgfqpoint{0.675523in}{1.183802in}}%
\pgfpathlineto{\pgfqpoint{0.677213in}{1.189696in}}%
\pgfpathlineto{\pgfqpoint{0.679747in}{1.207414in}}%
\pgfpathlineto{\pgfqpoint{0.683549in}{1.253565in}}%
\pgfpathlineto{\pgfqpoint{0.688618in}{1.348046in}}%
\pgfpathlineto{\pgfqpoint{0.696221in}{1.539433in}}%
\pgfpathlineto{\pgfqpoint{0.710160in}{1.894097in}}%
\pgfpathlineto{\pgfqpoint{0.715651in}{1.980807in}}%
\pgfpathlineto{\pgfqpoint{0.719453in}{2.012601in}}%
\pgfpathlineto{\pgfqpoint{0.721987in}{2.020075in}}%
\pgfpathlineto{\pgfqpoint{0.723254in}{2.019708in}}%
\pgfpathlineto{\pgfqpoint{0.724944in}{2.015070in}}%
\pgfpathlineto{\pgfqpoint{0.727478in}{1.999632in}}%
\pgfpathlineto{\pgfqpoint{0.731280in}{1.959373in}}%
\pgfpathlineto{\pgfqpoint{0.736771in}{1.873650in}}%
\pgfpathlineto{\pgfqpoint{0.753245in}{1.599254in}}%
\pgfpathlineto{\pgfqpoint{0.757046in}{1.570011in}}%
\pgfpathlineto{\pgfqpoint{0.759581in}{1.562777in}}%
\pgfpathlineto{\pgfqpoint{0.760848in}{1.563088in}}%
\pgfpathlineto{\pgfqpoint{0.762538in}{1.567667in}}%
\pgfpathlineto{\pgfqpoint{0.765072in}{1.583470in}}%
\pgfpathlineto{\pgfqpoint{0.768451in}{1.620727in}}%
\pgfpathlineto{\pgfqpoint{0.773098in}{1.699094in}}%
\pgfpathlineto{\pgfqpoint{0.780278in}{1.864590in}}%
\pgfpathlineto{\pgfqpoint{0.794640in}{2.203863in}}%
\pgfpathlineto{\pgfqpoint{0.799709in}{2.276564in}}%
\pgfpathlineto{\pgfqpoint{0.803510in}{2.304496in}}%
\pgfpathlineto{\pgfqpoint{0.805622in}{2.309434in}}%
\pgfpathlineto{\pgfqpoint{0.806890in}{2.308749in}}%
\pgfpathlineto{\pgfqpoint{0.808579in}{2.303651in}}%
\pgfpathlineto{\pgfqpoint{0.811114in}{2.287372in}}%
\pgfpathlineto{\pgfqpoint{0.814915in}{2.245279in}}%
\pgfpathlineto{\pgfqpoint{0.820406in}{2.155106in}}%
\pgfpathlineto{\pgfqpoint{0.838570in}{1.830991in}}%
\pgfpathlineto{\pgfqpoint{0.842371in}{1.801232in}}%
\pgfpathlineto{\pgfqpoint{0.844906in}{1.794033in}}%
\pgfpathlineto{\pgfqpoint{0.846173in}{1.794429in}}%
\pgfpathlineto{\pgfqpoint{0.847862in}{1.799154in}}%
\pgfpathlineto{\pgfqpoint{0.850397in}{1.815172in}}%
\pgfpathlineto{\pgfqpoint{0.853776in}{1.852506in}}%
\pgfpathlineto{\pgfqpoint{0.858845in}{1.938452in}}%
\pgfpathlineto{\pgfqpoint{0.866870in}{2.122307in}}%
\pgfpathlineto{\pgfqpoint{0.878698in}{2.386302in}}%
\pgfpathlineto{\pgfqpoint{0.883766in}{2.456597in}}%
\pgfpathlineto{\pgfqpoint{0.887568in}{2.483171in}}%
\pgfpathlineto{\pgfqpoint{0.889680in}{2.487432in}}%
\pgfpathlineto{\pgfqpoint{0.890947in}{2.486347in}}%
\pgfpathlineto{\pgfqpoint{0.892637in}{2.480707in}}%
\pgfpathlineto{\pgfqpoint{0.895171in}{2.463556in}}%
\pgfpathlineto{\pgfqpoint{0.898973in}{2.419909in}}%
\pgfpathlineto{\pgfqpoint{0.904464in}{2.326662in}}%
\pgfpathlineto{\pgfqpoint{0.923472in}{1.973221in}}%
\pgfpathlineto{\pgfqpoint{0.927274in}{1.942695in}}%
\pgfpathlineto{\pgfqpoint{0.929808in}{1.935162in}}%
\pgfpathlineto{\pgfqpoint{0.931075in}{1.935420in}}%
\pgfpathlineto{\pgfqpoint{0.932765in}{1.939975in}}%
\pgfpathlineto{\pgfqpoint{0.935299in}{1.955724in}}%
\pgfpathlineto{\pgfqpoint{0.938678in}{1.992581in}}%
\pgfpathlineto{\pgfqpoint{0.943747in}{2.077279in}}%
\pgfpathlineto{\pgfqpoint{0.951773in}{2.257083in}}%
\pgfpathlineto{\pgfqpoint{0.962755in}{2.495552in}}%
\pgfpathlineto{\pgfqpoint{0.967824in}{2.565497in}}%
\pgfpathlineto{\pgfqpoint{0.971626in}{2.592239in}}%
\pgfpathlineto{\pgfqpoint{0.973738in}{2.596667in}}%
\pgfpathlineto{\pgfqpoint{0.975005in}{2.595688in}}%
\pgfpathlineto{\pgfqpoint{0.976694in}{2.590177in}}%
\pgfpathlineto{\pgfqpoint{0.979229in}{2.573159in}}%
\pgfpathlineto{\pgfqpoint{0.983030in}{2.529456in}}%
\pgfpathlineto{\pgfqpoint{0.988522in}{2.435270in}}%
\pgfpathlineto{\pgfqpoint{1.008797in}{2.053858in}}%
\pgfpathlineto{\pgfqpoint{1.012598in}{2.026367in}}%
\pgfpathlineto{\pgfqpoint{1.014710in}{2.021245in}}%
\pgfpathlineto{\pgfqpoint{1.015978in}{2.021773in}}%
\pgfpathlineto{\pgfqpoint{1.017667in}{2.026691in}}%
\pgfpathlineto{\pgfqpoint{1.020202in}{2.042955in}}%
\pgfpathlineto{\pgfqpoint{1.024003in}{2.086191in}}%
\pgfpathlineto{\pgfqpoint{1.029072in}{2.173592in}}%
\pgfpathlineto{\pgfqpoint{1.038365in}{2.382726in}}%
\pgfpathlineto{\pgfqpoint{1.047658in}{2.573975in}}%
\pgfpathlineto{\pgfqpoint{1.052726in}{2.638827in}}%
\pgfpathlineto{\pgfqpoint{1.056106in}{2.660097in}}%
\pgfpathlineto{\pgfqpoint{1.058218in}{2.663697in}}%
\pgfpathlineto{\pgfqpoint{1.059485in}{2.662227in}}%
\pgfpathlineto{\pgfqpoint{1.061597in}{2.653787in}}%
\pgfpathlineto{\pgfqpoint{1.064554in}{2.629859in}}%
\pgfpathlineto{\pgfqpoint{1.068778in}{2.573624in}}%
\pgfpathlineto{\pgfqpoint{1.075114in}{2.453004in}}%
\pgfpathlineto{\pgfqpoint{1.090742in}{2.139705in}}%
\pgfpathlineto{\pgfqpoint{1.095389in}{2.089781in}}%
\pgfpathlineto{\pgfqpoint{1.098346in}{2.075494in}}%
\pgfpathlineto{\pgfqpoint{1.100035in}{2.073873in}}%
\pgfpathlineto{\pgfqpoint{1.101302in}{2.075824in}}%
\pgfpathlineto{\pgfqpoint{1.103414in}{2.085065in}}%
\pgfpathlineto{\pgfqpoint{1.106371in}{2.110181in}}%
\pgfpathlineto{\pgfqpoint{1.110595in}{2.168418in}}%
\pgfpathlineto{\pgfqpoint{1.116931in}{2.293165in}}%
\pgfpathlineto{\pgfqpoint{1.133405in}{2.636877in}}%
\pgfpathlineto{\pgfqpoint{1.138051in}{2.688055in}}%
\pgfpathlineto{\pgfqpoint{1.141008in}{2.702866in}}%
\pgfpathlineto{\pgfqpoint{1.142698in}{2.704737in}}%
\pgfpathlineto{\pgfqpoint{1.143965in}{2.702965in}}%
\pgfpathlineto{\pgfqpoint{1.146077in}{2.694029in}}%
\pgfpathlineto{\pgfqpoint{1.149034in}{2.669416in}}%
\pgfpathlineto{\pgfqpoint{1.153258in}{2.612223in}}%
\pgfpathlineto{\pgfqpoint{1.159594in}{2.490208in}}%
\pgfpathlineto{\pgfqpoint{1.175222in}{2.173681in}}%
\pgfpathlineto{\pgfqpoint{1.179869in}{2.122853in}}%
\pgfpathlineto{\pgfqpoint{1.182826in}{2.108003in}}%
\pgfpathlineto{\pgfqpoint{1.184515in}{2.106064in}}%
\pgfpathlineto{\pgfqpoint{1.185782in}{2.107779in}}%
\pgfpathlineto{\pgfqpoint{1.187894in}{2.116631in}}%
\pgfpathlineto{\pgfqpoint{1.190851in}{2.141209in}}%
\pgfpathlineto{\pgfqpoint{1.195075in}{2.198694in}}%
\pgfpathlineto{\pgfqpoint{1.201411in}{2.322347in}}%
\pgfpathlineto{\pgfqpoint{1.217885in}{2.663397in}}%
\pgfpathlineto{\pgfqpoint{1.222531in}{2.713869in}}%
\pgfpathlineto{\pgfqpoint{1.225488in}{2.728240in}}%
\pgfpathlineto{\pgfqpoint{1.227178in}{2.729864in}}%
\pgfpathlineto{\pgfqpoint{1.228445in}{2.727908in}}%
\pgfpathlineto{\pgfqpoint{1.230557in}{2.718668in}}%
\pgfpathlineto{\pgfqpoint{1.233514in}{2.693635in}}%
\pgfpathlineto{\pgfqpoint{1.237738in}{2.635855in}}%
\pgfpathlineto{\pgfqpoint{1.244074in}{2.512987in}}%
\pgfpathlineto{\pgfqpoint{1.259702in}{2.194483in}}%
\pgfpathlineto{\pgfqpoint{1.264349in}{2.143101in}}%
\pgfpathlineto{\pgfqpoint{1.267306in}{2.127907in}}%
\pgfpathlineto{\pgfqpoint{1.268995in}{2.125774in}}%
\pgfpathlineto{\pgfqpoint{1.270262in}{2.127344in}}%
\pgfpathlineto{\pgfqpoint{1.272374in}{2.135957in}}%
\pgfpathlineto{\pgfqpoint{1.275331in}{2.160206in}}%
\pgfpathlineto{\pgfqpoint{1.279555in}{2.217231in}}%
\pgfpathlineto{\pgfqpoint{1.285891in}{2.340215in}}%
\pgfpathlineto{\pgfqpoint{1.302365in}{2.679634in}}%
\pgfpathlineto{\pgfqpoint{1.307011in}{2.729674in}}%
\pgfpathlineto{\pgfqpoint{1.309968in}{2.743776in}}%
\pgfpathlineto{\pgfqpoint{1.311658in}{2.745248in}}%
\pgfpathlineto{\pgfqpoint{1.312925in}{2.743179in}}%
\pgfpathlineto{\pgfqpoint{1.315037in}{2.733753in}}%
\pgfpathlineto{\pgfqpoint{1.317994in}{2.708464in}}%
\pgfpathlineto{\pgfqpoint{1.322218in}{2.650324in}}%
\pgfpathlineto{\pgfqpoint{1.328554in}{2.526933in}}%
\pgfpathlineto{\pgfqpoint{1.344182in}{2.207219in}}%
\pgfpathlineto{\pgfqpoint{1.348829in}{2.155498in}}%
\pgfpathlineto{\pgfqpoint{1.351786in}{2.140093in}}%
\pgfpathlineto{\pgfqpoint{1.353475in}{2.137841in}}%
\pgfpathlineto{\pgfqpoint{1.354742in}{2.139322in}}%
\pgfpathlineto{\pgfqpoint{1.356854in}{2.147790in}}%
\pgfpathlineto{\pgfqpoint{1.359811in}{2.171838in}}%
\pgfpathlineto{\pgfqpoint{1.364035in}{2.228580in}}%
\pgfpathlineto{\pgfqpoint{1.370371in}{2.351154in}}%
\pgfpathlineto{\pgfqpoint{1.386845in}{2.689575in}}%
\pgfpathlineto{\pgfqpoint{1.391491in}{2.739351in}}%
\pgfpathlineto{\pgfqpoint{1.394448in}{2.753288in}}%
\pgfpathlineto{\pgfqpoint{1.396138in}{2.754667in}}%
\pgfpathlineto{\pgfqpoint{1.397405in}{2.752529in}}%
\pgfpathlineto{\pgfqpoint{1.399517in}{2.742989in}}%
\pgfpathlineto{\pgfqpoint{1.402474in}{2.717543in}}%
\pgfpathlineto{\pgfqpoint{1.406698in}{2.659183in}}%
\pgfpathlineto{\pgfqpoint{1.413034in}{2.535472in}}%
\pgfpathlineto{\pgfqpoint{1.428662in}{2.215017in}}%
\pgfpathlineto{\pgfqpoint{1.433309in}{2.163088in}}%
\pgfpathlineto{\pgfqpoint{1.436688in}{2.146521in}}%
\pgfpathlineto{\pgfqpoint{1.438378in}{2.145403in}}%
\pgfpathlineto{\pgfqpoint{1.439645in}{2.147735in}}%
\pgfpathlineto{\pgfqpoint{1.441757in}{2.157596in}}%
\pgfpathlineto{\pgfqpoint{1.444714in}{2.183485in}}%
\pgfpathlineto{\pgfqpoint{1.448938in}{2.242464in}}%
\pgfpathlineto{\pgfqpoint{1.455274in}{2.367076in}}%
\pgfpathlineto{\pgfqpoint{1.470902in}{2.689616in}}%
\pgfpathlineto{\pgfqpoint{1.475549in}{2.742129in}}%
\pgfpathlineto{\pgfqpoint{1.478928in}{2.759112in}}%
\pgfpathlineto{\pgfqpoint{1.480618in}{2.760434in}}%
\pgfpathlineto{\pgfqpoint{1.481885in}{2.758254in}}%
\pgfpathlineto{\pgfqpoint{1.483997in}{2.748644in}}%
\pgfpathlineto{\pgfqpoint{1.486954in}{2.723101in}}%
\pgfpathlineto{\pgfqpoint{1.491178in}{2.664607in}}%
\pgfpathlineto{\pgfqpoint{1.497514in}{2.540700in}}%
\pgfpathlineto{\pgfqpoint{1.513142in}{2.219792in}}%
\pgfpathlineto{\pgfqpoint{1.517789in}{2.167736in}}%
\pgfpathlineto{\pgfqpoint{1.521168in}{2.151078in}}%
\pgfpathlineto{\pgfqpoint{1.522858in}{2.149915in}}%
\pgfpathlineto{\pgfqpoint{1.524125in}{2.152214in}}%
\pgfpathlineto{\pgfqpoint{1.526237in}{2.162021in}}%
\pgfpathlineto{\pgfqpoint{1.529194in}{2.187835in}}%
\pgfpathlineto{\pgfqpoint{1.533418in}{2.246708in}}%
\pgfpathlineto{\pgfqpoint{1.539754in}{2.371166in}}%
\pgfpathlineto{\pgfqpoint{1.555382in}{2.693351in}}%
\pgfpathlineto{\pgfqpoint{1.560029in}{2.745765in}}%
\pgfpathlineto{\pgfqpoint{1.563408in}{2.762677in}}%
\pgfpathlineto{\pgfqpoint{1.565098in}{2.763965in}}%
\pgfpathlineto{\pgfqpoint{1.566365in}{2.761759in}}%
\pgfpathlineto{\pgfqpoint{1.568477in}{2.752106in}}%
\pgfpathlineto{\pgfqpoint{1.571434in}{2.726504in}}%
\pgfpathlineto{\pgfqpoint{1.575658in}{2.667928in}}%
\pgfpathlineto{\pgfqpoint{1.581994in}{2.543901in}}%
\pgfpathlineto{\pgfqpoint{1.597622in}{2.222715in}}%
\pgfpathlineto{\pgfqpoint{1.602269in}{2.170581in}}%
\pgfpathlineto{\pgfqpoint{1.605648in}{2.153868in}}%
\pgfpathlineto{\pgfqpoint{1.607338in}{2.152678in}}%
\pgfpathlineto{\pgfqpoint{1.608605in}{2.154957in}}%
\pgfpathlineto{\pgfqpoint{1.610717in}{2.164730in}}%
\pgfpathlineto{\pgfqpoint{1.613674in}{2.190498in}}%
\pgfpathlineto{\pgfqpoint{1.617898in}{2.249306in}}%
\pgfpathlineto{\pgfqpoint{1.624234in}{2.373671in}}%
\pgfpathlineto{\pgfqpoint{1.639862in}{2.695639in}}%
\pgfpathlineto{\pgfqpoint{1.644509in}{2.747991in}}%
\pgfpathlineto{\pgfqpoint{1.647888in}{2.764861in}}%
\pgfpathlineto{\pgfqpoint{1.649578in}{2.766127in}}%
\pgfpathlineto{\pgfqpoint{1.650845in}{2.763905in}}%
\pgfpathlineto{\pgfqpoint{1.652957in}{2.754226in}}%
\pgfpathlineto{\pgfqpoint{1.655914in}{2.728588in}}%
\pgfpathlineto{\pgfqpoint{1.660138in}{2.669961in}}%
\pgfpathlineto{\pgfqpoint{1.666474in}{2.545861in}}%
\pgfpathlineto{\pgfqpoint{1.682102in}{2.224505in}}%
\pgfpathlineto{\pgfqpoint{1.686749in}{2.172323in}}%
\pgfpathlineto{\pgfqpoint{1.690128in}{2.155576in}}%
\pgfpathlineto{\pgfqpoint{1.691818in}{2.154370in}}%
\pgfpathlineto{\pgfqpoint{1.693085in}{2.156636in}}%
\pgfpathlineto{\pgfqpoint{1.695197in}{2.166389in}}%
\pgfpathlineto{\pgfqpoint{1.698154in}{2.192128in}}%
\pgfpathlineto{\pgfqpoint{1.702378in}{2.250897in}}%
\pgfpathlineto{\pgfqpoint{1.708714in}{2.375204in}}%
\pgfpathlineto{\pgfqpoint{1.724342in}{2.697039in}}%
\pgfpathlineto{\pgfqpoint{1.728989in}{2.749355in}}%
\pgfpathlineto{\pgfqpoint{1.732368in}{2.766197in}}%
\pgfpathlineto{\pgfqpoint{1.734058in}{2.767450in}}%
\pgfpathlineto{\pgfqpoint{1.735325in}{2.765219in}}%
\pgfpathlineto{\pgfqpoint{1.737437in}{2.755524in}}%
\pgfpathlineto{\pgfqpoint{1.740394in}{2.729864in}}%
\pgfpathlineto{\pgfqpoint{1.744618in}{2.671206in}}%
\pgfpathlineto{\pgfqpoint{1.750954in}{2.547060in}}%
\pgfpathlineto{\pgfqpoint{1.766582in}{2.225600in}}%
\pgfpathlineto{\pgfqpoint{1.771229in}{2.173390in}}%
\pgfpathlineto{\pgfqpoint{1.774608in}{2.156622in}}%
\pgfpathlineto{\pgfqpoint{1.776298in}{2.155405in}}%
\pgfpathlineto{\pgfqpoint{1.777565in}{2.157664in}}%
\pgfpathlineto{\pgfqpoint{1.779677in}{2.167404in}}%
\pgfpathlineto{\pgfqpoint{1.782634in}{2.193126in}}%
\pgfpathlineto{\pgfqpoint{1.786858in}{2.251871in}}%
\pgfpathlineto{\pgfqpoint{1.793194in}{2.376143in}}%
\pgfpathlineto{\pgfqpoint{1.808822in}{2.697896in}}%
\pgfpathlineto{\pgfqpoint{1.813469in}{2.750189in}}%
\pgfpathlineto{\pgfqpoint{1.816848in}{2.767015in}}%
\pgfpathlineto{\pgfqpoint{1.818538in}{2.768261in}}%
\pgfpathlineto{\pgfqpoint{1.819805in}{2.766023in}}%
\pgfpathlineto{\pgfqpoint{1.821917in}{2.756318in}}%
\pgfpathlineto{\pgfqpoint{1.824874in}{2.730645in}}%
\pgfpathlineto{\pgfqpoint{1.829098in}{2.671968in}}%
\pgfpathlineto{\pgfqpoint{1.835434in}{2.547795in}}%
\pgfpathlineto{\pgfqpoint{1.851062in}{2.226271in}}%
\pgfpathlineto{\pgfqpoint{1.855709in}{2.174043in}}%
\pgfpathlineto{\pgfqpoint{1.859088in}{2.157262in}}%
\pgfpathlineto{\pgfqpoint{1.860778in}{2.156039in}}%
\pgfpathlineto{\pgfqpoint{1.862045in}{2.158294in}}%
\pgfpathlineto{\pgfqpoint{1.864157in}{2.168026in}}%
\pgfpathlineto{\pgfqpoint{1.867114in}{2.193737in}}%
\pgfpathlineto{\pgfqpoint{1.871338in}{2.252468in}}%
\pgfpathlineto{\pgfqpoint{1.877674in}{2.376718in}}%
\pgfpathlineto{\pgfqpoint{1.893302in}{2.698421in}}%
\pgfpathlineto{\pgfqpoint{1.897949in}{2.750700in}}%
\pgfpathlineto{\pgfqpoint{1.901328in}{2.767517in}}%
\pgfpathlineto{\pgfqpoint{1.903018in}{2.768757in}}%
\pgfpathlineto{\pgfqpoint{1.904285in}{2.766516in}}%
\pgfpathlineto{\pgfqpoint{1.906397in}{2.756805in}}%
\pgfpathlineto{\pgfqpoint{1.909354in}{2.731123in}}%
\pgfpathlineto{\pgfqpoint{1.913578in}{2.672434in}}%
\pgfpathlineto{\pgfqpoint{1.919914in}{2.548245in}}%
\pgfpathlineto{\pgfqpoint{1.935542in}{2.226682in}}%
\pgfpathlineto{\pgfqpoint{1.940189in}{2.174442in}}%
\pgfpathlineto{\pgfqpoint{1.943568in}{2.157654in}}%
\pgfpathlineto{\pgfqpoint{1.945258in}{2.156427in}}%
\pgfpathlineto{\pgfqpoint{1.946525in}{2.158679in}}%
\pgfpathlineto{\pgfqpoint{1.948637in}{2.168407in}}%
\pgfpathlineto{\pgfqpoint{1.951594in}{2.194112in}}%
\pgfpathlineto{\pgfqpoint{1.955818in}{2.252833in}}%
\pgfpathlineto{\pgfqpoint{1.962154in}{2.377070in}}%
\pgfpathlineto{\pgfqpoint{1.977782in}{2.698743in}}%
\pgfpathlineto{\pgfqpoint{1.982429in}{2.751013in}}%
\pgfpathlineto{\pgfqpoint{1.985808in}{2.767823in}}%
\pgfpathlineto{\pgfqpoint{1.987498in}{2.769061in}}%
\pgfpathlineto{\pgfqpoint{1.988765in}{2.766817in}}%
\pgfpathlineto{\pgfqpoint{1.990877in}{2.757103in}}%
\pgfpathlineto{\pgfqpoint{1.993834in}{2.731416in}}%
\pgfpathlineto{\pgfqpoint{1.998058in}{2.672720in}}%
\pgfpathlineto{\pgfqpoint{2.004394in}{2.548520in}}%
\pgfpathlineto{\pgfqpoint{2.020022in}{2.226933in}}%
\pgfpathlineto{\pgfqpoint{2.024669in}{2.174687in}}%
\pgfpathlineto{\pgfqpoint{2.028048in}{2.157894in}}%
\pgfpathlineto{\pgfqpoint{2.029738in}{2.156665in}}%
\pgfpathlineto{\pgfqpoint{2.031005in}{2.158915in}}%
\pgfpathlineto{\pgfqpoint{2.033117in}{2.168640in}}%
\pgfpathlineto{\pgfqpoint{2.036074in}{2.194341in}}%
\pgfpathlineto{\pgfqpoint{2.040298in}{2.253056in}}%
\pgfpathlineto{\pgfqpoint{2.046634in}{2.377285in}}%
\pgfpathlineto{\pgfqpoint{2.062262in}{2.698939in}}%
\pgfpathlineto{\pgfqpoint{2.066909in}{2.751205in}}%
\pgfpathlineto{\pgfqpoint{2.070288in}{2.768011in}}%
\pgfpathlineto{\pgfqpoint{2.071978in}{2.769247in}}%
\pgfpathlineto{\pgfqpoint{2.073245in}{2.767002in}}%
\pgfpathlineto{\pgfqpoint{2.075357in}{2.757285in}}%
\pgfpathlineto{\pgfqpoint{2.078314in}{2.731595in}}%
\pgfpathlineto{\pgfqpoint{2.082538in}{2.672895in}}%
\pgfpathlineto{\pgfqpoint{2.088874in}{2.548689in}}%
\pgfpathlineto{\pgfqpoint{2.104502in}{2.227087in}}%
\pgfpathlineto{\pgfqpoint{2.109149in}{2.174837in}}%
\pgfpathlineto{\pgfqpoint{2.112528in}{2.158041in}}%
\pgfpathlineto{\pgfqpoint{2.114218in}{2.156811in}}%
\pgfpathlineto{\pgfqpoint{2.115485in}{2.159059in}}%
\pgfpathlineto{\pgfqpoint{2.117597in}{2.168783in}}%
\pgfpathlineto{\pgfqpoint{2.120554in}{2.194481in}}%
\pgfpathlineto{\pgfqpoint{2.124778in}{2.253193in}}%
\pgfpathlineto{\pgfqpoint{2.131114in}{2.377417in}}%
\pgfpathlineto{\pgfqpoint{2.146742in}{2.699060in}}%
\pgfpathlineto{\pgfqpoint{2.151389in}{2.751322in}}%
\pgfpathlineto{\pgfqpoint{2.154768in}{2.768126in}}%
\pgfpathlineto{\pgfqpoint{2.156458in}{2.769360in}}%
\pgfpathlineto{\pgfqpoint{2.157725in}{2.767115in}}%
\pgfpathlineto{\pgfqpoint{2.159837in}{2.757397in}}%
\pgfpathlineto{\pgfqpoint{2.162794in}{2.731705in}}%
\pgfpathlineto{\pgfqpoint{2.167018in}{2.673002in}}%
\pgfpathlineto{\pgfqpoint{2.173354in}{2.548792in}}%
\pgfpathlineto{\pgfqpoint{2.188982in}{2.227182in}}%
\pgfpathlineto{\pgfqpoint{2.193629in}{2.174929in}}%
\pgfpathlineto{\pgfqpoint{2.197008in}{2.158131in}}%
\pgfpathlineto{\pgfqpoint{2.198698in}{2.156900in}}%
\pgfpathlineto{\pgfqpoint{2.199965in}{2.159148in}}%
\pgfpathlineto{\pgfqpoint{2.202077in}{2.168870in}}%
\pgfpathlineto{\pgfqpoint{2.205034in}{2.194567in}}%
\pgfpathlineto{\pgfqpoint{2.209258in}{2.253277in}}%
\pgfpathlineto{\pgfqpoint{2.215594in}{2.377498in}}%
\pgfpathlineto{\pgfqpoint{2.231222in}{2.699134in}}%
\pgfpathlineto{\pgfqpoint{2.235869in}{2.751394in}}%
\pgfpathlineto{\pgfqpoint{2.239248in}{2.768197in}}%
\pgfpathlineto{\pgfqpoint{2.240938in}{2.769430in}}%
\pgfpathlineto{\pgfqpoint{2.242205in}{2.767184in}}%
\pgfpathlineto{\pgfqpoint{2.244317in}{2.757465in}}%
\pgfpathlineto{\pgfqpoint{2.247274in}{2.731772in}}%
\pgfpathlineto{\pgfqpoint{2.251498in}{2.673068in}}%
\pgfpathlineto{\pgfqpoint{2.257834in}{2.548855in}}%
\pgfpathlineto{\pgfqpoint{2.273462in}{2.227239in}}%
\pgfpathlineto{\pgfqpoint{2.278109in}{2.174985in}}%
\pgfpathlineto{\pgfqpoint{2.281488in}{2.158186in}}%
\pgfpathlineto{\pgfqpoint{2.283178in}{2.156954in}}%
\pgfpathlineto{\pgfqpoint{2.284445in}{2.159202in}}%
\pgfpathlineto{\pgfqpoint{2.286557in}{2.168923in}}%
\pgfpathlineto{\pgfqpoint{2.289514in}{2.194619in}}%
\pgfpathlineto{\pgfqpoint{2.293738in}{2.253328in}}%
\pgfpathlineto{\pgfqpoint{2.300074in}{2.377548in}}%
\pgfpathlineto{\pgfqpoint{2.315702in}{2.699179in}}%
\pgfpathlineto{\pgfqpoint{2.320349in}{2.751438in}}%
\pgfpathlineto{\pgfqpoint{2.323728in}{2.768240in}}%
\pgfpathlineto{\pgfqpoint{2.325418in}{2.769473in}}%
\pgfpathlineto{\pgfqpoint{2.326685in}{2.767226in}}%
\pgfpathlineto{\pgfqpoint{2.328797in}{2.757507in}}%
\pgfpathlineto{\pgfqpoint{2.331754in}{2.731813in}}%
\pgfpathlineto{\pgfqpoint{2.335978in}{2.673108in}}%
\pgfpathlineto{\pgfqpoint{2.342314in}{2.548894in}}%
\pgfpathlineto{\pgfqpoint{2.357942in}{2.227275in}}%
\pgfpathlineto{\pgfqpoint{2.362589in}{2.175019in}}%
\pgfpathlineto{\pgfqpoint{2.365968in}{2.158220in}}%
\pgfpathlineto{\pgfqpoint{2.367658in}{2.156988in}}%
\pgfpathlineto{\pgfqpoint{2.368925in}{2.159235in}}%
\pgfpathlineto{\pgfqpoint{2.371037in}{2.168956in}}%
\pgfpathlineto{\pgfqpoint{2.373994in}{2.194652in}}%
\pgfpathlineto{\pgfqpoint{2.378218in}{2.253360in}}%
\pgfpathlineto{\pgfqpoint{2.384554in}{2.377578in}}%
\pgfpathlineto{\pgfqpoint{2.400182in}{2.699207in}}%
\pgfpathlineto{\pgfqpoint{2.404829in}{2.751464in}}%
\pgfpathlineto{\pgfqpoint{2.408208in}{2.768266in}}%
\pgfpathlineto{\pgfqpoint{2.409898in}{2.769499in}}%
\pgfpathlineto{\pgfqpoint{2.411165in}{2.767252in}}%
\pgfpathlineto{\pgfqpoint{2.413277in}{2.757532in}}%
\pgfpathlineto{\pgfqpoint{2.416234in}{2.731839in}}%
\pgfpathlineto{\pgfqpoint{2.420458in}{2.673133in}}%
\pgfpathlineto{\pgfqpoint{2.426794in}{2.548918in}}%
\pgfpathlineto{\pgfqpoint{2.442422in}{2.227296in}}%
\pgfpathlineto{\pgfqpoint{2.447069in}{2.175041in}}%
\pgfpathlineto{\pgfqpoint{2.450448in}{2.158240in}}%
\pgfpathlineto{\pgfqpoint{2.452138in}{2.157008in}}%
\pgfpathlineto{\pgfqpoint{2.453405in}{2.159255in}}%
\pgfpathlineto{\pgfqpoint{2.455517in}{2.168976in}}%
\pgfpathlineto{\pgfqpoint{2.458474in}{2.194671in}}%
\pgfpathlineto{\pgfqpoint{2.462698in}{2.253379in}}%
\pgfpathlineto{\pgfqpoint{2.469034in}{2.377596in}}%
\pgfpathlineto{\pgfqpoint{2.484662in}{2.699223in}}%
\pgfpathlineto{\pgfqpoint{2.489309in}{2.751481in}}%
\pgfpathlineto{\pgfqpoint{2.492688in}{2.768282in}}%
\pgfpathlineto{\pgfqpoint{2.494378in}{2.769515in}}%
\pgfpathlineto{\pgfqpoint{2.495645in}{2.767268in}}%
\pgfpathlineto{\pgfqpoint{2.497757in}{2.757548in}}%
\pgfpathlineto{\pgfqpoint{2.500714in}{2.731854in}}%
\pgfpathlineto{\pgfqpoint{2.504938in}{2.673148in}}%
\pgfpathlineto{\pgfqpoint{2.511274in}{2.548932in}}%
\pgfpathlineto{\pgfqpoint{2.526902in}{2.227310in}}%
\pgfpathlineto{\pgfqpoint{2.531549in}{2.175053in}}%
\pgfpathlineto{\pgfqpoint{2.534928in}{2.158253in}}%
\pgfpathlineto{\pgfqpoint{2.536618in}{2.157021in}}%
\pgfpathlineto{\pgfqpoint{2.537885in}{2.159268in}}%
\pgfpathlineto{\pgfqpoint{2.539997in}{2.168989in}}%
\pgfpathlineto{\pgfqpoint{2.542954in}{2.194683in}}%
\pgfpathlineto{\pgfqpoint{2.547178in}{2.253391in}}%
\pgfpathlineto{\pgfqpoint{2.553514in}{2.377608in}}%
\pgfpathlineto{\pgfqpoint{2.569142in}{2.699234in}}%
\pgfpathlineto{\pgfqpoint{2.573789in}{2.751491in}}%
\pgfpathlineto{\pgfqpoint{2.577168in}{2.768292in}}%
\pgfpathlineto{\pgfqpoint{2.578858in}{2.769525in}}%
\pgfpathlineto{\pgfqpoint{2.580125in}{2.767278in}}%
\pgfpathlineto{\pgfqpoint{2.582237in}{2.757558in}}%
\pgfpathlineto{\pgfqpoint{2.585194in}{2.731863in}}%
\pgfpathlineto{\pgfqpoint{2.589418in}{2.673157in}}%
\pgfpathlineto{\pgfqpoint{2.595754in}{2.548941in}}%
\pgfpathlineto{\pgfqpoint{2.611382in}{2.227318in}}%
\pgfpathlineto{\pgfqpoint{2.616029in}{2.175061in}}%
\pgfpathlineto{\pgfqpoint{2.619408in}{2.158261in}}%
\pgfpathlineto{\pgfqpoint{2.621098in}{2.157028in}}%
\pgfpathlineto{\pgfqpoint{2.622365in}{2.159275in}}%
\pgfpathlineto{\pgfqpoint{2.624477in}{2.168996in}}%
\pgfpathlineto{\pgfqpoint{2.627434in}{2.194691in}}%
\pgfpathlineto{\pgfqpoint{2.631658in}{2.253398in}}%
\pgfpathlineto{\pgfqpoint{2.637994in}{2.377615in}}%
\pgfpathlineto{\pgfqpoint{2.653622in}{2.699240in}}%
\pgfpathlineto{\pgfqpoint{2.658269in}{2.751497in}}%
\pgfpathlineto{\pgfqpoint{2.661648in}{2.768298in}}%
\pgfpathlineto{\pgfqpoint{2.663338in}{2.769531in}}%
\pgfpathlineto{\pgfqpoint{2.664605in}{2.767284in}}%
\pgfpathlineto{\pgfqpoint{2.666717in}{2.757564in}}%
\pgfpathlineto{\pgfqpoint{2.669674in}{2.731869in}}%
\pgfpathlineto{\pgfqpoint{2.673898in}{2.673162in}}%
\pgfpathlineto{\pgfqpoint{2.680234in}{2.548947in}}%
\pgfpathlineto{\pgfqpoint{2.695862in}{2.227323in}}%
\pgfpathlineto{\pgfqpoint{2.700509in}{2.175066in}}%
\pgfpathlineto{\pgfqpoint{2.703888in}{2.158266in}}%
\pgfpathlineto{\pgfqpoint{2.705578in}{2.157033in}}%
\pgfpathlineto{\pgfqpoint{2.706422in}{2.158230in}}%
\pgfpathlineto{\pgfqpoint{2.706422in}{2.158230in}}%
\pgfusepath{stroke}%
\end{pgfscope}%
\begin{pgfscope}%
\pgfpathrectangle{\pgfqpoint{0.594000in}{0.480000in}}{\pgfqpoint{2.112000in}{2.400000in}}%
\pgfusepath{clip}%
\pgfsetrectcap%
\pgfsetroundjoin%
\pgfsetlinewidth{1.505625pt}%
\definecolor{currentstroke}{rgb}{0.000000,0.000000,0.000000}%
\pgfsetstrokecolor{currentstroke}%
\pgfsetdash{}{0pt}%
\pgfpathmoveto{\pgfqpoint{0.639619in}{1.551374in}}%
\pgfpathlineto{\pgfqpoint{0.722410in}{2.020254in}}%
\pgfpathlineto{\pgfqpoint{0.806045in}{2.309508in}}%
\pgfpathlineto{\pgfqpoint{0.889680in}{2.487432in}}%
\pgfpathlineto{\pgfqpoint{0.973738in}{2.596667in}}%
\pgfpathlineto{\pgfqpoint{1.058218in}{2.663697in}}%
\pgfpathlineto{\pgfqpoint{1.142698in}{2.704737in}}%
\pgfpathlineto{\pgfqpoint{1.226755in}{2.729912in}}%
\pgfpathlineto{\pgfqpoint{1.311235in}{2.745334in}}%
\pgfpathlineto{\pgfqpoint{1.395715in}{2.754776in}}%
\pgfpathlineto{\pgfqpoint{1.480195in}{2.760557in}}%
\pgfpathlineto{\pgfqpoint{1.564675in}{2.764096in}}%
\pgfpathlineto{\pgfqpoint{1.733635in}{2.767590in}}%
\pgfpathlineto{\pgfqpoint{1.987075in}{2.769205in}}%
\pgfpathlineto{\pgfqpoint{2.662915in}{2.769676in}}%
\pgfpathlineto{\pgfqpoint{2.716000in}{2.769678in}}%
\pgfpathlineto{\pgfqpoint{2.716000in}{2.769678in}}%
\pgfusepath{stroke}%
\end{pgfscope}%
\begin{pgfscope}%
\pgfpathrectangle{\pgfqpoint{0.594000in}{0.480000in}}{\pgfqpoint{2.112000in}{2.400000in}}%
\pgfusepath{clip}%
\pgfsetrectcap%
\pgfsetroundjoin%
\pgfsetlinewidth{1.505625pt}%
\definecolor{currentstroke}{rgb}{0.000000,0.000000,1.000000}%
\pgfsetstrokecolor{currentstroke}%
\pgfsetdash{}{0pt}%
\pgfpathmoveto{\pgfqpoint{0.615120in}{1.497227in}}%
\pgfpathlineto{\pgfqpoint{0.642154in}{2.153778in}}%
\pgfpathlineto{\pgfqpoint{0.720298in}{2.673413in}}%
\pgfpathlineto{\pgfqpoint{0.803510in}{2.761517in}}%
\pgfpathlineto{\pgfqpoint{0.887990in}{2.770909in}}%
\pgfpathlineto{\pgfqpoint{1.141430in}{2.767746in}}%
\pgfpathlineto{\pgfqpoint{1.563830in}{2.767278in}}%
\pgfpathlineto{\pgfqpoint{2.716000in}{2.767274in}}%
\pgfpathlineto{\pgfqpoint{2.716000in}{2.767274in}}%
\pgfusepath{stroke}%
\end{pgfscope}%
\begin{pgfscope}%
\pgfsetrectcap%
\pgfsetmiterjoin%
\pgfsetlinewidth{0.803000pt}%
\definecolor{currentstroke}{rgb}{0.000000,0.000000,0.000000}%
\pgfsetstrokecolor{currentstroke}%
\pgfsetdash{}{0pt}%
\pgfpathmoveto{\pgfqpoint{0.594000in}{0.480000in}}%
\pgfpathlineto{\pgfqpoint{0.594000in}{2.880000in}}%
\pgfusepath{stroke}%
\end{pgfscope}%
\begin{pgfscope}%
\pgfsetrectcap%
\pgfsetmiterjoin%
\pgfsetlinewidth{0.803000pt}%
\definecolor{currentstroke}{rgb}{0.000000,0.000000,0.000000}%
\pgfsetstrokecolor{currentstroke}%
\pgfsetdash{}{0pt}%
\pgfpathmoveto{\pgfqpoint{2.706000in}{0.480000in}}%
\pgfpathlineto{\pgfqpoint{2.706000in}{2.880000in}}%
\pgfusepath{stroke}%
\end{pgfscope}%
\begin{pgfscope}%
\pgfsetrectcap%
\pgfsetmiterjoin%
\pgfsetlinewidth{0.803000pt}%
\definecolor{currentstroke}{rgb}{0.000000,0.000000,0.000000}%
\pgfsetstrokecolor{currentstroke}%
\pgfsetdash{}{0pt}%
\pgfpathmoveto{\pgfqpoint{0.594000in}{0.480000in}}%
\pgfpathlineto{\pgfqpoint{2.706000in}{0.480000in}}%
\pgfusepath{stroke}%
\end{pgfscope}%
\begin{pgfscope}%
\pgfsetrectcap%
\pgfsetmiterjoin%
\pgfsetlinewidth{0.803000pt}%
\definecolor{currentstroke}{rgb}{0.000000,0.000000,0.000000}%
\pgfsetstrokecolor{currentstroke}%
\pgfsetdash{}{0pt}%
\pgfpathmoveto{\pgfqpoint{0.594000in}{2.880000in}}%
\pgfpathlineto{\pgfqpoint{2.706000in}{2.880000in}}%
\pgfusepath{stroke}%
\end{pgfscope}%
\begin{pgfscope}%
\pgfsetbuttcap%
\pgfsetmiterjoin%
\definecolor{currentfill}{rgb}{1.000000,1.000000,1.000000}%
\pgfsetfillcolor{currentfill}%
\pgfsetfillopacity{0.800000}%
\pgfsetlinewidth{1.003750pt}%
\definecolor{currentstroke}{rgb}{0.800000,0.800000,0.800000}%
\pgfsetstrokecolor{currentstroke}%
\pgfsetstrokeopacity{0.800000}%
\pgfsetdash{}{0pt}%
\pgfpathmoveto{\pgfqpoint{1.333630in}{0.542500in}}%
\pgfpathlineto{\pgfqpoint{2.618500in}{0.542500in}}%
\pgfpathquadraticcurveto{\pgfqpoint{2.643500in}{0.542500in}}{\pgfqpoint{2.643500in}{0.567500in}}%
\pgfpathlineto{\pgfqpoint{2.643500in}{0.903611in}}%
\pgfpathquadraticcurveto{\pgfqpoint{2.643500in}{0.928611in}}{\pgfqpoint{2.618500in}{0.928611in}}%
\pgfpathlineto{\pgfqpoint{1.333630in}{0.928611in}}%
\pgfpathquadraticcurveto{\pgfqpoint{1.308630in}{0.928611in}}{\pgfqpoint{1.308630in}{0.903611in}}%
\pgfpathlineto{\pgfqpoint{1.308630in}{0.567500in}}%
\pgfpathquadraticcurveto{\pgfqpoint{1.308630in}{0.542500in}}{\pgfqpoint{1.333630in}{0.542500in}}%
\pgfpathlineto{\pgfqpoint{1.333630in}{0.542500in}}%
\pgfpathclose%
\pgfusepath{stroke,fill}%
\end{pgfscope}%
\begin{pgfscope}%
\pgfsetrectcap%
\pgfsetroundjoin%
\pgfsetlinewidth{1.505625pt}%
\definecolor{currentstroke}{rgb}{0.000000,0.000000,0.000000}%
\pgfsetstrokecolor{currentstroke}%
\pgfsetdash{}{0pt}%
\pgfpathmoveto{\pgfqpoint{1.358630in}{0.834861in}}%
\pgfpathlineto{\pgfqpoint{1.483630in}{0.834861in}}%
\pgfpathlineto{\pgfqpoint{1.608630in}{0.834861in}}%
\pgfusepath{stroke}%
\end{pgfscope}%
\begin{pgfscope}%
\definecolor{textcolor}{rgb}{0.000000,0.000000,0.000000}%
\pgfsetstrokecolor{textcolor}%
\pgfsetfillcolor{textcolor}%
\pgftext[x=1.708630in,y=0.791111in,left,base]{\color{textcolor}{\rmfamily\fontsize{9.000000}{10.800000}\selectfont\catcode`\^=\active\def^{\ifmmode\sp\else\^{}\fi}\catcode`\%=\active\def
\end{pgfscope}%
\begin{pgfscope}%
\pgfsetrectcap%
\pgfsetroundjoin%
\pgfsetlinewidth{1.505625pt}%
\definecolor{currentstroke}{rgb}{0.000000,0.000000,1.000000}%
\pgfsetstrokecolor{currentstroke}%
\pgfsetdash{}{0pt}%
\pgfpathmoveto{\pgfqpoint{1.358630in}{0.660556in}}%
\pgfpathlineto{\pgfqpoint{1.483630in}{0.660556in}}%
\pgfpathlineto{\pgfqpoint{1.608630in}{0.660556in}}%
\pgfusepath{stroke}%
\end{pgfscope}%
\begin{pgfscope}%
\definecolor{textcolor}{rgb}{0.000000,0.000000,0.000000}%
\pgfsetstrokecolor{textcolor}%
\pgfsetfillcolor{textcolor}%
\pgftext[x=1.708630in,y=0.616806in,left,base]{\color{textcolor}{\rmfamily\fontsize{9.000000}{10.800000}\selectfont\catcode`\^=\active\def^{\ifmmode\sp\else\^{}\fi}\catcode`\%=\active\def
\end{pgfscope}%
\end{pgfpicture}%
\makeatother%
\endgroup%

%% file: figures/1D2solids.tex
\tikzset{every picture/.style={line width=0.75pt}} 

\begin{tikzpicture}[x=0.75pt,y=0.75pt,yscale=-1,xscale=1]

\draw  [fill={rgb, 255:red, 164; green, 196; blue, 255 }  ,fill opacity=1 ] (204,96) -- (360,96) -- (360,156) -- (204,156) -- cycle ;
\draw [color={rgb, 255:red, 255; green, 0; blue, 30 }  ,draw opacity=1 ]   (556,126.67) -- (519,126.67) ;
\draw [shift={(516,126.67)}, rotate = 360] [fill={rgb, 255:red, 255; green, 0; blue, 30 }  ,fill opacity=1 ][line width=0.08]  [draw opacity=0] (8.93,-4.29) -- (0,0) -- (8.93,4.29) -- cycle    ;
\draw [color={rgb, 255:red, 255; green, 0; blue, 29 }  ,draw opacity=1 ]   (155.58,122.75) .. controls (158.38,106.92) and (160.49,114.54) .. (162.59,122.45) .. controls (164.75,130.56) and (166.89,138.97) .. (169.77,122.75) ;
\draw [color={rgb, 255:red, 255; green, 0; blue, 29 }  ,draw opacity=1 ]   (183.96,122.75) .. controls (188.16,102.3) and (187.33,123.35) .. (201.17,123.29) ;
\draw [shift={(204,123)}, rotate = 182.56] [fill={rgb, 255:red, 255; green, 0; blue, 29 }  ,fill opacity=1 ][line width=0.08]  [draw opacity=0] (8.93,-4.29) -- (0,0) -- (8.93,4.29) -- cycle    ;
\draw [color={rgb, 255:red, 255; green, 0; blue, 29 }  ,draw opacity=1 ]   (169.77,122.75) .. controls (175.44,90.69) and (178.28,154.8) .. (183.96,122.75) ;
\draw [color={rgb, 255:red, 255; green, 0; blue, 29 }  ,draw opacity=1 ]   (141.39,122.75) .. controls (144.19,106.92) and (146.3,114.54) .. (148.4,122.45) .. controls (150.56,130.56) and (152.7,138.97) .. (155.58,122.75) ;
\draw [color={rgb, 255:red, 255; green, 0; blue, 29 }  ,draw opacity=1 ]   (127.2,122.75) .. controls (130,106.92) and (132.11,114.54) .. (134.22,122.45) .. controls (136.37,130.56) and (138.52,138.97) .. (141.39,122.75) ;
\draw  [fill={rgb, 255:red, 186; green, 170; blue, 166 }  ,fill opacity=1 ] (360,96) -- (516,96) -- (516,156) -- (360,156) -- cycle ;
\draw [color={rgb, 255:red, 255; green, 0; blue, 0 }  ,draw opacity=1 ]   (379,118.33) -- (342,118.33) ;
\draw [shift={(339,118.33)}, rotate = 360] [fill={rgb, 255:red, 255; green, 0; blue, 0 }  ,fill opacity=1 ][line width=0.08]  [draw opacity=0] (8.93,-4.29) -- (0,0) -- (8.93,4.29) -- cycle    ;
\draw [color={rgb, 255:red, 255; green, 0; blue, 0 }  ,draw opacity=1 ]   (377.67,128) -- (340.67,128) ;
\draw [shift={(380.67,128)}, rotate = 180] [fill={rgb, 255:red, 255; green, 0; blue, 0 }  ,fill opacity=1 ][line width=0.08]  [draw opacity=0] (8.93,-4.29) -- (0,0) -- (8.93,4.29) -- cycle    ;
\draw    (204,168) -- (240,168) -- (360,168) ;
\draw [shift={(204,168)}, rotate = 180] [color={rgb, 255:red, 0; green, 0; blue, 0 }  ][line width=0.75]    (0,5.59) -- (0,-5.59)   ;
\draw    (360,168) -- (396,168) -- (516,168) ;
\draw [shift={(516,168)}, rotate = 180] [color={rgb, 255:red, 0; green, 0; blue, 0 }  ][line width=0.75]    (0,5.59) -- (0,-5.59)   ;
\draw [shift={(360,168)}, rotate = 180] [color={rgb, 255:red, 0; green, 0; blue, 0 }  ][line width=0.75]    (0,5.59) -- (0,-5.59)   ;
\draw    (360,84) -- (394,84) ;
\draw [shift={(396,84)}, rotate = 180] [color={rgb, 255:red, 0; green, 0; blue, 0 }  ][line width=0.75]    (10.93,-3.29) .. controls (6.95,-1.4) and (3.31,-0.3) .. (0,0) .. controls (3.31,0.3) and (6.95,1.4) .. (10.93,3.29)   ;
\draw [shift={(360,84)}, rotate = 180] [color={rgb, 255:red, 0; green, 0; blue, 0 }  ][line width=0.75]    (0,5.59) -- (0,-5.59)   ;

\draw (127.67,88.4) node [anchor=north west][inner sep=0.75pt]  [color={rgb, 255:red, 249; green, 255; blue, 0 }  ,opacity=1 ] [align=left] {\textcolor[rgb]{0.98,0,0.11}{q}\textcolor[rgb]{0.98,0,0.11}{$\displaystyle ''$}\textcolor[rgb]{0.98,0,0.11}{= F}\textcolor[rgb]{0.98,0,0.11}{$\displaystyle _{\text{ext}}$}};
\draw (525.67,102.33) node [anchor=north west][inner sep=0.75pt]   [align=left] {\textcolor[rgb]{1,0,0.13}{q}\textcolor[rgb]{1,0,0.13}{$\displaystyle ''$}\textcolor[rgb]{1,0,0.13}{ = 0}};
\draw (250,114.67) node [anchor=north west][inner sep=0.75pt]   [align=left] {$\displaystyle [ \kappa ,\rho c_{p}]_{f}$};
\draw (409,114.67) node [anchor=north west][inner sep=0.75pt]   [align=left] {$\displaystyle [ \kappa ,\rho c_{p}]_{s}$};
\draw (262,182) node [anchor=north west][inner sep=0.75pt]   [align=left] {0.1*L};
\draw (433,182) node [anchor=north west][inner sep=0.75pt]   [align=left] {L};
\draw (385,62) node [anchor=north west][inner sep=0.75pt]   [align=left] {x};

\end{tikzpicture}

%% file: figures/new_thin_layer.tex
 
\tikzset{
pattern size/.store in=\mcSize, 
pattern size = 5pt,
pattern thickness/.store in=\mcThickness, 
pattern thickness = 0.3pt,
pattern radius/.store in=\mcRadius, 
pattern radius = 1pt}
\makeatletter
\pgfutil@ifundefined{pgf@pattern@name@_sz9wqpgxc}{
\pgfdeclarepatternformonly[\mcThickness,\mcSize]{_sz9wqpgxc}
{\pgfqpoint{0pt}{0pt}}
{\pgfpoint{\mcSize+\mcThickness}{\mcSize+\mcThickness}}
{\pgfpoint{\mcSize}{\mcSize}}
{
\pgfsetcolor{\tikz@pattern@color}
\pgfsetlinewidth{\mcThickness}
\pgfpathmoveto{\pgfqpoint{0pt}{0pt}}
\pgfpathlineto{\pgfpoint{\mcSize+\mcThickness}{\mcSize+\mcThickness}}
\pgfusepath{stroke}
}}
\makeatother
\tikzset{every picture/.style={line width=0.75pt}} 

\begin{tikzpicture}[x=0.75pt,y=0.75pt,yscale=-1,xscale=1]
\hspace*{0.9cm}
\draw  [draw opacity=0][pattern=_sz9wqpgxc,pattern size=22.5pt,pattern thickness=5.25pt,pattern radius=0pt, pattern color={rgb, 255:red, 230; green, 240; blue, 218}] (183.68,194.52) -- (258.89,211.13) -- (367.53,194.52) -- (534.67,144.7) -- (546.35,181.08) -- (378.23,235.04) -- (259.06,252.15) -- (170.81,237.2) -- (150.48,226.92) -- (63.2,182.78) -- (78.87,134.11) -- cycle ;
\draw  [draw opacity=0][fill={rgb, 255:red, 139; green, 87; blue, 42 }  ,fill opacity=0.52 ] (258.89,211.13) -- (213.39,201.08) -- (209.72,200.27) -- (183.68,194.52) -- (130.09,164.94) -- (108.47,153) -- (58.33,119.79) -- (58.33,177.69) -- (58.33,179.69) -- (58.33,186.69) -- (58.33,352.29) -- (546.21,353.45) -- (568.1,128.09) -- (509.6,153) -- (444.98,171.89) -- (414.92,180.68) -- (410.18,182.06) -- (375.74,192.13) -- (367.53,194.52) -- (308.29,203.58) -- (300.95,204.7) -- (273.87,208.84) -- cycle ;
\draw    (108.47,153) -- (183.68,194.52) ;
\draw    (183.68,194.52) -- (258.89,211.13) ;
\draw    (258.89,211.13) -- (367.53,194.52) ;
\draw    (367.53,194.52) -- (509.6,153) ;
\draw  [dash pattern={on 0.84pt off 2.51pt}]  (108.47,153) -- (58.33,119.79) ;
\draw  [dash pattern={on 0.84pt off 2.51pt}]  (568.1,128.09) -- (509.6,153) ;
\draw  [color={rgb, 255:red, 208; green, 2; blue, 27 }  ,draw opacity=1 ][fill={rgb, 255:red, 12; green, 100; blue, 255 }  ,fill opacity=0.35 ][line width=1.5]  (258.89,211.13) -- (183.68,194.52) -- (108.47,153) -- (58.33,119.79) -- (70.19,78.44) -- (159.11,59.84) -- (219.28,58.51) -- (564.93,71.13) -- (568.1,128.09) -- (509.6,153) -- (367.53,194.52) -- cycle ;
\draw [color={rgb, 255:red, 0; green, 0; blue, 255 }  ,draw opacity=1 ]   (108.47,153) -- (108.47,111.49) ;
\draw [color={rgb, 255:red, 0; green, 0; blue, 255 }  ,draw opacity=1 ]   (183.68,194.52) -- (183.68,136.4) ;
\draw [color={rgb, 255:red, 0; green, 0; blue, 255 }  ,draw opacity=1 ]   (258.89,211.13) -- (258.89,153) ;
\draw [color={rgb, 255:red, 0; green, 0; blue, 255 }  ,draw opacity=1 ]   (367.53,194.52) -- (367.53,144.7) ;
\draw [color={rgb, 255:red, 0; green, 0; blue, 255 }  ,draw opacity=1 ]   (509.6,153) -- (509.6,103.18) ;
\draw [color={rgb, 255:red, 0; green, 0; blue, 255 }  ,draw opacity=1 ]   (108.47,144.7) -- (183.68,186.22) ;
\draw [color={rgb, 255:red, 0; green, 0; blue, 255 }  ,draw opacity=1 ]   (183.68,186.22) -- (258.89,202.83) ;
\draw [color={rgb, 255:red, 0; green, 0; blue, 255 }  ,draw opacity=1 ]   (258.89,202.83) -- (367.53,186.22) ;
\draw [color={rgb, 255:red, 0; green, 0; blue, 255 }  ,draw opacity=1 ]   (367.53,186.22) -- (509.6,144.7) ;
\draw [color={rgb, 255:red, 0; green, 0; blue, 255 }  ,draw opacity=1 ]   (108.47,128.09) -- (183.68,169.61) ;
\draw [color={rgb, 255:red, 0; green, 0; blue, 255 }  ,draw opacity=1 ]   (183.68,169.61) -- (258.89,186.22) ;
\draw [color={rgb, 255:red, 0; green, 0; blue, 255 }  ,draw opacity=1 ]   (258.89,186.22) -- (367.53,169.61) ;
\draw [color={rgb, 255:red, 0; green, 0; blue, 255 }  ,draw opacity=1 ]   (367.53,169.61) -- (509.6,128.09) ;
\draw [color={rgb, 255:red, 0; green, 0; blue, 255 }  ,draw opacity=1 ] [dash pattern={on 0.84pt off 2.51pt}]  (108.47,111.49) -- (108.47,94.88) ;
\draw [color={rgb, 255:red, 0; green, 0; blue, 255 }  ,draw opacity=1 ] [dash pattern={on 0.84pt off 2.51pt}]  (183.68,136.4) -- (183.68,107.33) ;
\draw [color={rgb, 255:red, 0; green, 0; blue, 255 }  ,draw opacity=1 ] [dash pattern={on 0.84pt off 2.51pt}]  (258.89,153) -- (258.89,123.94) ;
\draw [color={rgb, 255:red, 0; green, 0; blue, 255 }  ,draw opacity=1 ] [dash pattern={on 0.84pt off 2.51pt}]  (367.53,144.7) -- (367.53,115.64) ;
\draw [color={rgb, 255:red, 0; green, 0; blue, 255 }  ,draw opacity=1 ] [dash pattern={on 0.84pt off 2.51pt}]  (509.6,103.18) -- (509.6,74.12) ;
\draw [color={rgb, 255:red, 128; green, 128; blue, 128 }  ,draw opacity=1 ]   (178.5,215.61) -- (183.68,194.52) ;
\draw [color={rgb, 255:red, 128; green, 128; blue, 128 }  ,draw opacity=1 ]   (96.6,175.09) -- (108.47,153) ;
\draw [color={rgb, 255:red, 128; green, 128; blue, 128 }  ,draw opacity=1 ]   (259.06,227.9) -- (258.89,211.13) ;
\draw [color={rgb, 255:red, 128; green, 128; blue, 128 }  ,draw opacity=1 ]   (373.05,216.61) -- (367.53,194.52) ;
\draw [color={rgb, 255:red, 128; green, 128; blue, 128 }  ,draw opacity=1 ]   (106.46,156.99) -- (182.85,198.67) ;
\draw [color={rgb, 255:red, 128; green, 128; blue, 128 }  ,draw opacity=1 ]   (102.45,165.29) -- (179.84,206.98) ;
\draw [color={rgb, 255:red, 128; green, 128; blue, 128 }  ,draw opacity=1 ]   (182.85,198.67) -- (258.73,214.95) ;
\draw [color={rgb, 255:red, 128; green, 128; blue, 128 }  ,draw opacity=1 ]   (179.84,206.98) -- (259.4,223.92) ;
\draw [color={rgb, 255:red, 128; green, 128; blue, 128 }  ,draw opacity=1 ]   (258.73,214.95) -- (368.7,197.68) ;
\draw [color={rgb, 255:red, 128; green, 128; blue, 128 }  ,draw opacity=1 ]   (259.4,223.92) -- (371.04,205.65) ;
\draw [color={rgb, 255:red, 128; green, 128; blue, 128 }  ,draw opacity=1 ]   (518.79,171.77) -- (509.6,153) ;
\draw [color={rgb, 255:red, 128; green, 128; blue, 128 }  ,draw opacity=1 ]   (368.7,197.68) -- (510.77,156.16) ;
\draw [color={rgb, 255:red, 128; green, 128; blue, 128 }  ,draw opacity=1 ]   (371.04,205.65) -- (514.78,163.14) ;
\draw [color={rgb, 255:red, 128; green, 128; blue, 128 }  ,draw opacity=1 ] [dash pattern={on 0.84pt off 2.51pt}]  (170.81,237.2) -- (178.5,215.61) ;
\draw [color={rgb, 255:red, 128; green, 128; blue, 128 }  ,draw opacity=1 ] [dash pattern={on 0.84pt off 2.51pt}]  (85.91,194.69) -- (96.6,175.09) ;
\draw [color={rgb, 255:red, 128; green, 128; blue, 128 }  ,draw opacity=1 ] [dash pattern={on 0.84pt off 2.51pt}]  (259.06,252.15) -- (259.06,227.9) ;
\draw [color={rgb, 255:red, 128; green, 128; blue, 128 }  ,draw opacity=1 ] [dash pattern={on 0.84pt off 2.51pt}]  (378.23,235.04) -- (373.05,216.61) ;
\draw [color={rgb, 255:red, 128; green, 128; blue, 128 }  ,draw opacity=1 ] [dash pattern={on 0.84pt off 2.51pt}]  (525.48,189.37) -- (518.79,171.77) ;
\draw    (108.47,153) -- (75.04,236.04) ;
\draw    (183.68,194.52) -- (75.04,236.04) ;
\draw    (91.75,302.47) -- (75.04,236.04) ;
\draw    (183.68,260.95) -- (183.68,224.78) -- (183.68,194.52) ;
\draw    (183.68,260.95) -- (258.89,211.13) ;
\draw    (258.89,211.13) -- (309.04,260.95) ;
\draw    (367.53,194.52) -- (309.04,260.95) ;
\draw    (367.53,194.52) -- (409.32,236.04) ;
\draw    (367.53,194.52) -- (509.6,211.13) ;
\draw    (509.6,153) -- (509.6,211.13) ;
\draw    (409.32,236.04) -- (509.6,211.13) ;
\draw    (509.6,211.13) -- (517.96,269.26) ;
\draw    (409.32,236.04) -- (409.32,294.17) ;
\draw    (509.6,211.13) -- (409.32,294.17) ;
\draw    (517.96,269.26) -- (409.32,294.17) ;
\draw    (183.68,260.95) -- (75.04,236.04) ;
\draw    (91.75,302.47) -- (183.68,260.95) ;
\draw    (91.75,302.47) -- (166.97,327.38) ;
\draw    (183.68,260.95) -- (166.97,327.38) ;
\draw    (309.04,260.95) -- (166.97,327.38) ;
\draw    (309.04,260.95) -- (183.68,260.95) ;
\draw    (292.32,327.38) -- (166.97,327.38) ;
\draw    (309.04,260.95) -- (292.32,327.38) ;
\draw    (309.04,260.95) -- (409.32,294.17) ;
\draw    (409.32,294.17) -- (292.32,327.38) ;
\draw    (309.04,260.95) -- (409.32,236.04) ;
\draw  [dash pattern={on 0.84pt off 2.51pt}]  (292.32,327.38) -- (267.25,360.6) ;
\draw  [dash pattern={on 0.84pt off 2.51pt}]  (409.32,294.17) -- (392.6,343.99) ;
\draw  [dash pattern={on 0.84pt off 2.51pt}]  (409.32,294.17) -- (517.96,327.38) ;
\draw  [dash pattern={on 0.84pt off 2.51pt}]  (517.96,269.26) -- (517.96,327.38) ;
\draw  [dash pattern={on 0.84pt off 2.51pt}]  (356.34,356.78) -- (292.32,327.38) ;
\draw  [dash pattern={on 0.84pt off 2.51pt}]  (166.97,327.38) -- (183.68,352.29) ;
\draw  [dash pattern={on 0.84pt off 2.51pt}]  (91.75,302.47) -- (91.75,335.69) ;
\draw  [dash pattern={on 0.84pt off 2.51pt}]  (166.97,327.38) -- (100.11,360.6) ;
\draw  [draw opacity=0][fill={rgb, 255:red, 255; green, 255; blue, 255 }  ,fill opacity=1 ] (91.75,335.69) .. controls (111.39,345.23) and (131.03,352.71) .. (166.97,352.29) .. controls (202.9,351.88) and (504.09,366.74) .. (543.03,302.47) .. controls (581.97,238.2) and (531.75,150.1) .. (534.67,144.7) .. controls (537.6,139.3) and (578.96,122.7) .. (576.46,136.4) .. controls (573.95,150.1) and (689.44,359.93) .. (516.12,360.1) .. controls (342.8,360.26) and (187.86,372.91) .. (155.69,374.76) .. controls (123.51,376.61) and (82.56,376.5) .. (58.33,368.9) .. controls (34.09,361.3) and (42.45,308.28) .. (41.61,211.13) .. controls (40.78,113.98) and (49.22,144.87) .. (55.49,114.97) .. controls (61.75,85.08) and (77.13,129.34) .. (79.22,132.25) .. controls (81.31,135.15) and (57.07,180.82) .. (58.33,236.04) .. controls (59.58,291.26) and (72.12,326.14) .. (91.75,335.69) -- cycle ;
\draw  [color={rgb, 255:red, 255; green, 255; blue, 255 }  ,draw opacity=1 ][fill={rgb, 255:red, 255; green, 255; blue, 255 }  ,fill opacity=1 ] (576.46,119.79) .. controls (584.81,115.64) and (535.55,166.81) .. (534.67,144.7) .. controls (533.79,122.6) and (539.5,85.34) .. (509.6,78.27) .. controls (479.7,71.2) and (374.14,64.82) .. (303.52,63.16) .. controls (232.9,61.5) and (212.01,59.75) .. (168.14,63.49) .. controls (124.26,67.23) and (96.1,80.26) .. (88.25,97.04) .. controls (80.39,113.81) and (78.22,122.61) .. (79.22,132.25) .. controls (80.22,141.88) and (45.62,118.46) .. (60.17,93.05) .. controls (74.71,67.64) and (88.91,52.53) .. (112.98,53.86) .. controls (137.05,55.19) and (486.7,62.99) .. (553.56,67.15) .. controls (620.42,71.3) and (568.1,123.94) .. (576.46,119.79) -- cycle ;
\draw  [color={rgb, 255:red, 0; green, 0; blue, 255 }  ,draw opacity=1 ][fill={rgb, 255:red, 0; green, 0; blue, 255 }  ,fill opacity=1 ] (313.42,198.14) .. controls (313.42,197.15) and (314.11,196.35) .. (314.96,196.35) .. controls (315.81,196.35) and (316.5,197.15) .. (316.5,198.14) .. controls (316.5,199.12) and (315.81,199.92) .. (314.96,199.92) .. controls (314.11,199.92) and (313.42,199.12) .. (313.42,198.14) -- cycle ;
\draw  [color={rgb, 255:red, 0; green, 0; blue, 0 }  ,draw opacity=1 ][fill={rgb, 255:red, 0; green, 0; blue, 0 }  ,fill opacity=1 ] (181.83,260.95) .. controls (181.83,260.13) and (182.66,259.47) .. (183.68,259.47) .. controls (184.7,259.47) and (185.53,260.13) .. (185.53,260.95) .. controls (185.53,261.77) and (184.7,262.44) .. (183.68,262.44) .. controls (182.66,262.44) and (181.83,261.77) .. (181.83,260.95) -- cycle ;
\draw [color={rgb, 255:red, 65; green, 117; blue, 5 }  ,draw opacity=1 ][line width=1.5]    (85.91,194.69) -- (170.81,237.2) ;
\draw [color={rgb, 255:red, 65; green, 117; blue, 5 }  ,draw opacity=1 ][line width=1.5]    (170.81,237.2) -- (259.06,252.15) ;
\draw [color={rgb, 255:red, 65; green, 117; blue, 5 }  ,draw opacity=1 ][line width=1.5]    (259.06,252.15) -- (378.23,235.04) ;
\draw [color={rgb, 255:red, 65; green, 117; blue, 5 }  ,draw opacity=1 ][line width=1.5]    (378.23,235.04) -- (525.48,189.37) ;
\draw [color={rgb, 255:red, 65; green, 117; blue, 5 }  ,draw opacity=1 ][line width=1.5]  [dash pattern={on 1.69pt off 2.76pt}]  (548.47,182.45) -- (525.48,189.37) ;
\draw [color={rgb, 255:red, 65; green, 117; blue, 5 }  ,draw opacity=1 ][line width=1.5]  [dash pattern={on 1.69pt off 2.76pt}]  (85.91,194.69) -- (61.5,181.78) ;

\draw  [color={rgb, 255:red, 87; green, 87; blue, 87 }  ,draw opacity=1 ][fill={rgb, 255:red, 94; green, 92; blue, 92 }  ,fill opacity=1 ] (257.23,223.92) .. controls (257.23,222.91) and (258.2,222.08) .. (259.4,222.08) .. controls (260.59,222.08) and (261.56,222.91) .. (261.56,223.92) .. controls (261.56,224.93) and (260.59,225.75) .. (259.4,225.75) .. controls (258.2,225.75) and (257.23,224.93) .. (257.23,223.92) -- cycle ;
\draw [color={rgb, 255:red, 0; green, 0; blue, 255 }  ,draw opacity=1 ]   (108.47,103.18) -- (183.68,144.7) ;
\draw [color={rgb, 255:red, 0; green, 0; blue, 255 }  ,draw opacity=1 ]   (183.68,144.7) -- (258.89,161.31) ;
\draw [color={rgb, 255:red, 0; green, 0; blue, 255 }  ,draw opacity=1 ]   (258.89,161.31) -- (367.53,144.7) ;
\draw [color={rgb, 255:red, 0; green, 0; blue, 255 }  ,draw opacity=1 ]   (367.53,144.7) -- (509.6,103.18) ;
\draw [color={rgb, 255:red, 128; green, 128; blue, 128 }  ,draw opacity=1 ]   (96.6,175.09) -- (177.38,217.78) ;
\draw [color={rgb, 255:red, 128; green, 128; blue, 128 }  ,draw opacity=1 ]   (177.38,217.78) -- (258.72,234.11) ;
\draw [color={rgb, 255:red, 128; green, 128; blue, 128 }  ,draw opacity=1 ]   (258.72,234.11) -- (373.05,216.61) ;
\draw [color={rgb, 255:red, 128; green, 128; blue, 128 }  ,draw opacity=1 ]   (373.05,216.61) -- (518.79,171.77) ;
\draw  [dash pattern={on 0.84pt off 2.51pt}]  (75.04,236.04) -- (58.84,242.78) ;
\draw  [dash pattern={on 0.84pt off 2.51pt}]  (557.02,238.78) -- (509.6,211.13) ;

\draw (187.53,263.95) node [anchor=north west][inner sep=0.75pt]  [font=\large] [align=left] {$\displaystyle T_{MFE}$};
\draw (302.22,175.1) node [anchor=north west][inner sep=0.75pt]  [font=\large,color={rgb, 255:red, 0; green, 0; blue, 255 }  ,opacity=1 ] [align=left] {$\displaystyle T_{f}$};
\draw (419.26,313.44) node [anchor=north west][inner sep=0.75pt]  [font=\LARGE,color={rgb, 255:red, 112; green, 63; blue, 23 }  ,opacity=1 ] [align=left] {$\displaystyle \Omega _{s}$};
\draw (409.08,78.09) node [anchor=north west][inner sep=0.75pt]  [font=\LARGE,color={rgb, 255:red, 0; green, 0; blue, 255 }  ,opacity=1 ] [align=left] {$\displaystyle \Omega _{f}$};
\draw (63.44,108.23) node [anchor=north west][inner sep=0.75pt]  [font=\LARGE,color={rgb, 255:red, 208; green, 2; blue, 27 }  ,opacity=1 ] [align=left] {$\displaystyle \Gamma $};
\draw (235.87,226.72) node [anchor=north west][inner sep=0.75pt]  [font=\large,color={rgb, 255:red, 92; green, 91; blue, 91 }  ,opacity=1 ] [align=left] {$\displaystyle T_{FE}$};
\draw (39.77,166.16) node [anchor=north west][inner sep=0.75pt]  [font=\LARGE,color={rgb, 255:red, 65; green, 117; blue, 5 }  ,opacity=1 ] [align=left] {$\displaystyle \Xi $};
\draw (522.09,150.94) node [anchor=north west][inner sep=0.75pt]  [font=\LARGE,color={rgb, 255:red, 115; green, 149; blue, 85 }  ,opacity=1 ] [align=left] {$\displaystyle \Omega _{s,FE}$};

\end{tikzpicture}

%% file: timeintegration.tex
\subsection{Time integration}
Advancement in time for the solid (\textit{FE} and \textit{MFE} solution) and fluid equations is carried out using a formally second-order accurate semi-implicit scheme, with implicit iterations based on a dual-time stepping formulation, originally proposed by \citet{jameson2009assessment}. Convergence in pseudo-time is further accelerated using techniques such as  local time stepping, residual smoothing, low-Mach number preconditioning. More details on the acceleration techniques implemented can be found in \cite{JoshThesis}.

On top of the intrinsic advantages of a dual-time  scheme such as improved stability or increased efficiency, directly coming from the use of implicit time integration schemes while maintaining stability and convergence properties typically associated with explicit schemes.  The dual-time framework is advantageous for unsteady problems with multiple time scales, such as CHT problems. It allows for the separation of these scales, enabling each domain to be resolved appropriately without the need for excessively small physical time steps.
The physical time-step selection can be based on physical considerations alone, regardless of numerical stability considerations, as numerical stability is  managed by the pseudo-time integration process.

With this scheme, both domains are coupled within each inner iteration and the criterion for both system to have converged to the desired tolerance is enforced. This allows to reduce greatly the temperature discontinuity coming from the solid-fluid coupling, \cite{Giles}.

\subsection{Numerical implementation of the flow solver}
The flow equations in \autoref{eq:system_coupled} are solved using the code H4X \citep{HOPECOLLINS2023111858,10.1063/5.0136568,HAO2023124324,YANN}. It is a cell-centered
finite volume code based on a multiblock grid arrangement. The flow field is represented by the viscous variables: velocity, temperature and pressure and the equations of motion for a compressible fluid are solved in conservative form. The spatial discretisation is third-order accurate in space for the inviscid fluxes. Third-order accuracy is achieved on a compact stencil by using variable extrapolation. No limiter is applied to the vorticity and entropy fields. The extrapolation is based on weighted least-square gradients. The gradient stencil contains all the face neighbors of each cell.
For the purpose of variable extrapolation onto a cell interface, the gradient stencil is biased by removing the contributions from the neighbor on the other side of the interface. The numerical fluxes are adapted to low Mach numbers and a modified pressure flux is employed, \citep{HOPECOLLINS2023111858}. The viscous fluxes are evaluated using a second-order discretisation. The code is parallelised by partitioning the blocks of the multiblock grid among the available MPI ranks. Within each rank, block operations are parallelised using OpenMP. Computations and communications are overlaid to hide the latency of the network fabric.

%% file: results.tex
\begin{figure*}
    \centering
    \begin{minipage}[b]{0.58\textwidth}
        \centering
        \scalebox{0.9}{\input{figures/error_vs_thickness.pgf}}
        \caption{Relative error with the thin layer extend and the modal truncation level for the problem in \autoref{sec:1D}. The thin layer extend $\delta_{\text{FE}}$ is normalised by the characteristic penetration depth $\delta_P$}
        \label{fig:error_thin_layer}
    \end{minipage}
    \hfill
    \begin{minipage}[b]{0.39\textwidth}
        \centering
        \includegraphics[width=7cm]{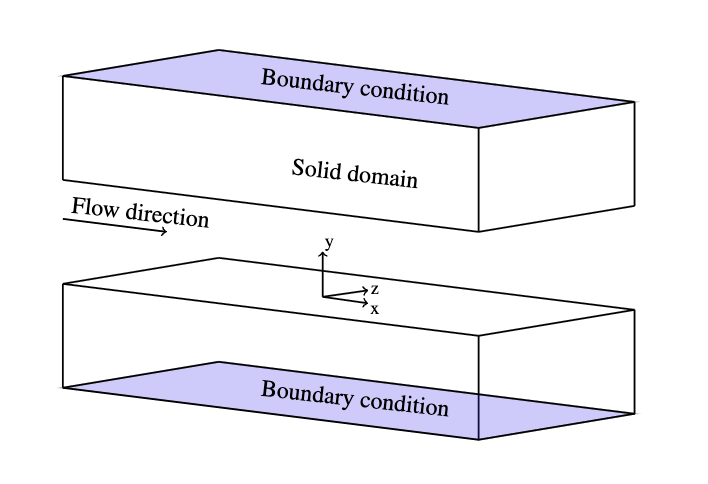}
        \vspace*{1cm}
        \caption{Turbulent channel flow case and coordinate system\\\quad }
        \label{fig:channel}
    \end{minipage}
\end{figure*}
In this section, the  unsteady CHT method described is applied to two three-dimensional test cases.
Both test cases are first validated with reference data and then the acceleration technique is anlysed. 

\subsection{Conjugate channel flow case} 
The first test case considered is a fully developed turbulent channel flow with conducting walls, illustrated in \autoref{fig:channel}. With turbulent channels being one of the most fundamental test cases in anisotropic turbulence, a great deal of knowledge has been gained on the flow dynamics over a large range of regimes. 
The numerical solver H4X has been validated for channel flows under different flow regimes in terms of mean and fluctuating velocity quantities, as detailed in \cite{10.1063/5.0136568}.
 High-resolution conjugate heat transfer simulations in channel flows have been conducted by \citet{flageul2015dns, 10.1115/1.1389060} and \citet{tiselj2012dns}, among others. This test case is used to validate the  different solution approaches for the solid domain—including a directly coupled finite element method, a coupled modal-finite element, and an accelerated procedure—against existing reference data.

\subsubsection{Numerical setup}
A channel flow is  characterised by the channel height, $2\delta$ as well as the streamwise and spanwise domain lengths, $L_x$ and $L_z$. 
The flow is simulated between two viscous walls, with periodic boundaries applied in the streamwise and spanwise directions. 
To maintain a constant mass flow in the channel, an artificial body force is introduced. 
The solid domains have a thickness of $\delta$ and with Neumann thermal boundary condition  applied at the far end to balance the energy dissipated by the fluid.
The Reynolds number $\text{Re}_\tau$, based on the friction velocity and the channel half-width investigated, is $ 150 $ at a Mach number  \textit{M} $= 0.1$. The Prandtl number is 0.71. The fluid-solid ratio of thermal conductivities and thermal diffusivities is set to 1, to have a significant effect of convection and conduction coupling at the interface, \cite{flageul2015dns}.

The computational domain size in the streamwise and spanwise directions is (8$\pi \delta$, 4/3$\pi\delta$), found by \citet{kim_moin_moser_1987} to be enough for the flow variables to become uncorrelated at maximum streamwise and spanwise separation. The grid size is (256$\times$144$\times$128) with a grid stretching in the wall-normal direction detailed in  \cite{PIROZZOLI2021110408}.  To capture the physics,  the first cell is in the viscous sub-layer,  $\Delta y^+ \mathcal{O} (1)$ and having  $\Delta x^{+}\mathcal{O} (10)$  and $\Delta z^{+}\mathcal{O} (5)$ is sufficient to resolve small-scale structures near the walls. 
The initial condition for the fluid was taken from an isothermal case and the solid was initialised from the steady-state solution.
The simulation was run until the first and second moment of the  temperature converged in both domains and then statistics have been computed for 10 flow-through times.

\begin{table}[h!]
    \caption{Computational parameters for the turbulent channel flow simulation}
    \centering
    \begin{tabular}{c|c|c|c}
        \hline
        Re$_\tau$ & $(N_x, N_y, N_z)$ & min($\Delta y^+$) &($\Delta x^+$,$\Delta z^+$)  \\
        \hline
        149.2  & (256,144,128) & 0.23 & (14.52, 4.71) \\
        \hline
        \hline
         $M$ &Pr&  $\kappa_f / \kappa_s$ & $\alpha_f/ \alpha_s$\\
         \hline 
         0.1 & 0.71 & 1 & 1 \\
    \end{tabular}
    
    \label{tab:drag_coefficient2}
\end{table}



The conjugate channel flow serves as a benchmark for the different modelling strategies used for the solid domain. 
Firstly, a finite element approach for the directly coupled conjugate solution is used, and this solution is referred to as the \textit{FE} solution. 
The finite element grid has the same resolution in the harmonic directions as the fluid mesh in \autoref{tab:drag_coefficient2} and the resolution is higher in the wall-normal direction, $N_{y,s} = 144$ for both domains. 
Then a combined solution strategy is also introduced,  labeled \textit{CFE} solution for combined finite element.
The \textit{CFE} has a local finite element  grid that spans  10\% of the overall thickness of the solid domain and maintains the  \textit{FE} mesh density, this thickness was chosen because, the amplitude of the fluctuations is expected to have diminished by a factor $1/e$ at that depth. 
 A modal mesh complements the finite element mesh. The resolution $N_{x,s} = 64$, $N_{y,s} = 36$, $N_{z,s} = 32$ and the truncation threshold is set to 70 \% of the total energy. 
 A third strategy is employed, labeled \textit{CFEA} solution, for combined finite element accelerated.
The \textit{CFEA} solution has the same grid arrangement and resolution as the \textit{CFE} solution, but the acceleration factors $\sigma$ and $\beta$ are set to 2.0 for the first twenty eigenmodes.

\begin{figure*}
\centering
    \scalebox{0.9}{\input{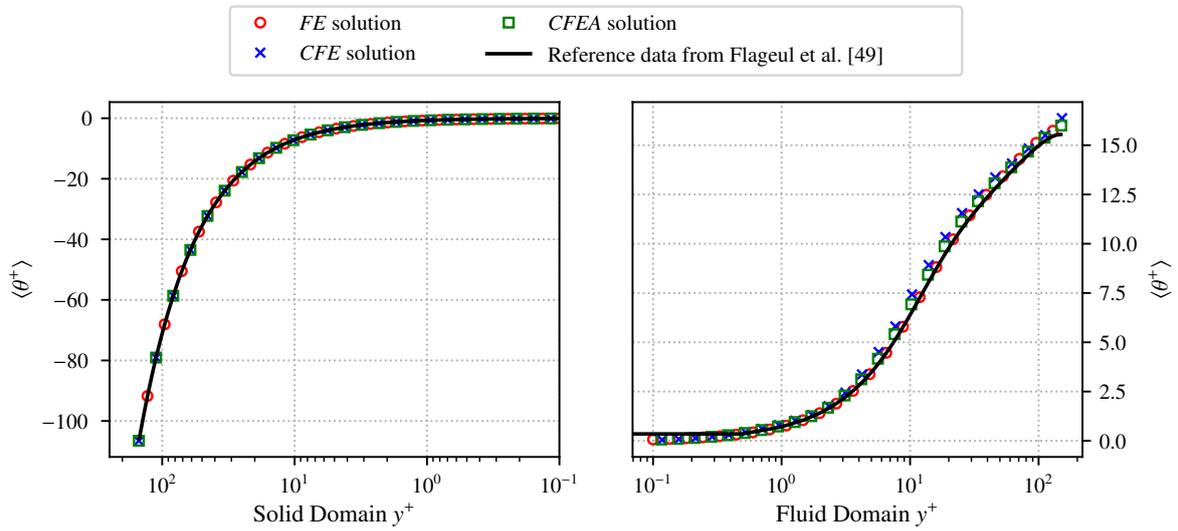}}
    \caption{Mean temperature profiles for the conjugate channel flow case. The symbols do not match  the mesh used. }
    \label{fig:meanT} 
\end{figure*}

\begin{figure*}
\centering
\scalebox{0.9}{\input{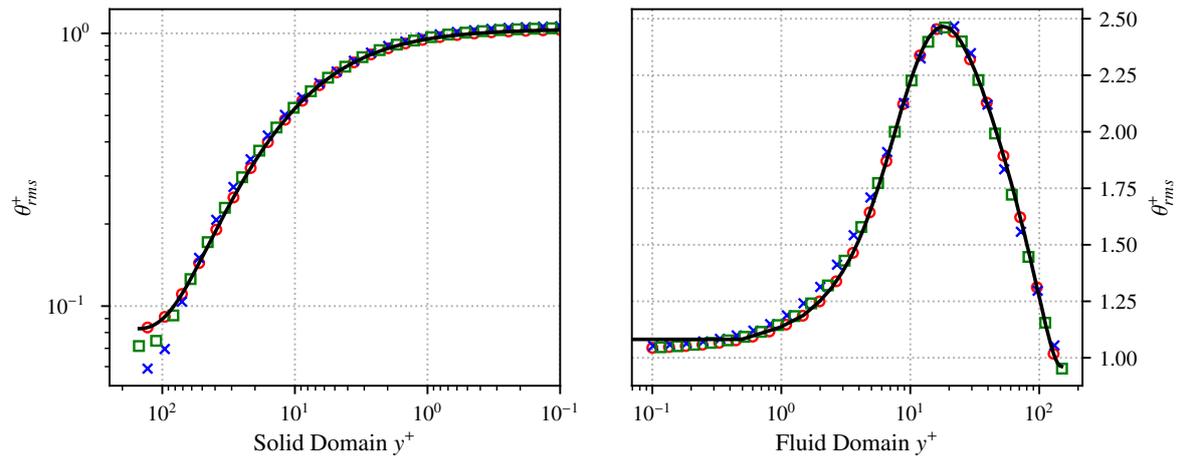}}
    \caption{Temperature fluctuation profiles for the conjugate channel flow case. The black line represents the reference data from \cite{flageul2015dns}. Symbols are defined in \autoref{fig:meanT} and do not correspond to the mesh used.}
    \label{fig:varT} 
\end{figure*}

\autoref{fig:meanT} and \autoref{fig:varT} show the mean temperature relative to the interface temperature in wall units, $\theta^+$, and standard deviation profiles, $\theta^+_\text{rms}$. The figures compare the results from \citet{flageul2015dns} with the three modeling strategies proposed. 
For the \textit{FE} solution,  although a small discrepancy is seen towards the middle of the channel for the mean flow and the RMS fluctuations are  slightly underpredicted at the interface, the general agreement in both domains validates the accuracy of the present code. 
The \textit{CFE} solution features the same discrepancies as the \textit{FE} solution and in addition the RMS fluctuations are underpredicted at  the far end of the solid domain. This is due to the modal truncation and coarser mesh used.  Overall, the alignment of this approach with other curves demonstrates its effective implementation. 
Finally, the \textit{CFEA} solution shows good agreement for the mean value and the RMS fluctuations. However, the modal fluctuations are underpredicted at the far end as a result of the acceleration factor making the eigenvectors stiffer and therefore the fluctuations' amplitude will decrease. 

\subsection{Pipe flow case}
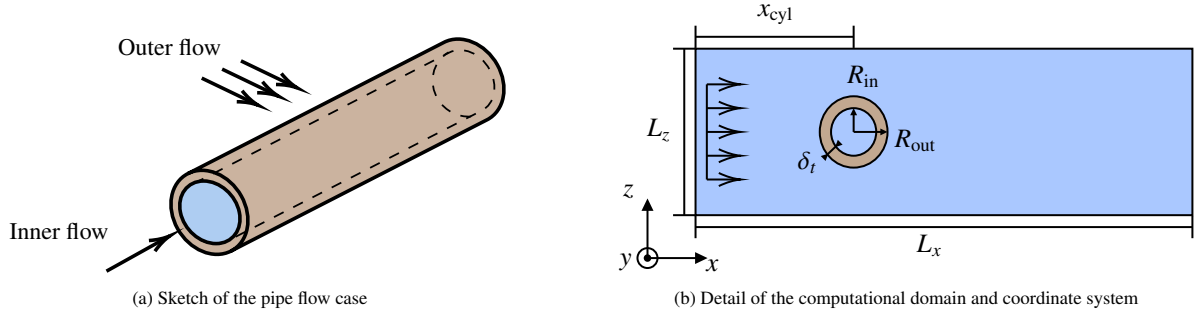
\begin{figure*}[t!]
    \centering
    \begin{subfigure}[b]{0.4\textwidth}
        \centering
        \scalebox{0.9}{\input{figures/cylinder_prob} }
        \caption{Sketch of the pipe flow case}
        \label{cv}
    \end{subfigure}
    \hfill
    \begin{subfigure}[b]{0.55\textwidth}
        \centering
        \input{figures/cv2}
        \caption{Detail of the computational domain and coordinate system}
        \label{detail}
    \end{subfigure}
    \caption{Pipe flow case and computational domain.} 
\end{figure*}
The turbulent channel flow case validated the ability of the solver to accurately handle conjugate heat transfer using different solution strategies. However, due to the limited large-scale temperature fluctuations, the transient behaviour of the temperature field was not very pronounced.

This second test case consists of a pipe in crossflow, as illustrated in \autoref{cv}. The pipe is subjected to an internal longitudinal flow and an external crossflow which are at different mean temperatures. This configuration serves as a fundamental example of crossflow heat exchanger operation and can provide a first-order approximation of the leading edge of an internally cooled turbine. 
This test case is particularly relevant to benchmark the proposed method because, when the pipe operates in the shedding regime, periodic fluctuations in heat transfer occur. To ensure the formation of a fully turbulent vortex street, the outer flow is characterised by a Reynolds number of 
$Re_D = 5000$, based on the tube's outer diameter. The Mach number is set to 0.1. The inner flow is characterised by a Reynolds number of Re$_{D, \text{in}} = 4500$ based on the pipe inner diameter while the temperature is set to be twice the outer temperature. The thickness of the pipe is set to 20\% of its radius.  
The Prandtl number is $0.71$ 
and the ratio of thermal diffusivities $\alpha_f/\alpha_s$ is  9. 

The main source of thermal excitation on the outside of the pipe will come from the vortex shedding. The dominant shedding frequency is set by the Strouhal frequency F$_{\text{Str}}$. The depth at which the amplitude of the temperature fluctuation is reduced to roughly 95\% of its surface value is:
\begin{align}
    \delta_{P,95\%} = 3\delta_P &= 3\sqrt{2}\sqrt{\frac{\alpha_s}{2\pi F_{\text{Str}}}} = 3\sqrt{2}\sqrt{\frac{2\alpha_sR_{\text{out}}}{2\pi\text{Str} U}} = 3\sqrt{2}\sqrt{ \frac{4R_{\text{out}}^2}{2\pi\text{Re Str Pr}}\frac{\alpha_s}{\alpha_f}} =  3\sqrt{2}\sqrt{ \frac{2(\delta_t/0.2)^2}{\pi\text{Re Str Pr}}\frac{\alpha_s}{\alpha_f}} = 0.215 \, \delta_t
\end{align}
The fluctuations are expected to extend approximately 20 \% of the pipe thickness, allowing for a significant penetration of the temperature fluctuations into the solid domain, to be captured by the modal $MFE$ solution. 

To understand how long the thermal transient is expected to last, the system can be simplified by considering  a one dimensional case with the limit case of Dirichlet boundary conditions on both sides, corresponding the best case scenario in terms of approach to steady state.  In dimensionless form, the first eigenvalue of the 
laplacian for the heat equation is  $\lambda_1 = \pi^2$. 
To determine the number of Strouhal periods $N_{\text{Str}}$ required for the transient of the first mode to decay by 95\%, the following condition must be satisfied:
$$ \lambda_1 \text{Fo} > 3 \implies \lambda_1 \frac{\alpha_s N}{\delta_t^2 F_{\text{Str}}} > 3 \implies N > 19.4$$
Highlighting that in practice the convergence of the lower modes will likely require many Strouhal periods. Furthermore, because the boundary conditions are not Dirichlet but Robin, the transient response will be longer.

\subsubsection{Computational setup}
Strictly speaking, the geometry of the problem is completely defined by the inner and outer pipe radii $R_{\text{in}}$ and $R_{\text{out}}$ along with the pipe length L$_y$. However, since the computational domain must be finite, additional parameters such as the streamwise and spanwise length $L_x$ and $L_z$ and the streamwise position of the cylinder within the domain, $x_{\text{cyl}}$ are introduced. These geometrical parameters are sketched in \autoref{detail}. 
The inner radius and the length of the cylinder are set to $R_{\text{in}} = 0.8 R_{\text{out}}$ and $L_y = 6 R_{\text{out}}$.  
To determine the optimal value of the parameters a sensitivity study on the drag coefficient of the cylinder is performed at a fixed mesh density. The range spanned by the parameters is $L_x / R_{\text{out}} \in [ 10, 50] $,  $L_z/ R_{\text{out}} \in [ 10, 50]$, $x_{\text{cyl}}/L_x \in [ 0.2, 0.5] $. Guiding values for scale resolving simulations of cylinders in crossflows at low Reynolds number can be found in \cite{Young2007ComparativeAO, Lysenko, 10.1063/1.870318}. A summary of the tested scenarios is available in \autoref{tab:drag_coefficient}. The final values of the parameters are $L_x = 30 R_{\text{out}}$, $L_z = 20 R_{\text{out}}$, $x_{\text{cyl}} = 0.35 L_x$. The fluid computational grid is a block-structured h-type with local refinement around the viscous surfaces.  

On the  solid side, two \textit{FE} meshes are extruded from the pipe surfaces on the inner and outer fluid domains. 
The \textit{FE} meshes extend across 10\% of the radial extend of the pipe and the radial resolution at the wall is approximately twice the resolution from the fluid side. 
A modal mesh \textit{MFE}, which spans the entire solid domain, is employed to thermally couple the two fluid domains using a coarser discretisation.
The mesh details are provided in \autoref{tab:meshdetails}, with a total node count of $8.4\times 10^6$.

\begin{table*}
\centering
\begin{minipage}{0.47\textwidth}
    \caption{Drag coefficient $C_D$ for different geometrical parameters } 
    \centering
    \begin{tabular}{|c|c|c|c|} 
        \hline
        $L_x/R_{\text{out}}$ & $L_z/R_{\text{out}}$ & $x_{\text{cyl}}/R_{\text{out}}$ & $C_D$ \\ 
        \hline
        10 & 5 & 0.4 & 1.120  \\
        20 & 10 & 0.3 & 1.110 \\
        20 & 20 & 0.4 & 1.075  \\ 
        30 & 20 & 0.35 & 1.073 \\ 
         50 & 40 & 0.35 & 1.073\\
        40 & 30 & 0.25 & 1.074\\
        \hline
    \end{tabular}
    
    \label{tab:drag_coefficient}
\end{minipage}
\hfill
\begin{minipage}{0.47\textwidth}
    \caption{Details of the different numerical grids used in the pipe flow case.}
    \centering
    \begin{tabular}{c|c|c|c|c}
        \hline
        Domain & Node count  & $N_y$ & $N_r$& max($\Delta r^+$) \\
        \hline
        Inner & 2.11 $\times10^6$& 120& - & 0.53 \\
        Outer & 5.04 $\times10^6$& 72 &  -&0.43 \\
        Solid \textit{FE}  & 6.15$\times10^5$ \footnotemark& 72 & 24&0.14 \\ 
        Solid \textit{MFE}  & 13824 & 24 & 18& 5.61 \\ 
        \hline
    \end{tabular}
    \vspace*{0.6cm}
    
    \label{tab:meshdetails}
\end{minipage}
\end{table*}
\footnotetext{Node count includes both $FE$ grids.}
At the inflow boundary of both inner and outer domains, synthetic turbulence is generated using the method described by  \citet{10.1063/5.0136568}.  This technique uses modified uniformly distributed random sequences to  construct divergence-free anisotropic random fields with sensible spectrum and complete complex correlation in space and time.
To allow for the turbulence to develop, the domain detailed above is  extended by 4 diameters before entering the mixing domain of interest. Each simulation is run for 40 Strouhal periods once the shedding regime has been established.

\subsubsection{Validation}
\begin{figure*}[h!]
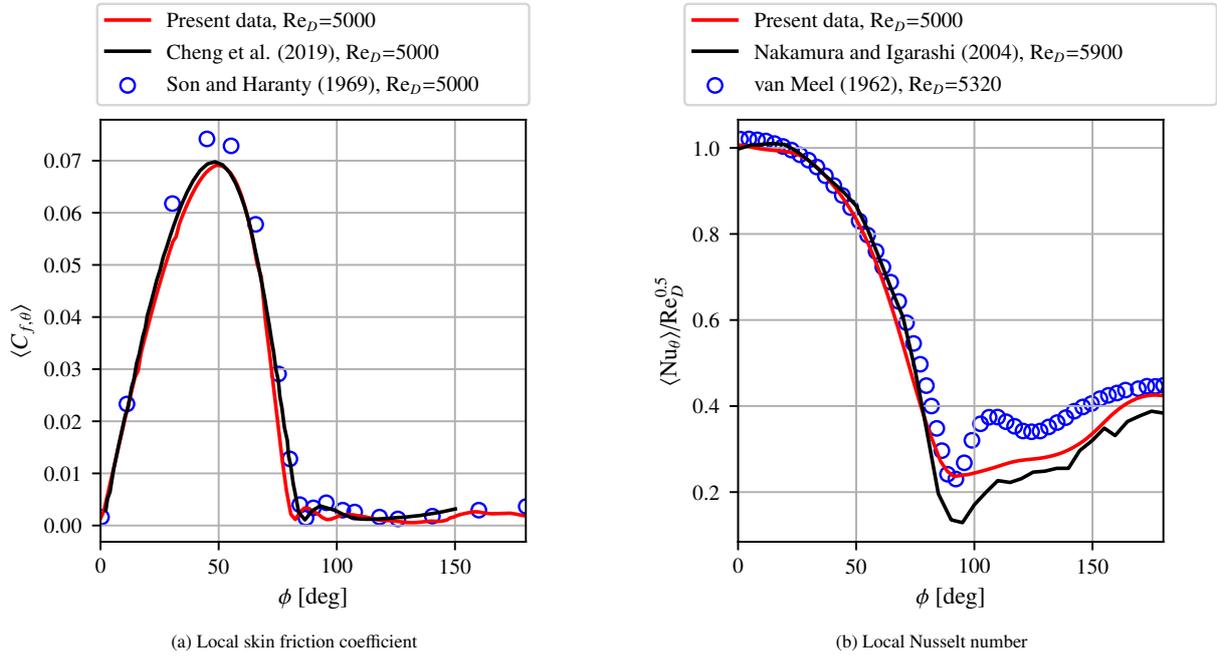

    \centering
    \begin{subfigure}[b]{0.49\textwidth}
        \centering
        \scalebox{.9}{\input{figures/Cf.pgf}}
        \caption{Local skin friction coefficient }
        \label{fig:cylinder_Cf}
    \end{subfigure}
    \hfill
    \begin{subfigure}[b]{0.49\textwidth}
        \centering
        \scalebox{.9}{\input{figures/local_nusselt.pgf}} 
        \caption{Local Nusselt number }
        \label{fig:cylinder_nusselt}
    \end{subfigure}
    \caption{Validation results for the pipe flow case }
    \label{fig:cylinder_valid}
\end{figure*}

The solver is validated against reference data of \cite{Cheng_Pullin_Samtaney_Zhang_Gao_2017,Son_Hanratty_1969,LOWERY19751229,NAKAMURA2004741,VANMEEL1962715}.  To match the boundary conditions of the reference data, these simulations only included the outer domain at the target Reynolds number with a constant heat flux or constant temperature on the cylinder surface for the validation of the skin friction and local Nusselt number respectively. \autoref{fig:cylinder_valid} shows the time-averaged skin friction coefficient $\langle C_{f_\theta} \rangle$ and the time-averaged Nusselt number $\langle Nu \rangle$. The local angle $\phi$ is starting from the stagnation point on the cylinder surface.  For the skin friction coefficient, the solver predicts separation for $\phi $ slightly below 90 degrees, which is in line with the experimental data at the studied Reynolds number, \cite{10.1063/1.5139479}. \\
Regarding the local Nusselt number in \autoref{fig:cylinder_nusselt}, the global trend is well captured, but the curves move away from each other on the rear side of the cylinder, multiple factors explain this discrepancy. First, the simulations are at slightly different Reynolds numbers. This will mostly impact the rear side of the cylinder, the greater the Reynolds number the greater the heat transfer coefficient on the rear side of the cylinder. 
Similarly, the local Nusselt number on the rear of the cylinder is also influenced by turbulence intensity, which is challenging to replicate accurately the wind tunnel conditions.
Additionally, the discrepancy at $\phi = 0$ is due to the difference in the ratio $k_s/k_f$. \citet{VANMEEL1962715} used a  ratio of approximately 50, whereas in the present case and in the experiment by \citet{NAKAMURA2004741}, the ratio is closer to 9000. As the ratio decreases, the local Nusselt number varies, particularly on the forward side of the cylinder. However, since the ratios are all greater than 20, their influence is confined to the forward side of the cylinder, as noted by \citet{SUNDEN19801359}.

\begin{figure}[htbp]
    \centering
    \begin{subfigure}[b]{0.23\textwidth}
        \centering
        \includegraphics[width=0.8\textwidth]{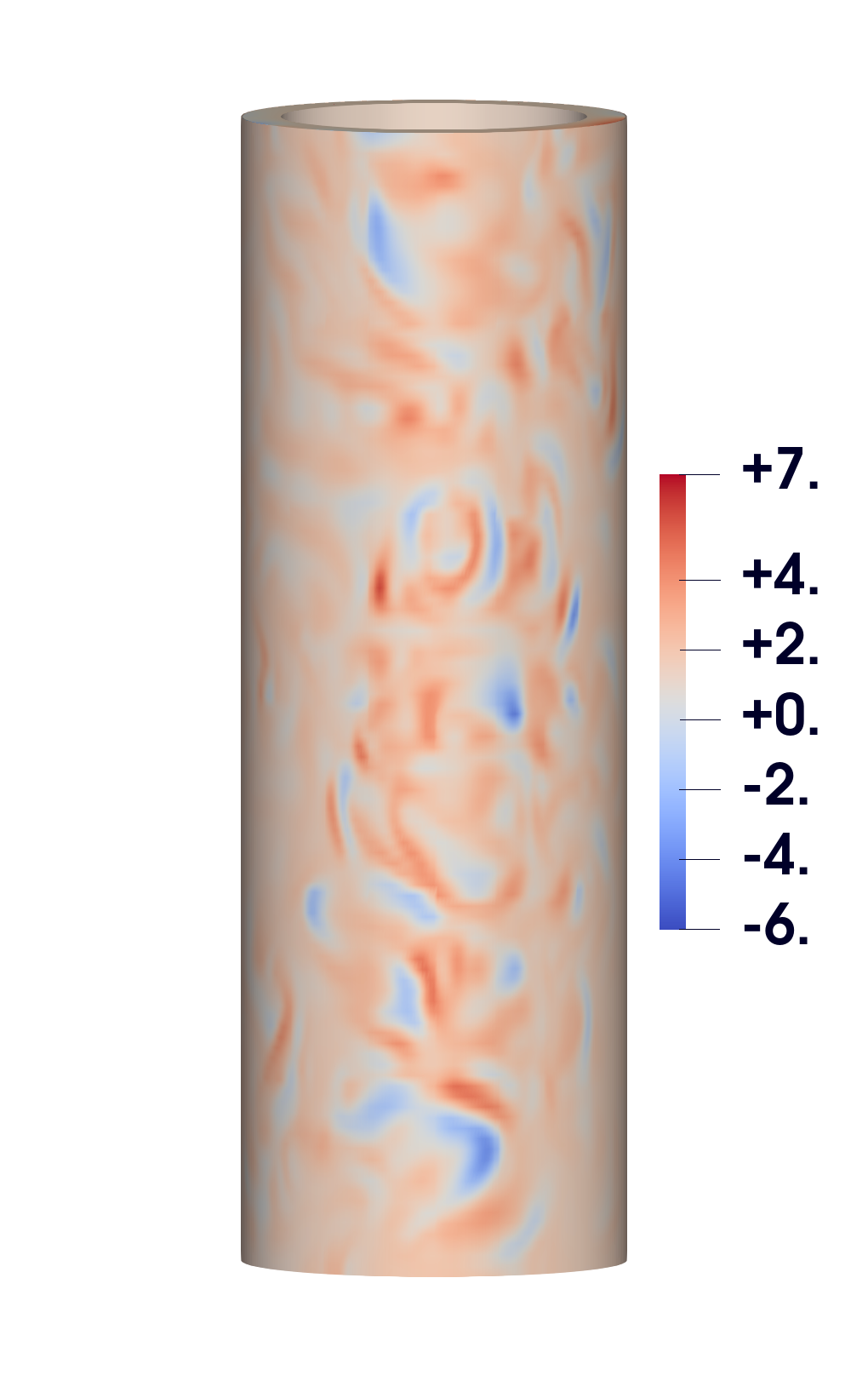}
        \caption{$r^+ = 3$}
        \label{fig:cyl_f}
    \end{subfigure}
    \begin{subfigure}[b]{0.23\textwidth}
        \centering
        \includegraphics[width=0.8\textwidth]{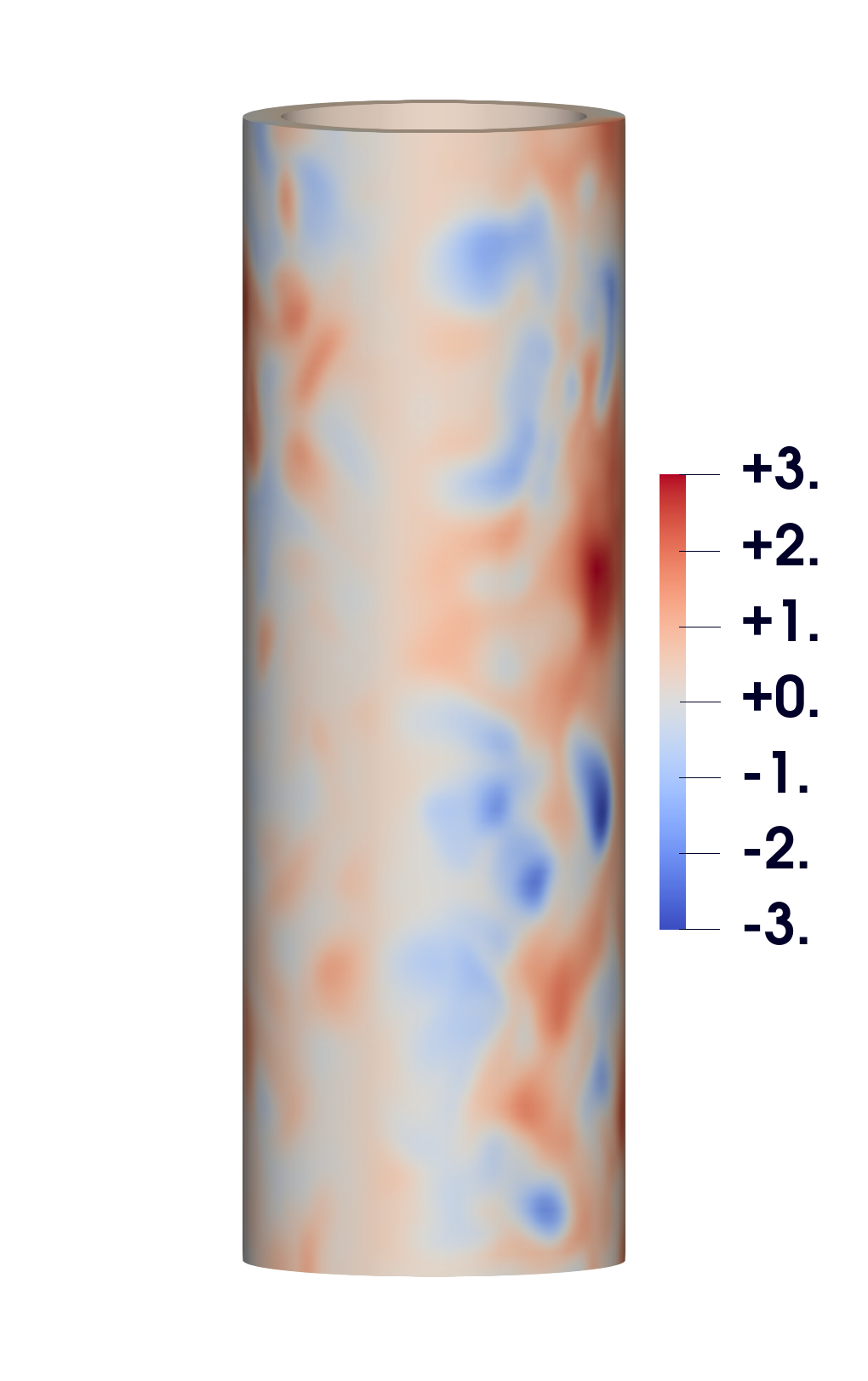}
        \caption{$r^+ =0 $, start of the $FE$ solution}
        \label{fig:cyl_sol}
    \end{subfigure}
    \begin{subfigure}[b]{0.23\textwidth}
        \centering
        \includegraphics[width=0.8\textwidth]{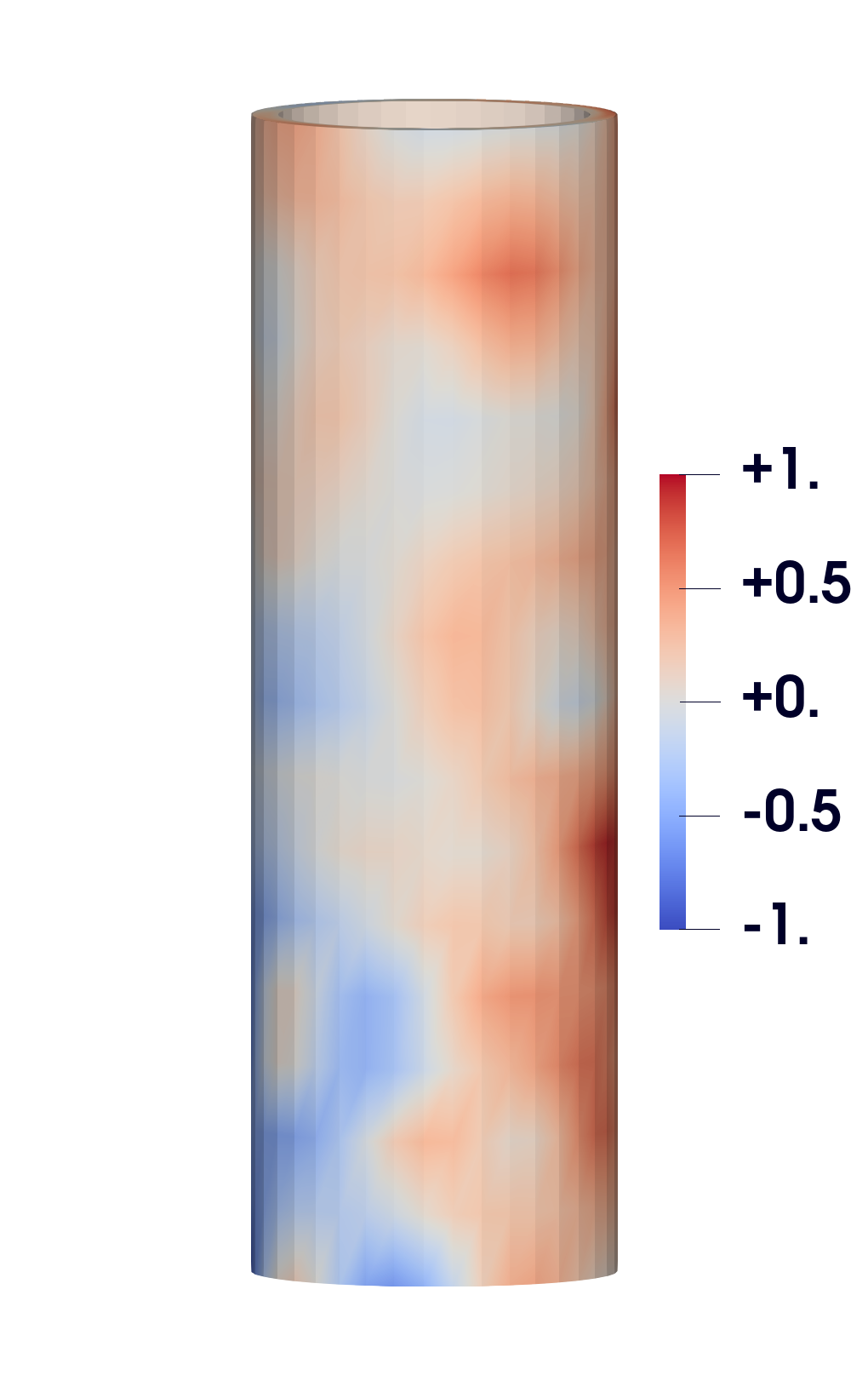}
        \caption{Edge of the $MFE$ solution}
        \label{fig:mode7}
    \end{subfigure}
    \begin{subfigure}[b]{0.23\textwidth}
        
        \centering
        \includegraphics[width=0.8\textwidth]{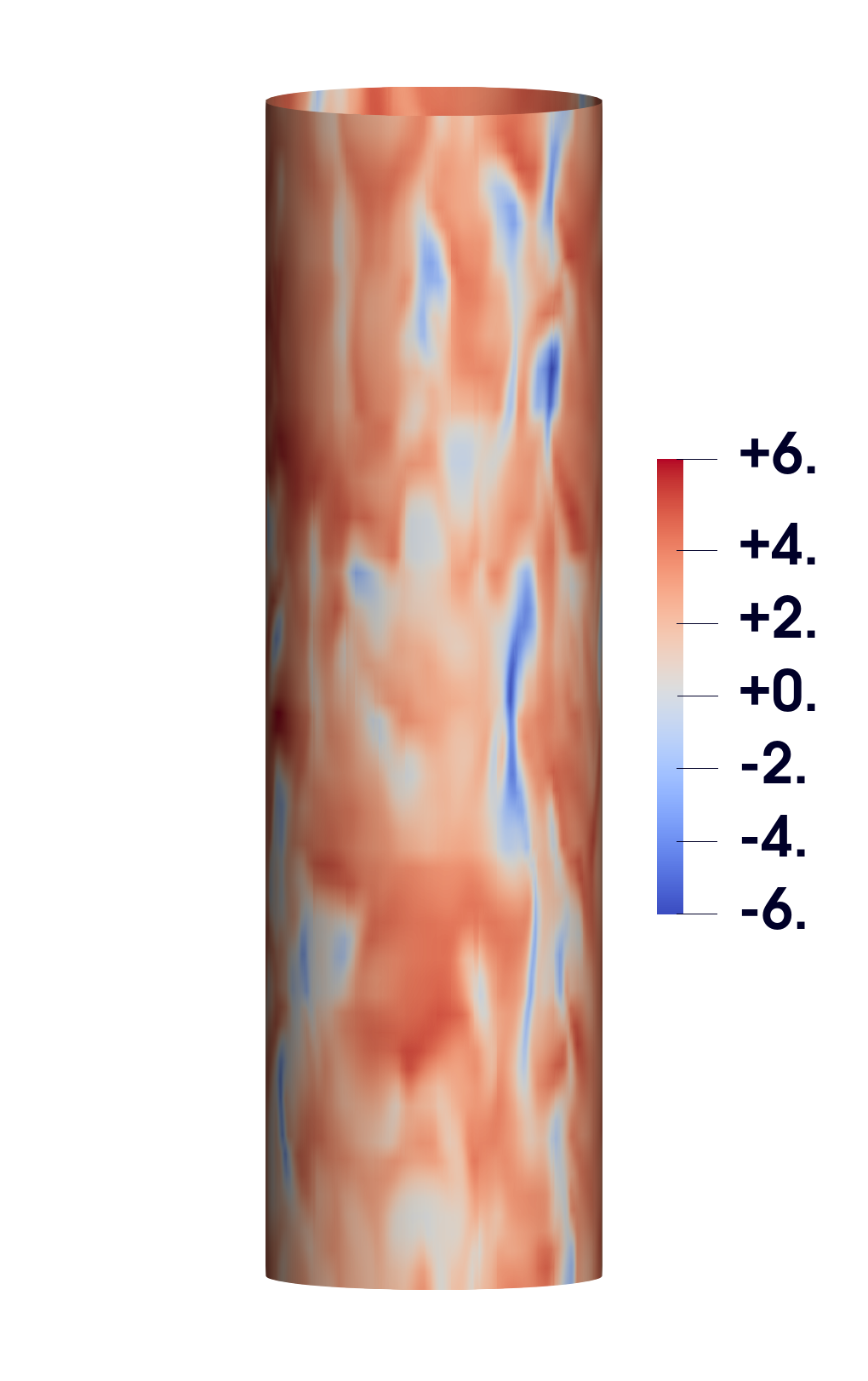}
        \caption{$r^+ = 5$ inside the pipe}
        \label{fig:mode283}
    \end{subfigure}
    \caption{Dimensionless temperature fluctuations at different depths in the domain}
    \label{fig:temperature_modes}
\end{figure}

Qualitatively, the thermal response of the pipe flow is displayed in \autoref{fig:temperature_modes}. 
It shows thermal slices at multiple radii within the domain, from left to right the slices correspond  to the fluid in the vicinity of the outer surface of the cylinder, the outer surface of the cylinder, the start of the \textit{MFE} solution at $0.9 R_{\text{out}}$ and finally the fluid in the vicinity of the inner section of the pipe $r^+ \approx 5$.  
The figure highlights that deeper into the solid, the temperature fluctuations exhibit larger scales, which are directly related to the penetration depth. Additionally, it reflects the difference in the nature of the perturbations originating from the inner and outer surfaces, duct streaks for the inner section and shedding regime for the outer region.
\subsubsection{Unsteady thermal behavior}

To study the unsteady  behavior, the thermal solution was initialised with a uniform temperature corresponding to the respective freestream temperature, while the solid pipe was initialised at the inner freestream temperature. 
The time trace of the modal amplitude has been recorded and is shown in \autoref{fig:cylinder_history}.
The figure presents the normalised modal amplitudes for selected modes against time normalised by the Strouhal frequency $f_{St}$.
The figure also illustrates the time required to reach a steady state. Mode 0 stabilises after approximately 0.4 Fo, while the other modes converge more rapidly as the mode index increases. 
Mode $0$ represents the constant mode characterised by a uniform amplitude in the circumferential direction and  participating mainly to the mean solution. 
In contrast, the other modes plotted have a zero mean along the circumferential direction and are therefore more associated with the fluctuating behavior of the temperature field, as seen by the more pronounced fluctuating behavior in their time trace matching the Strouhal frequency.

A second simulation is performed, this time initialising the flow field from an instantaneous snapshot of the shedding cylinder. The modal amplitudes are set initially to the mean values of the time traces from \autoref{fig:cylinder_history}, excluding the initial 10 Strouhal periods.  An example of a time trace when restarted from the mean value is shown in \autoref{fig:cylinder_history_ACC} with the black line. The curve exhibit similar behaviour than \autoref{fig:1Dtime}, even if the mode is initialised with the statistical steady state value the amplitude of the mode goes through a  transient before stabilising. 
It should be noted that not all the modes exhibit such a significant transient, it is a function of the phase between the initial state in the solid and the flow. 
Alongside, three additional simulation have been run with different acceleration factors. A summary of the performance of the different simulations is available in \autoref{tab:acc}. \autoref{tab:acc} shows, for multiple modes, the estimated time required to reach a steady state and the absolute percent error in standard deviation once the steady state is reached. It can be seen that for the lowest mode shapes, a time-to-steady state divided by four can be achieved with an error just above 6\%. This error is done on the modal amplitude of long-time low energy modes and  does not have a significant impact on the interface fluctuations on both sides of the pipe, which remain constant for all acceleration factors tested, however deeper in the solid, the behavior is dominated by these modes and therefore the error in the standard deviation directly correlates with the modal amplitude error. In \autoref{tab:acc}, it is interesting to note that for the modes that are close to the Strouhal frequency ($\lambda/\omega_{\text{Str}} = 0.7604 \, \&\,  2.606$),  the error in the standard deviation is significantly higher than for the other modes. This is because the main forcing frequency for these modes will be close to the pole of the transfer function and therefore the error in the amplitude will be larger with the SFD method used.


\begin{figure*}[h!]
    \centering
    \begin{minipage}[b]{0.49\textwidth}
        \centering
        \vspace*{-0.1cm}
        \scalebox{0.9}{\input{figures/modal_traces.pgf}}
        \caption{Normalised modal time trace with uniform temperature field initialization}
        \label{fig:cylinder_history}
    \end{minipage}
    \hfill
    \begin{minipage}[b]{0.49\textwidth}
        \centering
        \scalebox{.9}{\input{figures/cjg_history_ACC.pgf}}
        \caption{Modal time trace for mode $100$ with a temperature field initialised with steady-state solution}
        \label{fig:cylinder_history_ACC}
    \end{minipage}
   
\end{figure*}
\begin{table}
    \centering

    \caption{Strouhal periods to reach steady-state and aboslute percent error in standard deviation for different acceleration factors for selected modes}
    \begin{tabular}{c|c|c|c|c|c|c}
        \hline
        $\lambda/\omega_{\text{Str}}$ & 0.0242 & 0.0247 & 0.0872 & 0.7604& 2.606 & 10.331\\
        \hline
        $\beta = 1.0, \sigma = 1.0$ & 19.7, 0.00\% &  19.3, 0.00\% &5.47, 0.00\% & 0.63, 0.00\%  & 0.18, 0.00\% & 0.046, 0.00\% \\
        $\beta = 2.0, \sigma = 2.0$ & 9.86, 2.72\% &  9.66, 2.91\% &2.73, 3.32\% & 0.31, 6.87\%  & 0.09, 4.41\% & 0.023, 1.28\% \\
        $\beta = 4.0, \sigma = 4.0$ & 4.93, 6.27\% &  4.83, 5.48\% &1.36, 6.71\% & 0.16, 13.1\%  & 0.05, 9.39\% & 0.012, 2.32\% \\
        $\beta = 8.0, \sigma = 8.0$ & 2.47, 15.7\% &  2.42, 10.9\% &0.68, 12.8\% & 0.08, 27.2\%  & 0.03, 19.5\% & 0.006, 5.95\% \\
        \hline
    \end{tabular}

    \label{tab:acc}
\end{table}

%% file: figures/cylinder_prob.tex
\tikzset{every picture/.style={line width=0.75pt}} 

\begin{tikzpicture}[x=0.75pt,y=0.75pt,yscale=-1,xscale=1]

\draw  [fill={rgb, 255:red, 139; green, 87; blue, 42 }  ,fill opacity=0.45 ][line width=1.5]  (135.78,207.76) -- (277.78,134.41) .. controls (287.53,129.36) and (300.8,134.96) .. (307.41,146.91) .. controls (314.02,158.86) and (311.47,172.63) .. (301.71,177.67) -- (159.71,251.03) .. controls (149.96,256.07) and (136.69,250.47) .. (130.08,238.52) .. controls (123.47,226.57) and (126.02,212.8) .. (135.78,207.76) .. controls (145.54,202.72) and (158.81,208.32) .. (165.42,220.27) .. controls (172.03,232.21) and (169.47,245.99) .. (159.71,251.03) ;
\draw  [dash pattern={on 4.5pt off 4.5pt}][line width=0.75]  (139.05,213.31) -- (280.83,140.07) .. controls (288.15,136.29) and (298.1,140.49) .. (303.06,149.45) .. controls (308.02,158.41) and (306.1,168.75) .. (298.78,172.53) -- (157.01,245.77) .. controls (149.69,249.55) and (139.73,245.35) .. (134.77,236.38) .. controls (129.82,227.42) and (131.73,217.09) .. (139.05,213.31) .. controls (146.37,209.52) and (156.33,213.73) .. (161.28,222.69) .. controls (166.24,231.65) and (164.33,241.98) .. (157.01,245.77) ;
\draw  [draw opacity=0][fill={rgb, 255:red, 255; green, 255; blue, 255 }  ,fill opacity=1 ][line width=1.5]  (131.48,229.37) .. controls (129.57,219.71) and (135.38,211.88) .. (144.45,211.88) .. controls (153.52,211.88) and (162.42,219.71) .. (164.33,229.37) .. controls (166.24,239.03) and (160.44,246.86) .. (151.37,246.86) .. controls (142.3,246.86) and (133.39,239.03) .. (131.48,229.37) -- cycle ;
\draw  [fill={rgb, 255:red, 74; green, 144; blue, 226 }  ,fill opacity=0.45 ][line width=1.5]  (131.65,229.37) .. controls (129.74,219.71) and (135.5,211.88) .. (144.53,211.88) .. controls (153.56,211.88) and (162.42,219.71) .. (164.33,229.37) .. controls (166.24,239.03) and (160.48,246.86) .. (151.45,246.86) .. controls (142.43,246.86) and (133.56,239.03) .. (131.65,229.37) -- cycle ;
\draw  [draw opacity=0][dash pattern={on 4.5pt off 4.5pt}] (292.87,175.81) .. controls (285.86,176.4) and (278.61,172.92) .. (274.53,166.28) .. controls (268.89,157.1) and (271.42,145.35) .. (280.18,140.05) .. controls (280.24,140.01) and (280.31,139.97) .. (280.38,139.93) -- (290.39,156.67) -- cycle ; \draw  [dash pattern={on 4.5pt off 4.5pt}] (292.87,175.81) .. controls (285.86,176.4) and (278.61,172.92) .. (274.53,166.28) .. controls (268.89,157.1) and (271.42,145.35) .. (280.18,140.05) .. controls (280.24,140.01) and (280.31,139.97) .. (280.38,139.93) ;  
\draw [line width=1.5]    (90.65,262.28) -- (123.15,244.04) ;
\draw [shift={(125.77,242.57)}, rotate = 150.7] [color={rgb, 255:red, 0; green, 0; blue, 0 }  ][line width=1.5]    (11.37,-3.42) .. controls (7.23,-1.45) and (3.44,-0.31) .. (0,0) .. controls (3.44,0.31) and (7.23,1.45) .. (11.37,3.42)   ;
\draw [line width=1.5]    (143.48,154.38) -- (171.92,168.15) ;
\draw [shift={(174.62,169.46)}, rotate = 205.84] [color={rgb, 255:red, 0; green, 0; blue, 0 }  ][line width=1.5]    (11.37,-3.42) .. controls (7.23,-1.45) and (3.44,-0.31) .. (0,0) .. controls (3.44,0.31) and (7.23,1.45) .. (11.37,3.42)   ;
\draw [line width=1.5]    (154.78,149.35) -- (183.23,163.12) ;
\draw [shift={(185.93,164.43)}, rotate = 205.84] [color={rgb, 255:red, 0; green, 0; blue, 0 }  ][line width=1.5]    (11.37,-3.42) .. controls (7.23,-1.45) and (3.44,-0.31) .. (0,0) .. controls (3.44,0.31) and (7.23,1.45) .. (11.37,3.42)   ;
\draw [line width=1.5]    (165.23,144.74) -- (193.68,158.52) ;
\draw [shift={(196.38,159.82)}, rotate = 205.84] [color={rgb, 255:red, 0; green, 0; blue, 0 }  ][line width=1.5]    (11.37,-3.42) .. controls (7.23,-1.45) and (3.44,-0.31) .. (0,0) .. controls (3.44,0.31) and (7.23,1.45) .. (11.37,3.42)   ;

\draw (35.4,233.57) node [anchor=north west][inner sep=0.75pt]   [align=left] {Inner flow};
\draw (95.62,131.01) node [anchor=north west][inner sep=0.75pt]   [align=left] {Outer flow};

\end{tikzpicture}

%% file: figures/cv2
\tikzset{every picture/.style={line width=0.75pt}} 

\begin{tikzpicture}[x=0.75pt,y=0.75pt,yscale=-1,xscale=1]

\draw  [fill={rgb, 255:red, 12; green, 100; blue, 255 }  ,fill opacity=0.35 ] (81.98,132.68) -- (330,132.68) -- (330,216.42) -- (81.98,216.42) -- cycle ;
\draw  [draw opacity=0][fill={rgb, 255:red, 255; green, 255; blue, 255 }  ,fill opacity=1 ] (143.99,174.55) .. controls (143.99,164.64) and (151.56,156.61) .. (160.9,156.61) .. controls (170.24,156.61) and (177.81,164.64) .. (177.81,174.55) .. controls (177.81,184.46) and (170.24,192.5) .. (160.9,192.5) .. controls (151.56,192.5) and (143.99,184.46) .. (143.99,174.55) -- cycle ;
\draw  [fill={rgb, 255:red, 139; green, 87; blue, 42 }  ,fill opacity=0.52 ] (143.99,174.55) .. controls (143.99,164.64) and (151.56,156.61) .. (160.9,156.61) .. controls (170.24,156.61) and (177.81,164.64) .. (177.81,174.55) .. controls (177.81,184.46) and (170.24,192.5) .. (160.9,192.5) .. controls (151.56,192.5) and (143.99,184.46) .. (143.99,174.55) -- cycle ;
\draw    (87.62,174.55) -- (102.53,174.55) ;
\draw [shift={(104.53,174.55)}, rotate = 180] [color={rgb, 255:red, 0; green, 0; blue, 0 }  ][line width=0.75]    (10.93,-3.29) .. controls (6.95,-1.4) and (3.31,-0.3) .. (0,0) .. controls (3.31,0.3) and (6.95,1.4) .. (10.93,3.29)   ;
\draw    (87.62,186.52) -- (102.53,186.52) ;
\draw [shift={(104.53,186.52)}, rotate = 180] [color={rgb, 255:red, 0; green, 0; blue, 0 }  ][line width=0.75]    (10.93,-3.29) .. controls (6.95,-1.4) and (3.31,-0.3) .. (0,0) .. controls (3.31,0.3) and (6.95,1.4) .. (10.93,3.29)   ;
\draw    (87.62,162.59) -- (102.53,162.59) ;
\draw [shift={(104.53,162.59)}, rotate = 180] [color={rgb, 255:red, 0; green, 0; blue, 0 }  ][line width=0.75]    (10.93,-3.29) .. controls (6.95,-1.4) and (3.31,-0.3) .. (0,0) .. controls (3.31,0.3) and (6.95,1.4) .. (10.93,3.29)   ;
\draw    (87.62,198.48) -- (102.53,198.48) ;
\draw [shift={(104.53,198.48)}, rotate = 180] [color={rgb, 255:red, 0; green, 0; blue, 0 }  ][line width=0.75]    (10.93,-3.29) .. controls (6.95,-1.4) and (3.31,-0.3) .. (0,0) .. controls (3.31,0.3) and (6.95,1.4) .. (10.93,3.29)   ;
\draw    (87.62,150.62) -- (102.53,150.62) ;
\draw [shift={(104.53,150.62)}, rotate = 180] [color={rgb, 255:red, 0; green, 0; blue, 0 }  ][line width=0.75]    (10.93,-3.29) .. controls (6.95,-1.4) and (3.31,-0.3) .. (0,0) .. controls (3.31,0.3) and (6.95,1.4) .. (10.93,3.29)   ;
\draw  [draw opacity=0][fill={rgb, 255:red, 255; green, 255; blue, 255 }  ,fill opacity=1 ] (149.62,174.55) .. controls (149.62,167.94) and (154.67,162.59) .. (160.9,162.59) .. controls (167.12,162.59) and (172.17,167.94) .. (172.17,174.55) .. controls (172.17,181.16) and (167.12,186.52) .. (160.9,186.52) .. controls (154.67,186.52) and (149.62,181.16) .. (149.62,174.55) -- cycle ;
\draw  [fill={rgb, 255:red, 12; green, 100; blue, 255 }  ,fill opacity=0.35 ] (149.62,174.55) .. controls (149.62,167.94) and (154.67,162.59) .. (160.9,162.59) .. controls (167.12,162.59) and (172.17,167.94) .. (172.17,174.55) .. controls (172.17,181.16) and (167.12,186.52) .. (160.9,186.52) .. controls (154.67,186.52) and (149.62,181.16) .. (149.62,174.55) -- cycle ;
\draw    (87.62,150.62) -- (87.62,198.48) ;
\draw    (160.9,165.59) -- (160.9,174.55) ;
\draw [shift={(160.9,162.59)}, rotate = 90] [fill={rgb, 255:red, 0; green, 0; blue, 0 }  ][line width=0.08]  [draw opacity=0] (3.57,-1.72) -- (0,0) -- (3.57,1.72) -- cycle    ;
\draw    (174.81,174.55) -- (160.9,174.55) ;
\draw [shift={(177.81,174.55)}, rotate = 180] [fill={rgb, 255:red, 0; green, 0; blue, 0 }  ][line width=0.08]  [draw opacity=0] (3.57,-1.72) -- (0,0) -- (3.57,1.72) -- cycle    ;
\draw    (81.98,222.41) -- (330,222.41) ;
\draw [shift={(330,222.41)}, rotate = 180] [color={rgb, 255:red, 0; green, 0; blue, 0 }  ][line width=0.75]    (0,5.59) -- (0,-5.59)   ;
\draw [shift={(81.98,222.41)}, rotate = 180] [color={rgb, 255:red, 0; green, 0; blue, 0 }  ][line width=0.75]    (0,5.59) -- (0,-5.59)   ;
\draw    (76.35,132.68) -- (76.35,216.42) ;
\draw [shift={(76.35,216.42)}, rotate = 270] [color={rgb, 255:red, 0; green, 0; blue, 0 }  ][line width=0.75]    (0,5.59) -- (0,-5.59)   ;
\draw [shift={(76.35,132.68)}, rotate = 270] [color={rgb, 255:red, 0; green, 0; blue, 0 }  ][line width=0.75]    (0,5.59) -- (0,-5.59)   ;
\draw    (81.98,126.7) -- (160.9,126.7) ;
\draw [shift={(160.9,126.7)}, rotate = 180] [color={rgb, 255:red, 0; green, 0; blue, 0 }  ][line width=0.75]    (0,5.59) -- (0,-5.59)   ;
\draw [shift={(81.98,126.7)}, rotate = 180] [color={rgb, 255:red, 0; green, 0; blue, 0 }  ][line width=0.75]    (0,5.59) -- (0,-5.59)   ;
\draw    (58,238) -- (85,238) ;
\draw [shift={(88,238)}, rotate = 180] [fill={rgb, 255:red, 0; green, 0; blue, 0 }  ][line width=0.08]  [draw opacity=0] (6.25,-3) -- (0,0) -- (6.25,3) -- cycle    ;
\draw    (58,238) -- (58,211) ;
\draw [shift={(58,208)}, rotate = 90] [fill={rgb, 255:red, 0; green, 0; blue, 0 }  ][line width=0.08]  [draw opacity=0] (6.25,-3) -- (0,0) -- (6.25,3) -- cycle    ;
\draw   (53,238) .. controls (53,235.24) and (55.24,233) .. (58,233) .. controls (60.76,233) and (63,235.24) .. (63,238) .. controls (63,240.76) and (60.76,243) .. (58,243) .. controls (55.24,243) and (53,240.76) .. (53,238) -- cycle ;
\draw  [fill={rgb, 255:red, 0; green, 0; blue, 0 }  ,fill opacity=1 ] (56.62,238) .. controls (56.62,237.24) and (57.24,236.62) .. (58,236.62) .. controls (58.76,236.62) and (59.38,237.24) .. (59.38,238) .. controls (59.38,238.76) and (58.76,239.38) .. (58,239.38) .. controls (57.24,239.38) and (56.62,238.76) .. (56.62,238) -- cycle ;

\draw    (152.35,181.52) -- (147.54,186.14) ;
\draw [shift={(148.26,185.45)}, rotate = 136.17] [fill={rgb, 255:red, 0; green, 0; blue, 0 }  ][line width=0.08]  [draw opacity=0] (3.57,-1.72) -- (0,0) -- (3.57,1.72) -- cycle    ;
\draw [shift={(151.62,182.22)}, rotate = 316.17] [fill={rgb, 255:red, 0; green, 0; blue, 0 }  ][line width=0.08]  [draw opacity=0] (3.57,-1.72) -- (0,0) -- (3.57,1.72) -- cycle    ;

\draw (190.57,224.77) node [anchor=north west][inner sep=0.75pt]   [align=left] {$\displaystyle L_{x}$};
\draw (56.07,166.15) node [anchor=north west][inner sep=0.75pt]   [align=left] {$\displaystyle L_{z}$};
\draw (180,172) node [anchor=north west][inner sep=0.75pt]   [align=left] {$\displaystyle R_{\text{out}}$};
\draw (156.33,139.67) node [anchor=north west][inner sep=0.75pt]   [align=left] {$\displaystyle R_{\text{in}}$};
\draw (111,109) node [anchor=north west][inner sep=0.75pt]   [align=left] {$\displaystyle x_{\text{cyl}}$};
\draw (82,237) node [anchor=north west][inner sep=0.75pt]   [align=left] {$\displaystyle \ x$};
\draw (41,199.33) node [anchor=north west][inner sep=0.75pt]   [align=left] {$\displaystyle \ z$};
\draw (42.67,234.33) node [anchor=north west][inner sep=0.75pt]   [align=left] {$\displaystyle y$};
\draw (131.57,183.19) node [anchor=north west][inner sep=0.75pt]   [align=left] {$\displaystyle \delta _{t}$};

\end{tikzpicture}

%% file: thinlayer.tex
\subsection{Two  dimensions space}
\begin{figure}[t]
    \centering
    \includegraphics[width=0.45\textwidth]{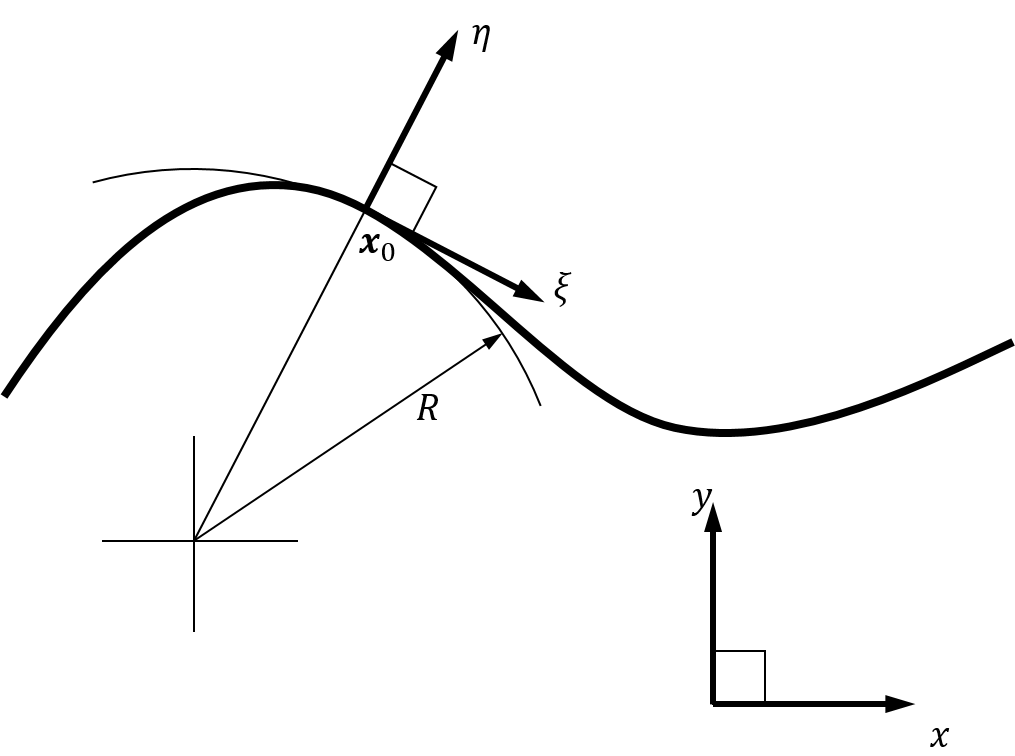}
    \caption{Surface frame of reference}\label{fig:surfaceframe}
\end{figure}
Consider a two-dimensional space with a curved boundary. The global coordinates are denoted by $\left(x,y\right)$ and the local surface bound coordinate are $\left(\xi,\eta\right)$, see \autoref{fig:surfaceframe}. Consider a position vector ${\bf x} = $ in the surface-bound frame of reference:
\begin{equation}
    {\bf x}={\bf x}_0\left(\xi\right)+ \eta {\bf n}
\end{equation}
Defining the unit tangent and normal vectors as
\begin{equation}
{\bf t}=\left[ \begin{matrix}t_x\\t_y\end{matrix}\right] \qquad  {\bf n}=\left[ \begin{matrix}-t_y\\t_x\end{matrix}\right]
\end{equation}
For a curve with radius of curvature $R$, the change of the tangent and normal vectors with respect to $\xi$ is:
\begin{equation}
\dfrac{\partial {\bf t}}{\partial \xi}=\dfrac{1}{R} \left[ \begin{matrix}t_y\\-t_x\end{matrix} \right] \qquad \dfrac{\partial{\bf n}}{\partial \xi}=\dfrac{1}{R}{\bf t}
\end{equation}
Differentiating the position vector $\mathbf{x}$ with respect to $\xi$ and $\eta$ yields
\begin{equation}
    d\mathbf{x} = \frac{d\mathbf{x}_0}{d\xi}\, d\xi + \eta\, \frac{d\mathbf{n}}{d\xi}\, d\xi + \mathbf{n}\, d\eta.
\end{equation}
Since the derivative of the boundary position is the tangent vector $\frac{d\mathbf{x}_0}{d\xi} = \mathbf{t}$ and using 
$\frac{d\mathbf{n}}{d\xi} = \frac{1}{R}\, \mathbf{t}$, 
we obtain
\begin{align}
    d\mathbf{x} &= \left[\mathbf{t} + \frac{\eta}{R}\, \mathbf{t}\right] d\xi + \mathbf{n}\, d\eta \nonumber\\[1mm]
    &= \left(1 + \frac{\eta}{R}\right) \mathbf{t}\, d\xi + \mathbf{n}\, d\eta.
\end{align}
Defining
\begin{equation}
    \varrho = 1 + \frac{\eta}{R},
\end{equation}
this can be written as
\begin{equation}
    d\mathbf{x} = \varrho\, \mathbf{t}\, d\xi + \mathbf{n}\, d\eta.
\end{equation}

\begin{equation}
    d\left[\begin{matrix} x \\ y \end{matrix}\right]=\left[\begin{matrix} \varrho t_x & -t_y \\ \varrho  t_y & t_x \end{matrix}\right]d\left[\begin{matrix}\xi \\ \eta\end{matrix}\right]
\end{equation}
The scaling factor
\[
\varrho = 1 + \frac{\eta}{R}
\]
accounts for the curvature of the boundary. At a distance \(\eta\) away from the boundary in the normal direction, the effective tangential length changes because the curves parallel to the boundary have different radii. At the boundary (i.e., when \(\eta = 0\)), we have \(\varrho = 1\) and the mapping is simply given by the tangent and normal directions. However, away from the boundary, the scaling factor \(\varrho\) modifies the \(d\xi\) component accordingly.\\
Inverting to get the derivatives of the local coordinates with respect to the global coordinates yields
\begin{equation}
    \left[\begin{matrix}\label{eq:metrics_last}
     \dfrac{\partial \xi}{\partial x}&
    \dfrac{\partial \xi}{\partial y}\\
    \vspace{0pt}\\
    \dfrac{\partial \eta}{\partial x}&
    \dfrac{\partial \eta}{\partial y}
    \end{matrix}\right]=
    \left[\begin{matrix}
     \dfrac{t_x}{\varrho}  & \dfrac{t_y}{\varrho}\\    
     \vspace{0pt}\\
     -t_y & t_x
    \end{matrix}\right]
\end{equation}
Differentiating each entry in \ref{eq:metrics_last} to get the second derivatives of $\xi$ and $\eta$ with respect to $x$ and $y$ yields
\begin{eqnarray}
    \dfrac{\partial^2\xi}{\partial x^2}&=&\dfrac{\partial}{\partial\xi}\left(\dfrac{\partial \xi}{\partial x}\right)\dfrac{\partial \xi}{\partial x}+\dfrac{\partial}{\partial\eta}\left(\dfrac{\partial \xi}{\partial x}\right)\dfrac{\partial \eta}{\partial x}
    =-\dfrac{\varrho'}{\varrho^3}t_x^2+\dfrac{2}{R\varrho^2}t_x t_y\label{eq:d2xidx2}\\
    \dfrac{\partial^2\xi}{\partial y^2}&=&\dfrac{\partial}{\partial\xi}\left(\dfrac{\partial \xi}{\partial y}\right)\dfrac{\partial \xi}{\partial y}+\dfrac{\partial}{\partial\eta}\left(\dfrac{\partial \xi}{\partial y}\right)\dfrac{\partial \eta}{\partial y}
    =-\dfrac{\varrho'}{\varrho^3}t_y^2-\dfrac{2}{R\varrho^2}t_x t_y\\
    \dfrac{\partial^2\eta}{\partial x^2}&=&\dfrac{\partial}{\partial\xi}\left(\dfrac{\partial \eta}{\partial x}\right)\dfrac{\partial \xi}{\partial x}+\dfrac{\partial}{\partial\eta}\left(\dfrac{\partial \eta}{\partial x}\right)\dfrac{\partial \eta}{\partial x}=\dfrac{1}{R\varrho}t_x^2\\
    \dfrac{\partial^2\eta}{\partial y^2}&=&\dfrac{\partial}{\partial\xi}\left(\dfrac{\partial \eta}{\partial y}\right)\dfrac{\partial \xi}{\partial y}+\dfrac{\partial}{\partial\eta}\left(\dfrac{\partial \eta}{\partial y}\right)\dfrac{\partial \eta}{\partial y}=\dfrac{1}{R\varrho}t_y^2\label{eq:d2etady2}
\end{eqnarray}
with $\varrho'$ denoting $\dfrac{d\varrho}{d\xi}$. Combining \ref{eq:d2xidx2}-\ref{eq:d2etady2} yields
\begin{eqnarray}  
\frac{\partial ^2 \xi }{\partial x^2}+ \frac{\partial ^2 \xi }{\partial y^2} =\mathcal{L}\left(\xi\right) &=& \dfrac{\varrho'}{\varrho^{3}}=\dfrac{\eta R'}{R^2\varrho^3}
\\
\frac{\partial ^2 \eta }{\partial x^2}+ \frac{\partial ^2 \eta }{\partial y^2}= \mathcal{L}\left(\eta\right) &=& \dfrac{1}{R\varrho}
\end{eqnarray}
Finally, using the chain rule for the Laplacian operator
\begin{eqnarray} 
\mathcal{L}\left(T\right)&=&\mathcal{L}\left(\xi\right) \dfrac{\partial T}{\partial \xi}+\mathcal{L}\left(\eta\right) \dfrac{\partial T}{\partial \eta}\nonumber+ 2\left(\dfrac{\partial \xi}{\partial x}\dfrac{\partial \eta}{\partial x}+\dfrac{\partial \xi}{\partial y}\dfrac{\partial \eta}{\partial y}\right)\dfrac{\partial ^2 T}{\partial \xi \partial \eta}\nonumber  
+\left(\left(\dfrac{\partial \xi}{\partial x}\right)^2+\left(\dfrac{\partial\xi}{\partial y}\right)^2\right)\dfrac{\partial^2 T}{\partial \xi^2}\\
&+&\left(\left(\dfrac{\partial \eta}{\partial x} \right)^2+\left(\dfrac{\partial \eta}{\partial y} \right)^2\right)\dfrac{\partial^2 T}{\partial \eta^2}\nonumber 
\end{eqnarray}
 yields
\begin{equation}
    \mathcal{L}\left(T\right)=\dfrac{1}{\varrho^2}\dfrac{\partial^2 T}{\partial \xi^2}+\dfrac{\partial^2 T}{\partial \eta^2}+\dfrac{1}{R\varrho}\dfrac{\partial T}{\partial \eta}
    +\dfrac{\eta R'}{R^2\varrho^3}\dfrac{\partial T}{\partial \xi}
\end{equation}
At small distances from a smooth surface $\eta \approx 0$, $\varrho\approx 1$ and $\eta R' \approx 0$, so that
\begin{equation} \label{eq:twodlaplacian}
    \mathcal{L}\left(T\right)=\dfrac{\partial^2 T}{\partial \xi^2}+\dfrac{\partial^2 T}{\partial \eta^2}+\dfrac{1}{R}\dfrac{\partial T}{\partial \eta}
\end{equation}
For a cylindrical surface of radius $R$, $\eta= r-R$,
$R\varrho=r$, $\xi=R\theta$, $R'=0$ hence the equation above reduces to the
familiar Laplacian in cylindrical coordinates
\begin{equation}
    \mathcal{L}\left(T\right)=\dfrac{1}{r^2}\dfrac{\partial^2 T}{\partial \theta^2}+\dfrac{\partial^2 T}{\partial r^2}+\dfrac{1}{r}\dfrac{\partial T}{\partial r}
\end{equation}
\subsection{Three  dimensions space}
\input{thinthreed.tex}

%% file: thinthreed.tex
\subsubsection*{Geodesic coordinates}
\begin{figure}
    \centering
    \includegraphics[width=0.4\textwidth]{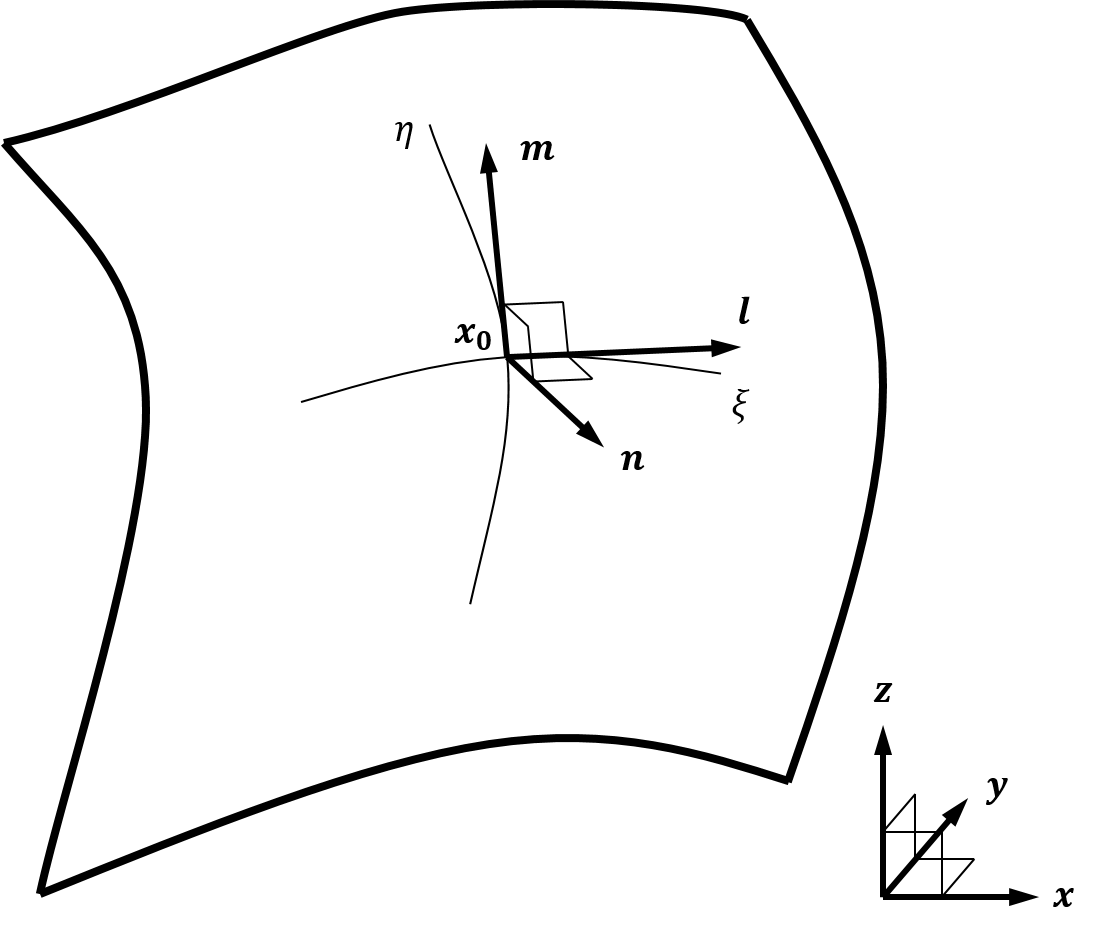}
    \caption{Geodesic coordinates on a surface}
    \label{fig:geodesics}
\end{figure}
For the three-dimensional case, a pair of geodesic coordinates $\xi$,$\eta$ is introduced on the surface so that a point $\mathbf{x}$ in a thin layer adjacent to the surface may be mapped as follows:
\begin{align}
    {\bf x}&={\bf x}_0\left(\xi,\eta\right)+ {\bf n}\zeta
\end{align}
On the surface, the tangent vectors are defined as:
\begin{equation}
    {\bf l}=\dfrac{\partial {\bf x}_0}{\partial \xi}\qquad {\bf m}=\dfrac{\partial {\bf x}_0}{\partial \eta}
\end{equation}
with ${\bf l}$,${\bf m}$ and ${\bf n}$ mutually orthogonal and $\left|{\bf l}\right|=\left|{\bf m}\right|=\left|{\bf n}\right|=1$ (See Figure \ref{fig:geodesics}).\\
Because of the normalization of the vectors 
\begin{align}
   {\bf l}\cdot\left(\dfrac{\partial {\bf l}}{\partial \xi}d\xi+\dfrac{\partial {\bf l}}{\partial \eta}d\eta\right)&=0\label{eq:lconstantlength}\\
 {\bf m}\cdot\left(\dfrac{\partial {\bf m}}{\partial \xi}d\xi+\dfrac{\partial {\bf m}}{\partial \eta}d\eta\right)&=0\label{eq:nconstantlength}
\end{align}
For equations \ref{eq:lconstantlength},\ref{eq:nconstantlength} to hold for any value of $d\xi$ and $d\eta$ the following must hold
\begin{align}
   {\bf l}\cdot\dfrac{\partial {\bf l}}{\partial \xi} &= {\bf l}\cdot\dfrac{\partial {\bf l}}{\partial\eta}= 0\label{eq:lorthogonal}\\
 {\bf m}\cdot\dfrac{\partial {\bf m}}{\partial \xi}&={\bf m}\cdot\dfrac{\partial {\bf m}}{\partial \eta}= 0\label{eq:northogonal}
\end{align}
By virtue of equations \ref{eq:lorthogonal},\ref{eq:northogonal}, the following representations are possible for the derivatives of ${\bf l}$ and ${\bf m}$:
\begin{align}
   \dfrac{\partial {\bf l}}{\partial \xi}&=\alpha {\bf m}+\beta {\bf n}\label{eq:dldxi}\\
   \dfrac{\partial {\bf l}}{\partial \eta}&=\delta {\bf m}+\epsilon{\bf n}\label{eq:dldeta}\\
   \dfrac{\partial {\bf m}}{\partial \xi}&=\phi {\bf l}+\chi {\bf n}\label{eq:dmdxi}\\
   \dfrac{\partial {\bf m}}{\partial \eta}&=\psi {\bf l}+\omega{\bf n}\label{eq:dmdeta}
\end{align}
The scalars $\alpha$, $\beta$, $\delta$, $\epsilon$, $\phi$, $\chi$, $\psi$ and $\omega$ describe how the local basis vectors change with respect to the geodesic coordinates.\\
Similar conditions are obeyed by ${\bf n}$ and its derivatives. By comparing equations \ref{eq:dldeta} and \ref{eq:dmdxi} one finds
\begin{equation}
   \dfrac{\partial {\bf l}}{\partial \eta}=\dfrac{\partial {\bf m}}{\partial \xi}=\dfrac{\partial^2 {\bf x_0}}{\partial \xi \partial \eta}
\end{equation}
\begin{equation}
   \delta {\bf m}+\epsilon{\bf n}=\phi {\bf l}+\chi {\bf n}
\end{equation}
which implies $\delta=\phi=0$ and $\epsilon=\chi$. \\
Furthermore :
\begin{equation}\label{eq:useorthogonality}
    \dfrac{\partial }{\partial \xi}\left({\bf l}\cdot{\bf m}\right)=
   {\bf l}\cdot \dfrac{{\bf m}}{\partial \xi}+\dfrac{\partial {\bf l}}{\partial \xi}\cdot{\bf m}=0
\end{equation}
Substituting equations \ref{eq:dldxi} and \ref{eq:dldeta} into \ref{eq:useorthogonality} yields $\alpha=0$. Similarly, combining orthogonality with equations \ref{eq:dmdxi} and \ref{eq:dmdeta} yields $\psi=0$. The derivatives of the base vectors ${\bf l}$ and ${\bf m}$ are therefore:
\begin{align}
   \dfrac{\partial {\bf l}}{\partial \xi}&=\beta {\bf n}\label{eq:dldxin}\\
   \dfrac{\partial {\bf l}}{\partial \eta}&=\epsilon{\bf n}\\
   \dfrac{\partial {\bf m}}{\partial \xi}&=\epsilon{\bf n}\\
   \dfrac{\partial {\bf m}}{\partial \eta}&=\omega{\bf n}\label{eq:dmdetan}
\end{align}
The functions $\beta$, $\epsilon$ and $\omega$ are the components of the curvature tensor. The selection $\epsilon=0$ identifies the geodesic coordinates $\xi$,$\eta$ as the unique pair of coordinates aligned with the principal directions of the curvature tensor.
\subsubsection*{Heat conduction in a thin surface layer}
\noindent Differentiating the position vector
\begin{equation}
    d{\bf x}={\bf l}d\xi + {\bf m}d\eta+ {\bf n}d\zeta+ \zeta\left(  \dfrac{\partial {\bf n}}{\partial \xi}d\xi+\dfrac{\partial {\bf n}}{\partial \eta}d\eta\right)
\end{equation}
Taking scalar products by ${\bf l}$, ${\bf m}$ and ${\bf n}$ yields
\begin{align}
{\bf l}\cdot d{\bf x}&=d\xi+ \zeta  {\bf l}\cdot\left( \dfrac{\partial {\bf n}}{\partial \xi}d\xi +\dfrac{\partial {\bf n}}{\partial \eta}d\eta\right)\\
{\bf m}\cdot d{\bf x}&=d\eta+ \zeta  {\bf m}\cdot\left( \dfrac{\partial {\bf n}}{\partial \xi}d\xi +\dfrac{\partial {\bf n}}{\partial \eta}d\eta\right)\\
{\bf n}\cdot d{\bf x}&=d\zeta \label{eq:ndx}
\end{align}
 Note that:
\begin{equation}
{\bf l}\cdot \dfrac{\partial{\bf n}}{\partial \xi}+{\bf n}\cdot \dfrac{\partial{\bf l}}{\partial \xi}=0
\end{equation}
therefore:
\begin{equation}
{\bf l}\cdot \dfrac{\partial{\bf n}}{\partial \xi}=-\beta
\end{equation}
by virtue of \ref{eq:dldxin}.\\ Similarly:
\begin{align}
{\bf l}\cdot \dfrac{\partial {\bf n}}{\partial \eta}=-\epsilon \qquad
{\bf m}\cdot \dfrac{\partial {\bf n}}{\partial \xi }=-\epsilon \qquad
{\bf m}\cdot \dfrac{\partial {\bf n}}{\partial \eta}=-\omega   
\end{align}
hence, with $\epsilon=0$
\begin{align}
   {\bf l}\cdot d{\bf x} &= \left(1-\zeta \beta\right)d\xi\\
   {\bf m}\cdot d{\bf x} &= \left(1-\zeta \omega\right)d\eta
\end{align}
The metrics are 
\begin{equation}\label{eq:metricsthreed}
    \left[\begin{matrix}
    \dfrac{\partial \xi}{\partial x}&
    \dfrac{\partial \xi}{\partial y}&  
    \dfrac{\partial \xi}{\partial z} \\ 
        \vspace{0pt}\\
    \dfrac{\partial \eta}{\partial x}& 
    \dfrac{\partial \eta}{\partial y}&
    \dfrac{\partial \eta}{\partial z}\\
            \vspace{0pt}\\
    \dfrac{\partial \zeta}{\partial x}& 
    \dfrac{\partial \zeta}{\partial y}& 
    \dfrac{\partial \zeta}{\partial z}
    \end{matrix}\right]=\left[\begin{matrix}
    \dfrac{l_x }{\varrho_\xi }&
    \dfrac{l_y }{\varrho_\xi }&
    \dfrac{l_z }{\varrho_\xi }\\
    \vspace{0pt}\\
    \dfrac{ m_x}{\varrho_\eta} &
    \dfrac{ m_y}{\varrho_\eta} &
    \dfrac{ m_z}{\varrho_\eta} \\
    \vspace{0pt}\\
     n_x &
     n_y &
     n_z
    \end{matrix}\right]
    \end{equation}
where $\varrho_\xi$ and $\varrho_\eta$ are defined as
\begin{align}
\varrho_\xi = 1-\zeta \beta \qquad
\varrho_\eta = 1-\zeta \omega
\end{align}
The Laplacian of a scalar field in the coordinates $\xi$, $\eta$, $\zeta$ requires the Laplacians of $\xi$, $\eta$ and $\zeta$ with respect to $x$,$y$,$z$. These can be found differentiating each entry in equation 
\ref{eq:metricsthreed}. As an example, for the coordinate $\xi$:
\begin{eqnarray}
\dfrac{\partial^2\xi}{\partial x^2}&=&\left(
\dfrac{\partial   \xi}{\partial x}\dfrac{\partial}{\partial   \xi}+
\dfrac{\partial  \eta}{\partial x}\dfrac{\partial}{\partial  \eta}+
\dfrac{\partial \zeta}{\partial x}\dfrac{\partial}{\partial \zeta}\right)\dfrac{\partial \xi}{\partial x}\\
 \dfrac{\partial^2\xi}{\partial y^2}&=&\left(
\dfrac{\partial   \xi}{\partial y}\dfrac{\partial}{\partial   \xi}+
\dfrac{\partial  \eta}{\partial y}\dfrac{\partial}{\partial  \eta}+
\dfrac{\partial \zeta}{\partial y}\dfrac{\partial}{\partial \zeta}\right)\dfrac{\partial \xi}{\partial y}\\
 \dfrac{\partial^2\xi}{\partial z^2}&=&\left(
\dfrac{\partial   \xi}{\partial z}\dfrac{\partial}{\partial   \xi}+
\dfrac{\partial  \eta}{\partial z}\dfrac{\partial}{\partial  \eta}+
\dfrac{\partial \zeta}{\partial z}\dfrac{\partial}{\partial \zeta}\right)\dfrac{\partial \xi}{\partial z}
\end{eqnarray}
Substituting 
\begin{align}
\dfrac{\partial^2\xi}{\partial x^2}&=
\dfrac{l_x}{\varrho_\xi^2           }\left(\dfrac{\partial l_x}{\partial   \xi}- \dfrac{l_x}{\varrho_\xi}\dfrac{\partial \varrho_\xi}{\partial   \xi} \right) +
\dfrac{m_x}{\varrho_\eta \varrho_\xi}\left(\dfrac{\partial l_x}{\partial  \eta}- \dfrac{l_x}{\varrho_\xi}\dfrac{\partial \varrho_\xi}{\partial  \eta} \right) \label{eq:d2xidx2threed}\\
 \dfrac{\partial^2\xi}{\partial y^2}&=  
\dfrac{l_y}{\varrho_\xi^2           }\left(\dfrac{\partial l_y}{\partial   \xi}- \dfrac{l_y}{\varrho_\xi}\dfrac{\partial \varrho_\xi}{\partial   \xi} \right) +
\dfrac{m_y}{\varrho_\eta \varrho_\xi}\left(\dfrac{\partial l_y}{\partial  \eta}- \dfrac{l_y}{\varrho_\xi}\dfrac{\partial \varrho_\xi}{\partial  \eta} \right) \\
 \dfrac{\partial^2\xi}{\partial z^2}&= 
\dfrac{l_z}{\varrho_\xi^2           }\left(\dfrac{\partial l_z}{\partial   \xi}- \dfrac{l_z}{\varrho_\xi}\dfrac{\partial \varrho_\xi}{\partial   \xi} \right) +
\dfrac{m_z}{\varrho_\eta \varrho_\xi}\left(\dfrac{\partial l_z}{\partial  \eta}- \dfrac{l_z}{\varrho_\xi}\dfrac{\partial \varrho_\xi}{\partial  \eta} \right) \label{eq:d2xidz2threed}
\end{align}
Adding equations \ref{eq:d2xidx2threed}-\ref{eq:d2xidz2threed} and performing similar operations on the variables $\eta$ and $\zeta$, finally yields
\begin{align}
   \mathcal{L}\left(\xi\right)&=
   -\dfrac{1}{\varrho_\xi^3}\dfrac{\partial\varrho_\xi}{\partial\xi}= \dfrac{\zeta \beta'}{\varrho_\xi^3}\\
   \mathcal{L}\left(\eta\right)&=
   -\dfrac{1}{\varrho_\eta^3}\dfrac{\partial\varrho_\eta}{\partial\eta}=\dfrac{\zeta \omega'}{\varrho_\eta^3}\\
   \mathcal{L}\left(\zeta\right)&=-\left(\dfrac{\beta}{\varrho_\xi}+\dfrac{\omega}{\varrho_\eta}\right)
\end{align}
The Laplacian of a scalar quantity $T$ with respect to the coordinates $\xi$,$\eta$,$\zeta$, therefore is
\begin{align}
   \mathcal{L}\left(T\right)&=
   \mathcal{L}\left(\xi\right)\dfrac{\partial T}{\partial \xi}+    \mathcal{L}\left(\eta\right)\dfrac{\partial T}{\partial \eta}+  \mathcal{L}\left(\zeta\right)\dfrac{\partial T}{\partial \zeta} 
 + \left(\left(\dfrac{\partial \xi}{\partial x}\right)^2+\left(\dfrac{\partial \xi}{\partial y}\right)^2+\left(\dfrac{\partial \xi}{\partial z}\right)^2\right)\dfrac{\partial^2 T}{\partial \xi^2} \nonumber \\
+ &\left(\left(\dfrac{\partial \eta}{\partial x}\right)^2+\left(\dfrac{\partial \eta}{\partial y}\right)^2+\left(\dfrac{\partial \eta}{\partial z}\right)^2\right)\dfrac{\partial^2 T}{\partial \eta^2} 
+ \left(\left(\dfrac{\partial \zeta}{\partial x}\right)^2+\left(\dfrac{\partial \zeta}{\partial y}\right)^2+\left(\dfrac{\partial \zeta}{\partial z}\right)^2\right)\dfrac{\partial^2 T}{\partial \zeta^2} \nonumber \\
+ &\left(
 \dfrac{\partial \xi}{\partial x}\dfrac{\partial \eta}{\partial x}+
 \dfrac{\partial \xi}{\partial y}\dfrac{\partial \eta}{\partial y}+
 \dfrac{\partial \xi}{\partial z}\dfrac{\partial \eta}{\partial z}\right)
 \dfrac{\partial^2 T}{\partial \xi \partial \eta}+ \left(
 \dfrac{\partial \xi}{\partial x}\dfrac{\partial \zeta}{\partial x}+
 \dfrac{\partial \xi}{\partial y}\dfrac{\partial \zeta}{\partial y}+
 \dfrac{\partial \xi}{\partial z}\dfrac{\partial \zeta}{\partial z}\right)
 \dfrac{\partial^2 T}{\partial \xi \partial \zeta}\nonumber\\&+
  \left(
 \dfrac{\partial \eta}{\partial x}\dfrac{\partial \zeta}{\partial x}+
 \dfrac{\partial \eta}{\partial y}\dfrac{\partial \zeta}{\partial y}+
 \dfrac{\partial \eta}{\partial z}\dfrac{\partial \zeta}{\partial z}\right)
 \dfrac{\partial^2 T}{\partial \eta \partial \zeta}
\end{align}
\begin{align}
    \mathcal{L}\left(T\right)=
    \dfrac{1}{\varrho_\xi^2}\dfrac{\partial^2 T}{\partial \xi^2}+
    \dfrac{1}{\varrho_\eta^2}\dfrac{\partial^2 T}{\partial \eta^2}+ \dfrac{\partial^2 T}{\partial \zeta^2}+
    \dfrac{\zeta \beta'}{\varrho_\xi^3}\dfrac{\partial T}{\partial \xi}+
    \dfrac{\zeta \omega'}{\varrho_\eta^3}\dfrac{\partial T}{\partial \eta}-
    \left(\dfrac{\beta}{\varrho_\xi}+\dfrac{\omega}{\varrho_\eta}\right)\dfrac{\partial T}{\partial \zeta}
\end{align}
At small distances from a smooth surface $\varrho_\xi \approx 1$, $\varrho_\eta \approx 1$ and $\zeta \beta' = \zeta \omega'\approx 0$ so that
\begin{equation}\label{eq:threedlaplacian}
    \mathcal{L}\left(T\right)= 
    \dfrac{\partial^2 T}{\partial \xi^2}+
    \dfrac{\partial^2 T}{\partial \eta^2}+ \dfrac{\partial^2 T}{\partial \zeta^2}+\dfrac{1}{R}
    \dfrac{\partial T}{\partial \zeta}
\end{equation}
where 
\begin{equation}
    \dfrac{1}{R}=\omega+\beta
\end{equation}
is the local harmonic mean curvature of the surface.